\documentclass[12pt]{article}
\usepackage[utf8]{inputenc}

\usepackage{lipsum}
\usepackage{amsmath}
\usepackage{color}
\usepackage{amsfonts}
\usepackage[mathscr]{euscript}
\usepackage{graphicx}
\usepackage{geometry}
\usepackage{amssymb,epsfig}
\usepackage{subfigure}
\usepackage{comment}
\usepackage{tabularx}
\usepackage{bm}
\usepackage[dvipsnames]{xcolor}
\usepackage{exscale}
\usepackage{amsbsy}
\usepackage{textcomp}
\usepackage{hyperref}
\usepackage{slashed}
\usepackage{authblk}
\usepackage{tabularx}
\usepackage{color}
\usepackage{exscale}
\usepackage{doi}
\usepackage{tensor}
\usepackage{wrapfig}
\usepackage[font=footnotesize,labelfont=bf]{caption}
\usepackage{ulem} 
\usepackage{tikz-feynman}
\usepackage{slashed}
\usepackage{array,multirow}
\usepackage{ascmac}

\usepackage{makecell}

\def\ba#1\ea{\begin{align}#1\end{align}}
\def\bg#1\eg{\begin{gather}#1\end{gather}}
\def\bm#1\em{\begin{multline}#1\end{multline}}
\def\bmd#1\emd{\begin{multlined}#1\end{multlined}}

\newcommand{\be}{\begin{equation}}
	\newcommand{\ee}{\end{equation}}
\newcommand{\bea}{\begin{eqnarray}}
	\newcommand{\eea}{\end{eqnarray}}
\newcommand{\Lag}{\mathcal{L}}
\renewcommand{\d}{\partial}
\newcommand{\bs}{\boldsymbol}
\newcommand{\U}{\mathrm{U}}
\newcommand{\SU}{\mathrm{SU}}
\newcommand{\CS}{\mathrm{CS}}
\newcommand{\Spin}{\mathrm{Spin}}
\newcommand{\SO}{\mathrm{SO}}
\renewcommand{\O}{\mathrm{O}}

\newcommand{\ct}{\mathrm{ct}}
\newcommand{\eff}{\mathrm{eff}}
\newcommand{\cA}{\mathcal{A}}
\newcommand{\matleft}{\left(\begin{array}}
	\newcommand{\matright}{\end{array}\right)}
\newcommand{\Tr}{\operatorname{Tr}}

\newcommand{\sgn}{\operatorname{sgn}}

\usepackage[numbers,sort&compress]{natbib}
\def\simge{
	\mathrel{\rlap{\raise 0.511ex 
			\hbox{$>$}}{\lower 0.511ex \hbox{$\sim$}}}}

\def\simle{
	\mathrel{\rlap{\raise 0.511ex 
			\hbox{$<$}}{\lower 0.511ex \hbox{$\sim$}}}}

\makeatletter
\renewcommand\section{\@startsection {section}{1}{\z@}%
	{-3.5ex \@plus -1ex \@minus -.2ex}
	{2.3ex \@plus.2ex}%
	{\normalfont\large\bfseries}}
\renewcommand\subsection{\@startsection{subsection}{2}{\z@}%
	{-3.25ex\@plus -1ex \@minus -.2ex}%
	{1.5ex \@plus .2ex}%
	{\normalfont\bfseries}}
\renewcommand\subsubsection{\@startsection{subsubsection}{3}{\z@}%
	{-3.25ex\@plus -1ex \@minus -.2ex}%
	{1.5ex \@plus .2ex}%
	{\normalfont\itshape}}
\makeatother

\def\pplogo{\vbox{\kern-\headheight\kern -29pt
		\halign{##&##\hfil\cr&{\ppnumber}\cr\rule{0pt}{2.5ex}&\ppdate\cr}}}
\makeatletter
\def\ps@firstpage{\ps@empty \def\@oddhead{\hss\pplogo}%
	\let\@evenhead\@oddhead 
}
\hypersetup{
	unicode=false,          
	pdftoolbar=true,        
	pdfmenubar=true,        
	pdffitwindow=false,     
	pdfstartview={FitH},    
	pdftitle={non-Abelian anyon SC},    
	pdfauthor={},     
	pdfsubject={Subject},   
	pdfcreator={},   
	pdfproducer={}, 
	pdfkeywords={keyword1} {key2} {key3}, 
	pdfnewwindow=true,      
	colorlinks=true,       
	linkcolor=OliveGreen, 
	citecolor=NavyBlue,        
	filecolor=magneta,      
	urlcolor=cyan           
}

\numberwithin{equation}{section}

\newcommand*\samethanks[1][\value{footnote}]{\footnotemark}

\newcommand\beal{\begin{equation}\begin{aligned}}
		\newcommand\eeal{\end{aligned}\end{equation}}

\begin{document}

\normalem

\setcounter{page}0
\def\ppnumber{\vbox{\baselineskip14pt
}}

\def\ppdate{
} 
\date{}

\title{\Large\bf Coloring in anyon superconductivity}
\author{Umang Mehta, Yuto Nakajima, and Hart Goldman}
\affil{\it\small School of Physics and Astronomy, University of Minnesota, Minneapolis, MN 55455, USA}
\maketitle\thispagestyle{firstpage}

\begin{abstract}
The recently observed signatures of superconductivity proximate to a fractional quantum anomalous Hall (FQAH) state in a twisted MoTe$_2$ bilayer has revitalized interest in quantum phases of matter induced by anyon dynamics. Here we show how a panoply of anyon-driven phases associated with doping the lattice ${\nu=2/3}$ FQAH state can be realized as competing instabilities of a Fermi surface of charge-$e/3$ ``quarks'' coupled to a SU$(3)_{-1}$ Chern-Simons gauge field, which is dual to the more conventional U$(1)_3$ Chern-Simons-Ginzburg-Landau theory of quasiholes. For example, a range of electronic superconductors emerge from \emph{color superconductivity}, under which the Fermi surface experiences a pairing instability mediated by gauge fluctuations. These include SC$\star$ phases -- where superconductivity coexists with topological order -- as well as topological superconductors displaying half-integer chiral central charges when the quarks are weakly paired. One example is a $p+ip$ ``color-valley-locked'' superconductor, a topological analogue of the color superconductor familiar in  quantum chromodynamics. On the other hand, both superconducting and non-Fermi liquid phases can emerge when the quarks form an itinerant ferromagnet, polarizing the Fermi surface to a particular combination of colors. Finally, our framework naturally accommodates the possibility of anyonic bound state formation, allowing access to phases induced by doping anyons of charge $2e/3$ as opposed to $e/3$ within the same model. Our work unifies many earlier proposed anyonic phases as instabilities of a single parent \emph{quark metal} phase, distilling their emergence into a competition between superconductivity and itinerant color ferromagnetism.
\end{abstract}

\pagebreak
\tableofcontents 
\pagebreak 


\section{Introduction}

The advent of fractional quantum anomalous Hall (FQAH) phases in moir\'{e} materials~\cite{Cai2023,Zeng2023,Xu2023,Park2023,Lu2024} has set a new stage for the interplay of fractionalization with electronic correlations. Unlike traditional fractional quantum Hall systems in strong magnetic field, lattice FQAH systems can host dispersing anyons~\cite{Tang2013,Schleith2025,Gonccalves2025,Yan2025,Iyer2026,Wang:2026nsc}. Hence on doping a FQAH system, dynamical anyons can collude to produce a range of correlated phases, running the gamut from superconductors (SCs) to charge orders to (non-)Fermi liquid metals~\cite{Kim2025,Shi2025c,Divic2025,Shi2025a,Zhang2025a,Nakajima2025,Pichler2025,Shi:2025arn,Seo2026,lotric2026,Fan:2026kay,Wang:2026ziw,Shi2026,senthil2026fractionalizedmetalsdopedanyons,han2026orthogonal,zhang2026colorsuperconductorsholonmetals}. The possibility of superconductivity, in particular, has led to a revival of old concepts of ``anyon superconductivity'' dating back to the 1980s~\cite{Laughlin1988,Fetter1989,Chen1989,Lee1989,Fradkin1990}, with contemporary theoretical work elaborating on those early ideas by introducing the modern understanding of lattice symmetry enrichment and filling constraints  essential for application to FQAH systems.

On the experimental front, anyon-driven phases may already have been realized in twisted MoTe$_2$ bilayers~\cite{Xu2025a}. In these systems, signatures of superconductivity have been observed on hole doping from the $\nu=2/3$ FQAH plateau, with an intervening resistivity peak and a normal state carrying a large Hall effect. This physics is broadly consistent with mechanisms for superconductivity based on doping anyons~\cite{Shi2025b,Nosov2026,Fan:2026kay,senthil2026fractionalizedmetalsdopedanyons,han2026orthogonal}, although more conventional mechanisms based on BCS pairing of holes remain possible~\cite{guerci2025,guerci2026topologicalsuperconductivityemergentvortex}. Determining how an anyonic mechanism for superconductivity may be realistically distinguished from an electronic one in  this system remains a key challenge. 

This challenge has deepened recently, with the growing understanding that  numerous \emph{distinct} chiral superconductors are available on doping the $\nu=2/3$ state (labeled by their chiral central charges, $c_-$). Further complicating matters, these superconductors often arise under completely different physical mechanisms depending on one's choice of variables for the anyon fluctuations. Put differently, the family of anyon-driven superconductors does not share an obvious set of unifying normal state properties, reflecting the inherent complexity of anyon dynamics. Indeed, different parton decompositions typically exhibit particular superconducting ground states (or other correlated phases) in mean field theory, with most others invisible. A useful table collecting all of the known options to this point can be found in Ref.~\cite{Shi2026}.

For example, in the earliest proposal for uniform anyon superconductivity on doping the $\nu=2/3$ FQAH state~\cite{Shi2025c}, a novel mechanism was introduced where the charge-$2e/3$ anyon becomes lighter than a pair of isolated charge-$e/3$ anyons. This mechanism closely resembles the proposals from the 1980s: Charge-$2e/3$ anyons view each other's fractional statistics as emergent flux, causing them to fill an integer number of Landau levels in mean field theory. On integrating them out, one finds that the microscopic Cooper pair is liberated and condenses, leading to a $c_-=-2$ chiral superconductor. In notable contrast, contemporary proposals based on charge-$e/3$ anyons leverage mean field theories where fractionally charged composite fermions form a Fermi surface, which undergoes a Cooper instability, producing topological superconductors with half-integer $c_-$~\cite{Fan:2026kay,Wang:2026ziw,Shi2026,senthil2026fractionalizedmetalsdopedanyons,han2026orthogonal}.   One nice feature of these proposals is a naturally available order parameter built from the composite fermion variables, whereas constructing a tangible order parameter for the earlier types of anyon superconductors has been thought to require introduction of extra auxiliary degrees of freedom~\cite{Shi2025a,Seo2026}. Other topological superconductors with half-integer $c_-$ were also seen in parton constructions with U$(2)$ gauge symmetry, with physics closer to the traditional mechanism~\cite{Shi:2025arn,lotric2026}.

In this work, we develop a parent theory unifying the vast majority\footnote{There is one notable exception: The $c_-=3/2$ topological SC found in Ref.~\cite{Shi:2025arn} is not directly accessible in our approach.} of proposed anyon-induced phases at $\nu=2/3+\delta$ -- while also predicting new ones -- enabling us to classify families of ground states produced by different physics using local order parameters. Our strategy leverages \emph{level-rank duality}~\cite{Naculich1990,Naculich1990a,Camperi1990,Aharony2015,Hsin2016}, which relates the traditional U$(1)_3$ Chern-Simons-Ginzburg-Landau (CS-GL) theory of quasiholes~\cite{Zhang1989} to a dual theory of charge-$e/3$ fermions, which we term ``quarks,'' coupled to a SU$(3)_{-1}$ gauge field, ${b_\mu=b_\mu^aT^a}$, with $T^a$ the generators of SU$(3)$. The three colors, $\alpha=\mathrm{cyan},\mathrm{magenta},\mathrm{yellow}$, of quarks, $\chi_{\alpha,I}$, occupy three valleys, ${I=1,2,3}$, in momentum space due to the projective action of lattice translations. Schematically, our parent theory is described by the Lagrangian, 
\begin{align}
\mathcal{L}=\sum_{I=1}^3\mathcal{L}_{\chi_I}[b+A\,\bs{1}_3/3]-\frac{1}{4\pi}\Tr\left[bdb+\frac{2}{3}b\wedge b\wedge b\right]+\dots\,,
\end{align}
where $A_\mu$ is the background electromagnetic (EM) gauge field. Although the gauge group in this presentation is non-abelian, the unit Chern-Simons level means it nonetheless constitutes a theory of \emph{abelian} anyons with $-\pi/3$ statistics. 

Doping electric charge via a chemical potential, $A_t=\mu$, causes the quarks to fill a Fermi sea in mean field theory, with the non-abelian gauge fluctuations destroying quasiparticles near the Fermi surface. We refer to this state as the \emph{quark metal}. An interplay of gauge fluctuations with electrostatic interactions can then produce  instabilities of the quark Fermi surface, from which anyon-driven phases emerge. The valley-preserving instabilities to expect are those familiar in non-Fermi liquids: BCS superconductivity and itinerant ferromagnetism, with the latter being in the gauge SU$(3)$ \emph{color} space. Although the quark metal may in principle be a stable phase at zero temperature, given that SU$(3)$ gauge fluctuations are strong and are attractive at tree level, we find it more natural to view the quark metal as the theory of a finite temperature parent state out of which the breadth of possible zero temperature anyon-driven phases can emerge. Nevertheless, signatures of a novel metal of charge-$e/3$ quasiparticles could be visible experimentally in shot noise or tunneling measurements, and we note that the expected transport features of such a state (up to possible non-Fermi liquid corrections) are generally consistent with the normal state measurements in Ref.~\cite{Xu2025a}. The possibility of the quark Fermi surface, as well as certain descendant phases, were also recognized in Refs.~\cite{lotric2026,Wang:2026ziw,zhang2026colorsuperconductorsholonmetals}, with Ref.~\cite{Wang:2026ziw} taking a different approach to translation symmetry enrichment.

\begin{figure}[t]
    \centering
    \includegraphics[width=0.6\linewidth]{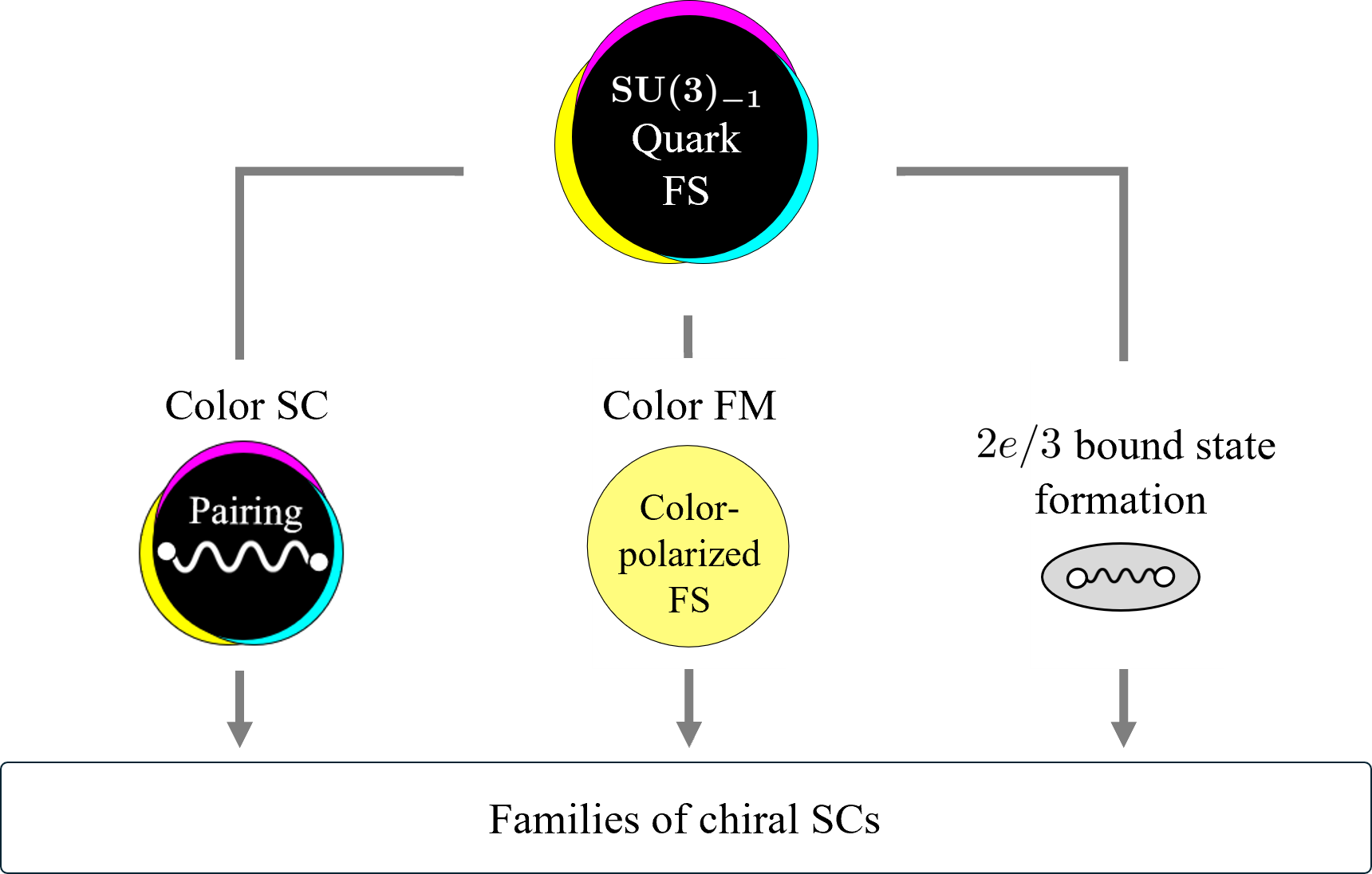}
    \caption{The ``black" (an unpolarized mixture of cyan, magenta, and yellow) SU$(3)_{-1}$ quark Fermi surface as the parent of $\nu=2/3+\delta$ anyon-induced phases. The three possibilities we study are
    (1) \textit{color superconductivity:} The quarks form pairs on the black Fermi surface; (2) \textit{Color ferromagnetism}: The quarks are polarized to one or more colors spontaneously; (3) $2e/3$ \textit{bound state formation} -- the quarks form charge-$2e/3$ bound states, which can exhibit dynamics of their own. Each possibility leads to different (but overlapping) families of superconducting phases, some of which have Majorana edge modes (topological SC) or coexist with a background topological order (SC$\star$).}
    \label{fig:summary}
\end{figure}

Starting from the quark metal, we identify three primary possibilities for anyon-driven phases (Figure~\ref{fig:summary}), each of which can lead to superconductivity of the underlying holes:
\begin{enumerate}
\item \emph{Color superconductors}, ${\Delta_{\alpha\beta}=\langle\chi^\dagger_\alpha\chi^\dagger_\beta\rangle\neq 0}$. In these cases, a Cooper instability Higgses the gauge group, in a manner analogous to the onset of color superconductivity in quantum chromodynamics. These phases are invisible to the abelian CS-GL dual, and until now a systematic treatment of them has been absent.
\item \emph{Color ferromagnets}, ${C^a=\langle \chi^\dagger T^a\chi\rangle\neq 0}$. Here color polarization of the Fermi surface Higgses the gauge group. As a result, the quarks generally (but do not always) experience a net color flux and fill Landau levels. Surprisingly, we will see that phases accessible to doping the abelian dual theory (or other presentations using abelian flux attachment) can all be embedded as color ferromagnets. These comprise the majority of proposed phases of itinerant charge-$e/3$ anyons.
\item \emph{Bound state formation.} Another possibility is for quarks to locally bind to produce anyons of higher charge. These composites may have their own dynamics -- they may form color ferromagnets themselves -- and they correspond to the scenarios for charge-$2e/3$ anyons becoming ``lighter'' than charge-$e/3$ anyons.
\end{enumerate}
 Because of the breadth of possible ground states, we choose to focus primarily on superconducting orders possessing a uniform component, which may or may not coexist with topological order to form a SC$\star$ phase.  Under each mechanism, superconductivity may arise directly from the quark dynamics, or through the dynamics of an intermediate compressible state such as a secondary composite Fermi liquid~\cite{Shi2025c,Zhang2025a,Fan:2026kay} or an orthogonal metal~\cite{Shi2026,senthil2026fractionalizedmetalsdopedanyons,han2026orthogonal}.

One especially notable superconductor arises from a mechanism we term ``color-valley-locking'' (CVL), in analogy with the color-flavor-locked superconductor studied in quantum chromodynamics~\cite{Son1998,Rajagopal2001,AlfordRMP}. On pairing different colors between the three valleys, the SU$(3)$ gauge group is Higgsed entirely, resulting in charge-$2e$ superconductivity of electrons. The smallest quark angular momentum with attraction in this channel corresponds to $p+ip$. Integrating out the quarks and combining with the FQAH background leads to a final chiral central charge of $c_-=5/2$. Hence the CVL state corresponds to a topological superconductor of the underlying electric charges and hosts Majorana zero modes! Furthermore, in the presence of an approximate SU$(3)$ valley symmetry, the quark metal  theory is invariant, and the CVL order parameter respects a global SU$(3)$ mutually rotating valley and color.

The quark metal parent state provides a unifying tool for mapping the landscape of anyon-driven phases and clarifying their physics, as well as a potential starting point for understanding universal aspects of anyon dynamics that are independent of which from among the great many zero temperature phases is ultimately chosen by energetics. In particular, it could be leveraged toward analytic studies of thermodynamic and transport properties for anyon gases valid at sufficiently high temperature and/or doping regimes (under weak coupling approximations), or potentially even lower doping regimes through strong coupling techniques like the 't Hooft large-$N$ limit.

We proceed as follows. In Section~\ref{sec:level_rank}, we introduce level-rank duality and motivate the SU$(k)_1$ quark metal for general $\nu=1/k$ Laughlin states. In Section~\ref{sec:nu=2/3_doping}, we present our analysis of the color superconducting, ferromagnetic, and bound state-driven phases.

\emph{Note added.} While this work was being prepared, independent but related constructions of topological superconductors with half-integer chiral central charge appeared based on related theories of charge-$e/3$ anyon gases~\cite{Fan:2026kay,Wang:2026ziw,senthil2026fractionalizedmetalsdopedanyons,han2026orthogonal}. An overlapping manuscript~\cite{zhang2026colorsuperconductorsholonmetals} with  complementary results also appeared during the same week as this work. The constructions in Refs.~\cite{Fan:2026kay,senthil2026fractionalizedmetalsdopedanyons,han2026orthogonal} can be embedded as color ferromagnetic descendants of our parent model. In the case of Ref.~\cite{Wang:2026ziw}, an alternate implementation of lattice translation symmetry was used in which only a single quark valley appears, and this inspired us to consider color-symmetric pairing channels. We have written the Introduction to reflect the contemporary literature, and we reference these works in the main text.


\section{Motivating the quark metal}\label{sec:level_rank}

To study the dynamics of anyons in lattice FQAH systems, we leverage the formalism of \emph{level-rank duality}~\cite{Naculich1990,Naculich1990a,Camperi1990}. Level-rank dualities equate different topological quantum field theories (TQFTs), which are the long-wavelength descriptions of FQAH phases. In abelian TQFTs, which are commonly expressed as U$(1)_{-k}$ Chern-Simons gauge theories, they allow one to encode the fractional statistics and fusion properties of the anyons into the ``color'' degree of freedom of a SU$(k)_{1}$ gauge theory. The purpose of this section is to first develop concrete intuition for this superficially abstract mapping in the context of Chern-Simons theories coupled to matter, which describe itinerant anyons. We then leverage this to develop the mean field theory of the quark Fermi surface which forms the basis of our analysis.

\subsection{Flux attachment and its discontents}

We begin by revisiting the traditional effective field theory description of a $\nu=1/k$ Laughlin state, with an eye toward anyon superfluidity. Ignoring the implementation of lattice translation symmetry for now, the fluctuations of charge-$e/k$ anyons can be described by coupling a non-relativistic scalar field, $\phi$, to a fluctuating U$(1)$ Chern-Simons gauge field, $a_\mu$~\cite{Zhang1989},
\begin{align}
\label{eq: abelian CSGL}
\mathcal{L}&=\Lag_\phi + \Lag_{\U(1)_{-k}}\, ,\\
\label{eq:abelian phi lagrangian}
\Lag_\phi[a] &= \phi^\dagger \left[ i D^a_t + \frac{1}{2m_\phi}(D^a_i)^2 \right] \phi-V[\phi]\, ,\\
\Lag_{\U(1)_{-k}}[a;A] &= - \frac{k}{4\pi} ada + \frac{1}{2\pi} adA\, .
\end{align}
Here $A_\mu$ is the background electromagnetic (EM) connection, $i=x,y$ are spatial indices, and we use the notation, $AdB=\varepsilon^{\mu\nu\lambda} A_\mu\partial_\nu B_\lambda$. ${D_\mu^a=\partial_\mu-ia_\mu}$ denotes the covariant derivative with respect to $a_\mu$. 
For $k$ odd (even), this theory describes anyons in a fractional quantum Hall phase of microscopic fermions (bosons), and $A$ is to be regarded as a background $\Spin_c$ (U$(1)$) connection.

To see that $\phi$ describes the anyon fluctuations, one may inspect the equation of motion for $a_t$ (setting $\varepsilon_{ij}\partial_iA_j=0$), 
\begin{equation}
    \phi^\dagger\phi = k \frac{\varepsilon_{ij}\partial_i a_j}{2\pi}\ , 
    \label{eq: abelian Gauss general}
\end{equation}
from which it is evident that a single quantum of $\phi$ activates a $2\pi/k$ flux in $a$. Other anyons in the vicinity will thus experience this as an abelian Aharanov-Bohm phase, resulting in anyons' fractional exchange statistics, or braiding. From the standpoint of this abelian model, the basic physics of anyons arises because they encounter one another simultaneously as fractional charge and flux.

How should we think of composites of anyons or, for that matter, local electrons or bosons starting from this field theory description? Indeed, information about the \emph{fusion} of anyons is not apparent at the classical field theory level in the same manner as fractional charge and braiding statistics. Most glaringly, a state with a local electron or boson is not simply described by acting on the vacuum with $k$ quanta of $\phi$, $(\phi^\dagger)^k$, as this object is not even gauge invariant. To rectify this, we must introduce the local monopole operator, $\mathcal{M}_a$, which carries charge $+1$ under $A_\mu$ and $-k$ under $a_\mu$ due to the Chern-Simons term. A stack of $k$ quanta of $\phi$ can then be rendered gauge invariant by adding a monopole, so that the local electron or boson can be represented as
\begin{equation}
    c^\dagger = (\phi^\dagger)^k \mathcal{M}_a\, .
    \label{eq: monopole screening}
\end{equation}
One cannot over-state the physical importance of this monopole insertion: When a local charged particle is inserted into the system, it splits into $k$ individual $\phi$ quanta \emph{and} a monopole, whose role is to \emph{screen} the emergent gauge charge. This screening guarantees that local particles do not braid with anyons and do not experience them as flux. However, because of the monopole's status as an inherently quantum object, which cannot be expressed as a local function of $\phi$ and $a_\mu$, this physics is invisible to any mean field analysis of the abelian effective field theory~\eqref{eq: abelian CSGL}. 

Now we imagine doping itinerant anyons into the system by turning on a chemical potential, $A_t=\mu$, leading to a finite gauge flux, ${\langle\varepsilon_{ij}\partial_i a_j\rangle\neq 0}$. From Gauss' Law, Eq.~\eqref{eq: abelian Gauss general}, exactly $k$ anyons will be nucleated per flux quantum, placing them at integer filling, $\nu_\phi=k$. When $k$ is odd and the microscopic degrees of freedom are fermions, the $\phi$ degrees of freedom cannot immediately establish a gapped phase. Rather, their immediate fate is to form a compressible state -- either a secondary composite Fermi liquid (CFL)~\cite{Shi2025c,Zhang2025a,Fan:2026kay} or a $\mathbb{Z}_3$ orthogonal metal~\cite{Shi2026,senthil2026fractionalizedmetalsdopedanyons,han2026orthogonal}. The secondary CFL may in turn undergo a pairing instability and lead to electronic superconductivity or other intriguing gapped and gapless phases. The orthogonal metal, in contrast, is in principle stable at zero temperature but may exhibit superconductivity induced through repulsive electrostatic interactions.

On the other hand, when $k$ is even, say $k=2$, the $\phi$ particles can form a gapped, bosonic integer quantum Hall phase with response precisely cancelling the native Chern-Simons term,
\begin{align}
\mathcal{L}&=\frac{2}{4\pi}ada-\frac{2}{4\pi}ada+\frac{1}{2\pi}adA=\frac{1}{2\pi}adA\,.
\end{align}
The fluctuations of $a$ then Higgs $A$. The result is an anyon superfluid, in which the microscopic charge-1 bosons (recall $k=2$ is \emph{even}) return to life and condense. This is the classic scenario for anyon superfluidity originating with Laughlin~\cite{Laughlin1988,Fetter1989,Chen1989,Lee1989,Fradkin1990}.

Recently, this mechanism has been adapted to systems with microscopic fermions through doping of higher-charge anyon composites that form local bosons on fusion~\cite{Shi2025c,Shi2025a}. One salient example is doping the charge-$2e/3$ anyon in the $\nu=2/3$ state, which was argued to lead to a charge-$2e$ superconductor. However, it is unclear from the standpoint of the abelian U$(1)_{-k}$, theory, Eq.~\eqref{eq: abelian CSGL}, how to describe the formation of these bound states. To evade this issue, such approaches utilize $K$-matrix theories that treat $e/k, 2e/k,\dots$ anyons as distinct members of a multi-component anyon fluid.

In each of the above scenarios, the driving physics is based on every anyon seeing its brethren as uniform flux. But as we noted around Eq.~\eqref{eq: monopole screening}, it is possible for anyons to have their flux screened by monopoles, forming local particles that are immune to the flux attachment constraint. This suggests that the dynamics of doped anyons may be much richer than what is suggested classically by abelian Chern-Simons-matter theories. Such richness has already been suggested by several recent analyses of itinerant anyons based on projective parton constructions~\cite{Shi:2025arn,lotric2026,Fan:2026kay,Wang:2026ziw,Shi2026,senthil2026fractionalizedmetalsdopedanyons,han2026orthogonal} and categorical approaches~\cite{Nakajima2025,Seo2026}.


\subsection{Basics of level-rank duality for Laughlin anyons}

Level-rank dualities provide a broader window into anyon dynamics. The quintessential level-rank dualities relate U$(N)_{-k}$ to SU$(k)_N$ TQFTs, meaning that the U$(1)_{-k}$ description of the $\nu=1/k$ Laughlin state above contains the same topological information as a SU$(k)_1$ Chern-Simons gauge theory. That level-rank dualities of TQFTs could be coupled to matter (at long wavelengths) was fully appreciated a little over a decade ago~\cite{Aharony2011,Giombi2012,Aharony2013b,Aharony2016a,Hsin2016,Radicevic2016}, although examples of applications of level-rank duality in the fractional quantum (anomalous) Hall context extend back to the 1990s and include Refs.~\cite{Balatsky1991,Fradkin1999,Goldman2019,Goldman2020a,Ma2020,Goldman2021a,Song2021,Shi2025a,Shi:2025arn,lotric2026,Wang:2026ziw}.

Applying level-rank duality to the anyon theory, Eq.~\eqref{eq: abelian CSGL}, the bosonic matter fields, $\phi$, are exchanged with $k$ colors of fermionic ``quarks'', $\chi_\alpha$,
\begin{equation}
    \Lag_\phi + \Lag_{\U(1)_{-k}} \longleftrightarrow \Lag_\chi + \Lag_{\SU(k)_1}\, ,
\end{equation}
where the dual theory is
\begin{align}
\label{eq: SU(k) dual Lag}
\widetilde{\mathcal{L}}&=\mathcal{L}_\chi+\mathcal{L}_{\SU(k)_1}\,,\\
\mathcal{L}_\chi&=\chi^{\dagger}\left[iD^b_t+\frac{1}{2m_\chi}(D^b_i)^2\right]\chi\,,\qquad D^b_\mu\chi=\partial_\mu\chi-ib_\mu\cdot\chi\,, \\
 \Lag_{\SU(k)_1}[b,c;A] &= \frac{1}{4\pi} \Tr \left( bdb + \frac{2}{3} b^3 \right) + \frac{1}{2\pi} c\,d(\Tr b - A) + \Lag_\ct + \dots\, .
\end{align}
Here $\chi^\alpha$, $\alpha=1,\ldots,k$, is a non-relativistic fermion with effective mass $m_\chi$ living in the fundamental representation of $\U(k)$, ${b_\mu=\Tr b\,\bs{1}_k/k +b_{\SU(k)}^a T^a}$ is a $\U(k)$ gauge field, and $c_\mu$ is an auxiliary $\U(1)$ gauge field. The matrices, $T^a$, are the $k^2-1$ generators of SU$(k)$, which are normalized such that ${\Tr[T^aT^b]=\delta^{ab}/2}$. The dot notation, $\cdot$, in the covariant derivative is simply meant to emphasize matrix multiplication. $\Lag_\ct$ denotes additional counterterms necessary to match responses on both sides of the duality (see Appendix~\ref{app:duality_counterterms}), while the ellipses contain higher-derivative terms (like Yang-Mills) and electrostatic interactions.

The purpose of $c_\mu$ is to act as a Lagrange multiplier Higgsing U$(k)$ down to SU$(k)$ by pinning $\Tr b$ to the background U$(1)$ connection\footnote{Coupling these theories properly to background Spin$_c$ connections is subtle, since the trace part of the dynamical U$(k)$ gauge field should not be Spin$_c$. As a result, the Spin$_c$ part of the background connection enters through (1) couplings to transparent fermion lines, as well as (2) direct coupling to the quarks~\cite{Hsin2016}. We present proper couplings in Appendix~\ref{app:duality_counterterms}, although this subtlety will not affect any of our conclusions.}, $A$. One may then seek to solve the constraint by writing 
\begin{align}
D_\mu^b\chi\rightarrow \partial_\mu \chi-i\,b_{\SU(k),\mu}\cdot \chi-i\,k^{-1}A_\mu\,\chi\,.
\end{align}
In other words, quanta of $\chi$ describe particles with EM charge $e/k$, which are the desired anyons! The reason for embedding the SU$(k)$ gauge theory into U$(k)$ is simply to allow the background EM field to couple with integer charge -- this will be convenient in many of our calculations below. Nevertheless, one may interpret Eq.~\eqref{eq: SU(k) dual Lag} as a SU$(k)$ gauge theory coupled to quarks with fractional EM charge. 

\begin{figure}[t]
    \centering
    \includegraphics[width=1\linewidth]{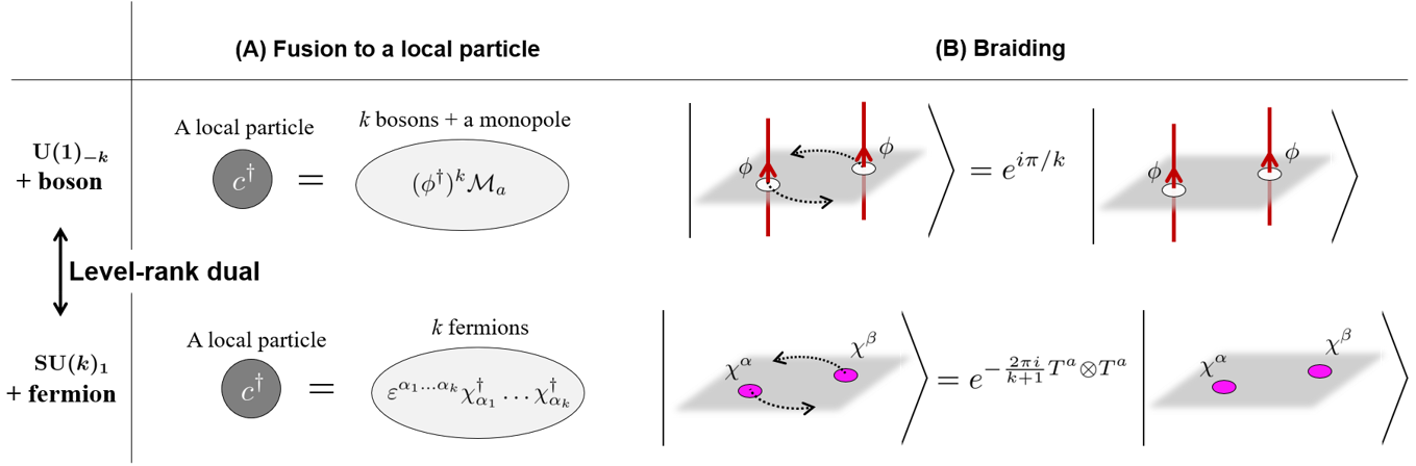}
    \caption{Level-rank duality provides two equivalent descriptions of $\nu=1/k$ Laughlin anyons: a $\U(1)_{-k}$ Chern-Simons theory coupled to a boson, and an $\SU(k)_1$ Chern-Simons theory coupled to a fermion. We highlight how local particles are represented in each description (A), as well as the action of braiding (B).}
    \label{fig:level_rank_dual}
\end{figure}

A conceptually attractive feature of the SU$(k)_1$ dual is that each anyon corresponds to a different representation of the gauge group. A field creating an anyon of charge $ne/k$ (under the constraint, $\Tr b = A$) can be  represented as an antisymmetrized bound state of $n$ quarks,
\begin{align}
\label{eq: anyon bound state general}
\mathcal{O}_n^{\alpha_{n+1}\dots\alpha_k}=\varepsilon^{\alpha_1\dots\alpha_k}\chi^\dagger_{\alpha_1}\dots\chi^\dagger_{\alpha_n}\,,
\end{align}
furnishing the $(k-n)^{\mathrm{th}}$ antisymmetric representation. Although one may in principle construct symmetric representations as well, the fact that the Chern-Simons theory is level 1 means that those representations do not correspond to distinct anyon types on quantizing the theory. In other words, \emph{anyon fusion is abelian} even though the gauge group is non-abelian. 

Hence, fusing $k$ quarks to a \emph{baryon} produces a SU$(k)$ singlet with charge $e$. In other words, a local particle may be assembled as (see Figure~\ref{fig:level_rank_dual})
\begin{equation}
\label{eq: local particle baryon}
    c^\dagger = \varepsilon^{\alpha_1\ldots\alpha_k} \chi^\dagger_{\alpha_1} \ldots \chi^\dagger_{\alpha_k}\, .
\end{equation}
This object is invariant under SU$(k)$, meaning that it is neutral under the Chern-Simons gauge fluctuations: It manifestly braids trivially with anyons! Therefore, in the SU$(k)_1$ presentation, fusion to local particles does not require screening from monopoles and can be captured using the matter degrees of freedom alone, unlike in the abelian model. 

As a technical aside, one might notice that while Eq.~\eqref{eq: local particle baryon} is invariant under SU$(k)$, it transforms under the determinant part of the full U$(k)$ gauge group. To make $c^\dagger$ fully gauge invariant requires attaching a monopole operator for the Lagrange multiplier gauge field, $\mathcal{M}_c$. However, this monopole operator is much more benign than the one we met in the abelian model. In fact, it condenses, acquiring a vacuum expectation value, $\langle \mathcal{M}_c\rangle\neq 0$, to Higgs U$(k)$ to SU$(k)$. The result is therefore only a field strength renormalization to $c^\dagger$ after appropriate gauge fixing. The monopole $\mathcal{M}_c$ is the IR avatar of the bosonic parton field in Ref.~\cite{Wang:2026ziw}.

How is braiding realized in the dual theory? The equation of motion for $b_{\SU(N)}$ tells us that each unit of color charge is accompanied by a unit of \emph{non-abelian} flux,
\begin{equation}
\label{eq: non abelian Gauss general}
    \begin{split}
        \chi^\dagger T^a \chi &= -\frac{1}{2\pi} \mathcal{F}^a_{xy}\, ,\qquad
        \mathcal{F}^a_{\mu\nu} = \partial_\mu b_\mu^a-\partial_\nu b_\mu^a+f^{abc}\, b_{\mu,b}\,b_{\nu,c}\,,
    \end{split}
\end{equation}
where $f^{abc}$ are the SU$(k)$ structure constants. We may wish to follow the procedure used in abelian Chern-Simons theory, attempting to compute braiding statistics for two quarks by viewing $\chi_\alpha$ as a two-particle spinor wave function and thinking of the flux as imparting an Aharonov-Bohm phase. But how do we do this if the flux is non-abelian? 

The wave function of a single quark is a $k$-dimensional column, meaning that under exchange a generic two-particle wave function undergoes a unitary transformation,
\begin{align}
U_{\mathrm{braid}}=\exp\left(-\frac{2\pi i}{k+1}\,T^a\otimes T^a\right)\,,
\end{align}
which should be interpreted as carrying four color indices\footnote{The $k+1$ appearing in the denominator is related to a choice of UV regularization~\cite{Witten1989b,Chen1992}. We will assume that gauge fluctuations at short distances are always regulated by a Yang-Mills term, which leads to this one-loop-exact shift by $k$.} (see Figure~\ref{fig:level_rank_dual}). Despite being an intimidatingly large tensor object, it only has two distinct eigenvalues:  $e^{i\theta_s}$ and $e^{i\theta_a}$, corresponding respectively to the symmetric ($S_2$) and antisymmetric ($A_2$) representations. Because the Chern-Simons level is $1$, physical multi-anyon states can only reside in the antisymmetric representation. As a result, a physical two-anyon wave function acquires a single \emph{abelian} phase, $e^{i\theta_a}$ , under quark exchange. One can compute~\cite{Witten1989b},
\begin{align}
\theta_a=\frac{\pi }{k+1}[2\,c_2(F)-c_2(A_2)]=\frac{\pi}{k}\,,
\end{align}
where $c_2(F)=(k^2-1)/2k$ and $c_2(A_2)=(k-2)(k+1)/k$ are respectively the coefficients of the quadratic Casimir operators in the fundamental and antisymmetric representations of SU$(k)$. This is the correct braiding phase for the desired anyons. 

The discussion above provides a physical sense for what level-rank duality offers us. In the abelian U$(1)_{-k}$ presentation, anyon \emph{braiding} statistics are an obvious consequence of the charge-flux constraint, Eq.~\eqref{eq: abelian Gauss general}, but anyon fusion is completely obscure prior to quantization. In contrast, in the SU$(k)_{1}$ dual, anyon \emph{fusion} can naturally be understood in terms of binding of matter particles charged under the Chern-Simons gauge field. But braiding information is hidden in the structure of the quantum Hilbert space. 

One might therefore expect that these two dual descriptions would lead to completely distinct mean field phase diagrams, as they each prioritize only one of the key elements of anyon dynamics, leaving the other below the surface in the fully quantum theory.
This does not make the theories any less dual to one another. Instead, it does mean that in general mean field phases apparent in one choice of variables should not be privileged energetically over those in the other.

In this work, we will consider the menagerie of phases that arise on doping the quark theory, Eq.~\eqref{eq: SU(k) dual Lag}. Surprisingly, we will find that the same physics encountered in the abelian description -- where anyons simultaneously regard each other as flux and form Landau levels -- can be transparently embedded into the dual quark model and competes with other possibilities for the anyon dynamics where the flux is screened. We thus expect the quark description to offer an even playing field for studying the competition between many distinct anyonic phases, despite its shortcomings in capturing abelian braiding statistics.    


\subsection{The quark metal}

We are now prepared to consider doping anyons into the system. Deep in the FQAH phase, the quarks may be regarded as forming a gapped band insulator, with chemical potential below the conduction band. Because the quarks carry EM charge $e/k$, doping the system  by ${\rho_e=\langle\delta \Lag/\delta A_t\rangle=\delta\neq0}$ leads to a finite quark density,    
\begin{align}
\label{eq: quarks at finite density}
\frac{1}{k}\,\rho_\chi=\frac{1}{k}\,\langle\chi^\dagger\chi\rangle=\delta\,,
\end{align}
which follows from directly plugging in the constraint, ${\Tr[b]=A}$.

How do the quarks react to life at finite density? To answer this, it is instructive to consider the Gauss' Law constraints for all of the gauge fields. The equation of motion for the trace component, $\Tr[b]$, and the Lagrange multiplier field, $c$, are respectively
\begin{align}
0&=\chi^\dagger\chi+\frac{1}{2\pi}\varepsilon_{ij}\partial_i\Tr[b_j]+\frac{k}{2\pi}\varepsilon_{ij}\partial_i c_j\,,\qquad \frac{1}{2\pi}\varepsilon_{ij}\partial_i\Tr[b_j]=\frac{1}{2\pi}\varepsilon_{ij}\partial_i A_j\,.
\end{align}
The second equation pins the abelian flux felt by the quarks to the background EM field, which for the FQAH phases of interest to us vanishes. Plugging this back into the first equation and recalling that the EM charge density is $\rho_e=\langle \varepsilon_{ij}\partial_i c_j/2\pi\rangle$ reproduces Eq.~\eqref{eq: quarks at finite density}.

In contrast, the flux in the non-abelian gauge components is fixed to the color charge densities by the Chern-Simons constraint, ${\rho_\chi^a=-\mathcal{F}^a_{xy}/2\pi}$, as in Eq.~\eqref{eq: non abelian Gauss general}. Excess color charge would be accompanied by color flux, in analogy with the abelian charge-flux constraint encountered in Eq.~\eqref{eq: abelian Gauss general}. But because the non-abelian gauge charges and fluxes transform as adjoints under SU$(k)$, gauge invariance mandates that they be divorced from any background fields, as they are betrothed to one another. Any net color charge or flux would have to arise spontaneously, Higgsing the gauge group.

We are therefore inexorably led to a single symmetry-preserving possibility: The quarks find themselves at finite charge density but zero flux, forming a Fermi sea without any color polarization. In this mean field state, which we dub the \emph{quark metal}, the quarks carry fractional electric charge $e/k$ and couple to the SU$(k)$ Chern-Simons gauge field.

The physics of the quark metal shares many qualitative features with the famous composite Fermi liquid (CFL) metal of the half-filled Landau level~\cite{Halperin1993}, although there are a few crucial distinctions. First, the quarks here carry charge $e/k$, as they model the underlying anyon fluctuations of the $\nu=1/k$ Laughlin state, whereas the composite fermions of the CFL carry unit electric charge in the Halperin-Lee-Read (HLR) theory. This fractional electromagnetic charge could in principle be visible in tunneling and shot noise.

Second, the quark and electronic conductivity tensors are related as 
\begin{align}
\sigma_e^{ij}=\frac{1}{k^2}\,\sigma_\chi^{ij}+\overline{\sigma}_{H}\,\varepsilon^{ij}\,,\qquad \sigma_\chi^{ij}=\frac{i}{\omega}\,G^R_{j_\chi^i\, j_\chi^j}(\omega)\,,
\end{align}
where $\overline\sigma_H$ is the Hall conductivity of the background FQAH phase being doped, $\sigma_e^{ij}$ is the conductivity tensor of the physical EM charges, and $G^R_{j_\chi^i\, j_\chi^j}(\omega)$ denotes the retarded Green's function of the quark EM current, ${j_\chi^i=i\,\frac{1}{m_\chi}\,\chi^\dagger D_i^b\chi+\mathrm{h.c.}}$, evaluated at ${\bs{q}=0}$. The above relation is very different from the Ioffe-Larkin composition rule connecting the composite fermion resistivity to the electron resistivity, as the background FQAH and quark responses are decoupled. Consequently, the quark metal should generally exhibit \emph{both} a large DC Hall angle \emph{and} a nonvanishing Drude weight, whereas the CFL only hosts the former~\cite{Shi2022,yue2026}. 

A further distinguishing feature with the CFL is that $1/r$ electrostatic interactions cannot be recast as a linear-in-momentum  $\sim|\bs{q}|$ contribution to the gauge kinetic term, since gauge flux is not identified with the global electric charge density. Consequently, at finite density the gauge dynamics are controlled by the Yang-Mills term,
\begin{align}
\mathcal{L}_{\mathrm{YM}}=-\frac{1}{2g_{\mathrm{YM}}^2}\Tr\mathcal{F}^2\,.
\end{align}
Deep in the gapped FQAH phase, the gauge theory is at strong coupling, $g_{\mathrm{YM}}^2\rightarrow\infty$. When a finite density of itinerant anyons is introduced, the Yang-Mills term becomes necessary to control the dynamics. If the doped charge density is very small, $\delta\ll g_{\mathrm{YM}}^2$, any gauge dynamics is projected to high energy scales: There are no gapless ``gluons.'' This does not mean the quarks are free. To the contrary, the quarks remain strongly coupled to the Chern-Simons gauge field, and any electrostatic interactions should also have outsize importance as well. The physics in this regime may therefore depart drastically from expectations based on limits where the gauge coupling is treated as perturbative. However, one context where the strong coupling regime has been studied is the 't Hooft large-$N$ limit, which is non-perturbative and can be taken to strong coupling. Remarkably, transport and thermodynamic observables can be calculated in this limit even at very low densities~\cite{Geracie2016,Gurari2016,Nakajima2025}. 

As the quark density is increased, it becomes legitimate to consider gapless transverse gauge fluctuations, which mediate strong interactions near the Fermi surface, and the system may be better approximated by studying perturbations to the weak coupling fixed point. At low temperatures, one may then conclude that SU$(k)$ gauge fluctuations are overdamped and possess dynamical exponent $z=3$, i.e. they disperse as $\omega(\bs{q})\sim |\bs{q}|^3$, with no dependence on the exponent of the Coulomb potential. In this regime, the physics of the quarks veers closest to the that of a Fermi surface coupled to gapless bosons\footnote{It is important to emphasize that as the density approaches $g_{\mathrm{YM}}^2$, level-rank duality is expected to break down, and the dynamical properties of the SU$(k)_1$ quark theory may differ from the U$(1)_{-k}$ CS-GL theory. This means that detailed energetics in this regime will differ depending on one's choice of anyon variables, although we expect the landscape of possible low-temperature phases to remain unaltered.}.

In either regime, due to the fluctuations of the SU$(k)$ gauge field, it is unlikely that quark Fermi surface is stable all the way to zero temperature -- indeed, SU$(k)$ gauge fluctuations are generically attractive! -- and we focus on its possible instabilities through the remainder of this work. We are specifically interested in the instabilities of the quark Fermi surface model in the experimentally salient context of the $\nu=2/3$ FQAH state, corresponding to $k=3$. As discussed in the Introduction, we find that this phase may be considered a metallic parent to the plethora of descendant anyon-driven phases discussed in the literature thus far~\cite{Shi2025c,Shi:2025arn,Zhang2025a,lotric2026,Fan:2026kay,Wang:2026ziw,Shi2026,senthil2026fractionalizedmetalsdopedanyons,han2026orthogonal}, along with some new ones. To limit our scope, our primary interest is in superconducting descendants possessing a uniform component.   

Before continuing, we note that we have elided a second symmetry-preserving possibility for the quarks, which was also discussed in Ref.~\cite{lotric2026}. Because at weak coupling SU$(k)$ gauge fluctuations are attractive in the singlet channel (see Appendix~\ref{app:attr_vs_repuls_tree_level}), there may be a regime where the quarks choose to form tight, SU$(k)$-neutral bound states -- the baryons in Eq.~\eqref{eq: local particle baryon} -- with a gap to single-quark excitations (note there is no genuine confinement, owing to the Chern-Simons term in the Lagrangian). This phenomenon has a simple interpretation: When charge is doped into the system, local electrons or bosons are energetically ``lighter'' than anyons. The resulting physics is that of a gas of un-fractionalized charges that do not couple to the underlying FQAH fluid except through electrostatic interactions. If $k$ is even, these charges are bosons, which condense to form a superfluid. This is the most trivial example of a SF$\star$ state, where local bosons condense and spontaneously break charge conservation without harming the topologically ordered background. If $k$ is odd, the baryons are local fermions, which will behave as a Fermi liquid metal. While the formation of such states is always a possibility -- especially in cases where electrostatic interactions are heavily screened -- in the discussion below we will largely assume that the formation of local charges is always more energetically costly than fractionalization into anyons.


\subsection{Enrichment by lattice translation symmetry}

Because we are interested in doping FQAH phases, it is crucial to consider constraints from the underlying lattice symmetries. For a lattice FQAH system, translation invariance generally requires that the physical ground state wave function involve background anyons with charge $e/k$ in each unit cell. Itinerant charge-$e/k$ anyons will thus be acted upon projectively under lattice translations, since they will acquire fractional Aharonov-Bohm phases on circling a unit cell. If $\tau_x,\tau_y$ are the translation generators~\cite{Cheng2016,Bultinck2018,Lu2020},
\begin{align}
\tau_x\tau_y\tau_x^{-1}\tau_y^{-1}=e^{i\vartheta}\,,\qquad\vartheta=\frac{2\pi}{k}\,,
\end{align}
resulting in a $k$-fold enhancement of the unit cell. The anyon band structure will  be $k$-fold degenerate in the resulting folded Brillouin zone.

At the level of the quark effective field theory, Eq.~\eqref{eq: SU(k) dual Lag}, this degeneracy implies that each quark color forms $k$ valleys, $\chi_{\alpha,J}$, $I=1,\dots,k$, meaning that the charge-$e/k$ quark Fermi surface has $k$ momentum space pockets, each of which hosts all $k$ colors. Lattice translations are then embedded into the low energy effective theory as a $\mathbb{Z}_k\times \mathbb{Z}_k$ symmetry,
\begin{align}
\label{eq: translation action}
\tau_x:\chi_{\alpha,J}\rightarrow \chi_{\alpha,J-1}\,\qquad \tau_y:\chi_{\alpha,J}\rightarrow \left(e^{i\vartheta}\right)^J\chi_{\alpha,J}\,.
\end{align}
Indeed, we find that the interplay of valley with color degrees of freedom leads to a number of exciting possibilities for anyonic phases inaccessible to the single-valley case. 

One may worry about the extension of the duality between Eq.~\eqref{eq: abelian CSGL} and Eq.~\eqref{eq: SU(k) dual Lag} to multiple matter species. Luckily, level-rank dualities are expected to hold in such cases as well~\cite{Aharony2016a}, in some cases corroborated by derivations in Euclidean lattice models~\cite{Zimet2018}. One way to roughly argue this is to derive Eq.~\eqref{eq: SU(k) dual Lag} microscopically using a projective parton construction where the underlying charges of the lattice Hamiltonian are fractionalized into fermionic chargons with a SU$(k)$ gauge redundancy, $f^\alpha$, $\alpha=1,\dots,k$, as ${e\sim \varepsilon_{\alpha_1\dots\alpha_k}f^{\alpha_1}\dots f^{\alpha_k}}$, with each color filling an identical $C=1$ band preserving SU$(k)$. Integrating out the filled Chern bands leads to the SU$(k)_1$ Chern-Simons action describing the FQAH phase, while doping the system further leads to $k$ valleys of charge-$e/k$ fermionic quarks charged under the emergent gauge field, which describe the anyon fluctuations. 

We remark that while the duality should hold at long wavelengths in the non-relativistic limit -- doping from deep within the FQAH phase -- a problem arises if the system is tuned to a quantum phase transition where the quarks form Dirac cones and their total Chern number changes by $k$. Under a na\"{i}ve interpretation of the duality with Eq.~\eqref{eq: abelian CSGL}, this transition would correspond to condensation of $k$ species of bosonic fields, $\phi$, which would \emph{both} Higgs the U$(1)$ gauge group and break the valley symmetry. The result is a charge density wave phase that is seemingly absent from the quark side of the duality, which preserves lattice translations. The only way out of this issue while preserving the duality, then, is to conjecture that the quarks' evolution toward the Chern number changing transition is  arrested by spontaneous breaking of the valley symmetry. If the $\mathbb{Z}_k$ permutation symmetry is enhanced to a continuous SU$(k)$ flavor symmetry, this gives rise to an intermediate Goldstone phase proposed in Ref.~\cite{Komargodski2017}. The same dilemma was recently noticed from the standpoint of a parton construction in Ref.~\cite{Pichler2025}, which suggested a similar conjecture in the context of an abelian duality.

If the $\mathbb{Z}_k\times\mathbb{Z}_k$ lattice translation symmetry in indeed enhanced to a continuous $\U(k)$ flavor symmetry, it is, strictly speaking, important to distinguish between upper and lower valley indices. We will not do so in this manuscript, and keep all valley indices in the subscript, while noting that all the translation-invariant phases that we consider can be enhanced to $\U(k)$ invariant phases.


\section{Doping the lattice $\nu=2/3$ FQAH state}\label{sec:nu=2/3_doping}


\subsection{Expectations from abelian flux attachment}

We now turn our attention to doping the experimentally relevant $\nu=2/3$ FQAH state, applying the general insights from level-rank duality developed above. The abelian topological order with Hall conductivity $\overline{\sigma}_{H}=2/3$ can be expressed as the particle-hole conjugate of the $\nu=1/3$ Laughlin state, in the form of a U$(1)_{+3}$ Chern-Simons TQFT stacked with a $\nu=1$ integer quantum Hall state. It has two anyons. We label their charges $e/3$ and $2e/3$. These two anyons can fuse to form local particles built out of microscopic charge-$1$ holes. Note that while we focus on the doped $\nu=2/3$ state from here on, level-rank duality can also be applied to the case of a doped semion gas on a lattice, yielding a SU$(2)_1$ quark metal. We survey the basic instabilities for this case in Appendix~\ref{app:semion_superconductors}.

The charge-$e/3$ anyon fluctuations may be modeled using the particle-hole conjugate of Eq.~\eqref{eq: abelian CSGL} with $k=-3$ and three boson species, $\phi_I$, $I=1,2,3$, transforming projectively under lattice translation symmetry,
\begin{align}
\label{eq: abelian e/3 Lagrangian}
    \Lag &= \Lag_\phi[a] + \frac{3}{4\pi} ada - \frac{1}{2\pi} Ada + \frac{1}{4\pi}AdA+2\,\mathrm{CS}_g\,,\\
    \Lag_\phi[a] &= \phi_I^\dagger \left[ i D^a_t + \frac{1}{2m_\phi}(D^a_i)^2 \right] \phi_I-V[\phi_I]\, ,
\end{align}
where $2\mathrm{CS}_g$ denotes the gravitational Chern-Simons term encoding the $c_-=+1$ chiral central charge of the stacked integer quantum Hall state. The interaction potential, $V[\phi]$, may include short-ranged boson self-interactions as well as longer-ranged repulsive density-density interactions inherited from the microscopic holes. It need only be invariant under transformations generated by the discrete symmetry generators, Eq.~\eqref{eq: translation action}, although it is possible at low doping for the Lagrangian to enjoy an approximate SU$(3)$ flavor symmetry.

One can imagine doping itinerant, charge-$e/3$ anyons into the system, $\nu=2/3+\delta\nu$. From the standpoint of the abelian Ginzburg-Landau presentation, the bosons, $\phi_I$,  mutually feel each other as $-1/3$ flux quanta, per Eq.~\eqref{eq: abelian Gauss general}, placing them at total filling, ${\sum_I\nu_{\phi,I}=-3}$. If translation invariance is preserved, each boson valley will find itself at $\nu_{\phi,I}=-1$. One natural expectation, at least at intermediate energy scales, would then be for the bosons to form a gapless, ``secondary'' composite Fermi liquid metal, which can suffer pairing instabilities to generate various metallic and superconducting descendant phases~\cite{Shi2025c,Shi:2025arn,Shi2026,Fan:2026kay}. Alternatively, a metallic state of emergent fermionic quasiparticles coupled to a fluctuating $\mathbb{Z}_3$ gauge field -- dubbed a $\mathbb{Z}_3$ orthogonal metal -- has also been proposed based on a parton decomposition of $\phi$ (or of analogous composite fermion degrees of freedom). This phase is stable at zero temperature but can exhibit superconductivity through a Kohn-Luttinger-like mechanism~\cite{Shi2026,senthil2026fractionalizedmetalsdopedanyons,han2026orthogonal}.

On the other hand, if a separate abelian theory of the charge-$2e/3$ anyon fluctuations is used -- corresponding to fermions coupled to a U$(1)_3$ Chern-Simons gauge field -- a chiral superconducting phase with $c_-=-2$ is obtained immediately as the sole translation-invariant option~\cite{Shi2025c}. Although one may naturally wish to access this physics by considering bound pairs of charge-$e/3$ anyons, this possibility appears completely invisible to mean field theories starting from the Lagrangian in Eq.~\eqref{eq: abelian e/3 Lagrangian}.


\subsection{The SU$(3)_{-1}$ quark metal: Facts and conventions}

These phases and more can be realized from the quark metal parent state. Particle-hole conjugating Eq.~\eqref{eq: SU(k) dual Lag} and introducing three quark valleys $\chi_I$, $I=1,2,3$, transforming projectively under lattice translations as in Eq.~\eqref{eq: translation action} gives  
\begin{align}\label{eq:SU(3) full Lagrangian}
    \Lag &= \Lag_\chi\left[ b\right] - \frac{1}{4\pi} \Tr \left( bdb + \frac{2}{3} b^3 \right) + \frac{1}{2\pi} cd(\Tr b - A) + \frac{1}{4\pi}AdA-4\mathrm{CS}_g\, , \\
    \Lag_\chi &= \chi_I^{\dagger}\left[iD^b_t+\frac{1}{2m_\chi}(D^b_i)^2\right]\chi_I-U_C[\rho_{I}]\,,\qquad D^b_\mu\chi_I=\partial_\mu\chi_I-ib_\mu\cdot\chi_I\,.
\end{align}
Here $U_C[\rho_I]$ denotes only the electrostatic interactions among the intra-valley charge densities, ${\rho_I=\sum_\alpha\chi^\dagger_{I,\alpha}\chi^\alpha_I}$ (defined with no implicit sum over $I$). As noted in the previous Section, this action is not formally correct because the probe field $A$ cannot be regarded as a Spin$_c$ connection itself, even though the local degrees of freedom are fermions. A correct account of how to couple this theory to a background Spin$_c$ connection is provided in Appendix~\ref{app:nu=2/3_counterterms}. The practical consequence is acquisition of the correct $-4\mathrm{CS}_g$ counter-term. The above form of the action suffices for computing all local responses by varying the partition function with respect to $A$. Note that we have suppressed the Yang-Mills term.

On doping, the quarks form three Fermi pockets without net color charge, and our goal is to study the possible instabilities of this non-Fermi liquid metallic state. The possible phases involve an intricate interplay of the (gauge) color and (global) valley degrees of freedom, making it useful to establish some naming conventions. Each quark valley comes in three colors, which we term \textit{cyan} (c), \textit{magenta} (m), and \textit{yellow} (y). A gauge invariant state must consist of equal parts cyan, magenta, and yellow, and we will hence describe it as \textit{black} in adherence to color theory. 

The charge-$2e/3$ anyons are realized as locally bound pairs of quarks in the \emph{antifundamental} representation (which for SU$(3)$ is dual to the antisymmetric representation, $A_2$),
\begin{align}
\Phi_{\alpha,IJ} =\varepsilon_{\alpha\beta\gamma}\chi^\beta_I\chi^\gamma_J\,.
\end{align}
From this vantage, $\Phi$ carries \emph{two} valley indices, $I$ and $J$, which are \emph{symmetric} due to the Fermi statistics of $\chi$ and the antisymmetry of the Levi-Civita symbol. That charge-$2e/3$ anyons here appear to furnish a higher-dimensional representation of lattice translations reflects the fact that a charge-$2e/3$ composite can in principle be assembled from any combination of two quark valleys. Similarly, local fermions carry three valley indices,
\begin{align}
c_{IJK}=\varepsilon_{\alpha\beta\gamma}\chi_I^\alpha\chi_J^\beta\chi_K^\gamma\,,
\end{align}
where we leave attachment of the monopole, $\mathcal{M}_c$, implicit here and below, since we assume it to have a spatially uniform vacuum expectation value, ${\langle\mathcal{M}_c\rangle\neq0}$. It is easy to see that all nonvanishing components of $c_{IJK}$ are acted upon faithfully by lattice translations, with ${\tau_x\tau_y\tau_x^{-1}\tau_y^{-1}[c]=c}$. 

We now proceed to study the fate of the SU$(3)_{-1}$ quark metal under two instabilities familiar in the context of (non-)Fermi liquid metals: BCS superconductivity and itinerant ferromagnetism, which here we take to occur in the space of colors. We also  consider the phases that can arise upon the formation of the charge-$2e/3$ bound states described by $\Phi$. Interestingly, in some cases the same phase can even be found to arise from distinct mechanisms. Each type of mechanism has the potential to lead to superconductivity of the microscopic holes, and their competition is determined by energetics.


\subsection{Color superconductivity}\label{sec:color_sc}

The most exciting opportunities for new physics come from Cooper pairing of quarks and the formation of \emph{color superconductors}. These anyonic states closely resemble the eponymous phases studied in QCD~\cite{Son1998,Rajagopal2001,AlfordRMP}; in both cases three quark flavors, each with three possible colors, may pair through fluctuations of the SU$(3)$ color gauge field. In QCD, color superconductivity is expected at high densities due to the march toward confinement at low energy scales. Likewise, very close to the FQAH plateau we also find ourselves in a strong coupling regime (in the sense that the Yang-Mills coupling is large), but confinement is arrested by the Chern-Simons term. So while we expect color superconductivity to primarily occur at higher anyon densities where the gauge fluctuations are more weakly coupled, it is still possible in intermediate density and coupling regimes. 

\begin{figure}[t]
    \centering
    \includegraphics[width=0.9\linewidth]{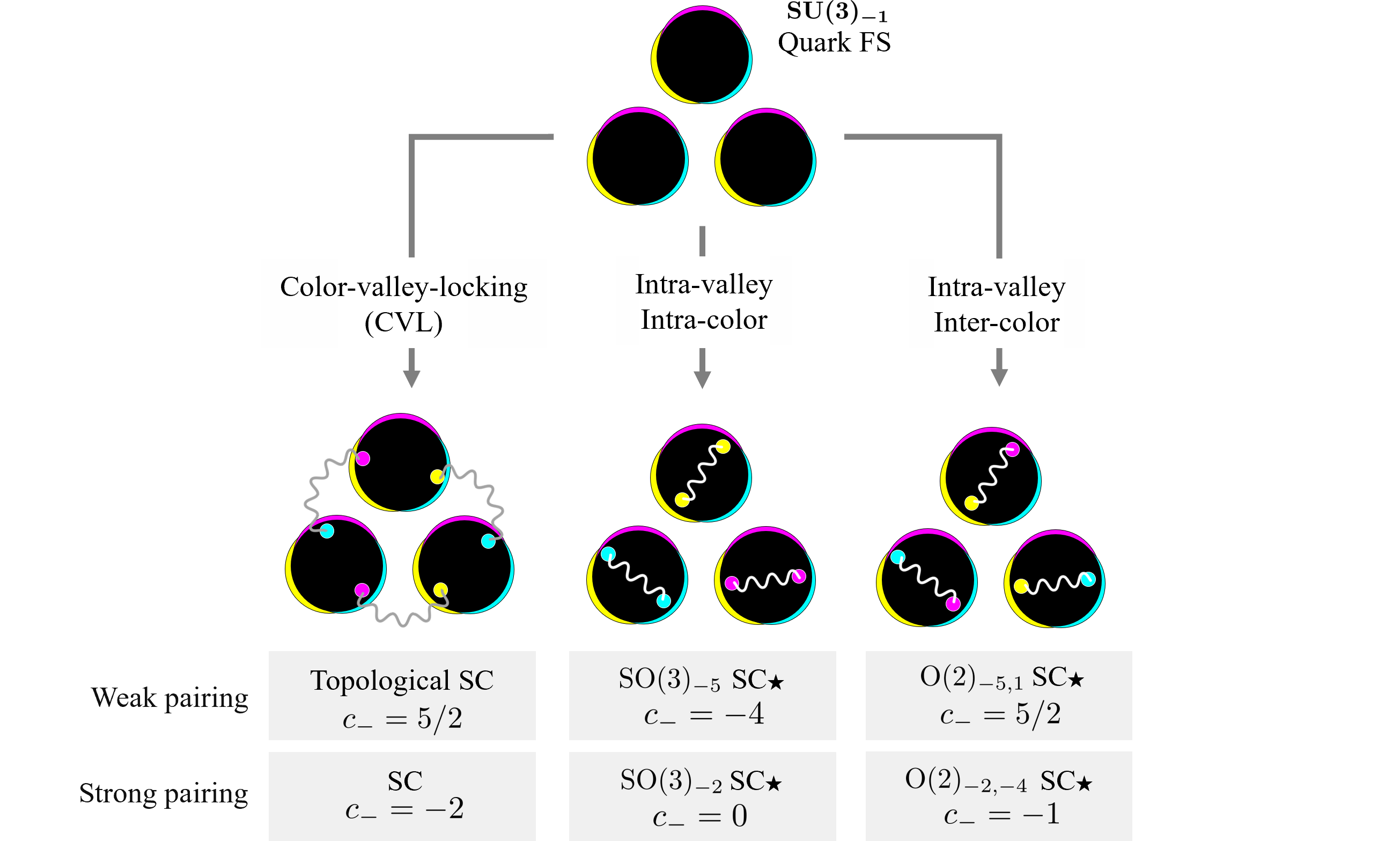}
    \caption{Three color superconducting instabilities of the black Fermi surface:
    (1) \textit{color-valley-locking (CVL)}, (2) \textit{intra-valley and intra-color pairing}, and (3) \textit{intra-valley and inter-color pairing}. The CVL pairs in the $p+ip$ channel, and the others pair in $p-ip$. All superconductors have a uniform component, with the SC$\star$ phases being accompanied by spatial modulation in general.}
    \label{fig:color_SC}
\end{figure}

The presence of the Chern-Simons term also leads to unconventional pairing possibilities beyond what is normally expected in the quantum chromodynamics (QCD) literature, which tends to focus on $s$-wave instabilities. While the $s$-wave color-antisymmetric channel is also attractive here -- perhaps reflecting a natural tendency for anyons to fuse that has been suggested by microscopic approaches recently~\cite{Munoz2020,Gattu2025,Yang2025,Gonccalves2025,Li2026a} -- the SU$(3)$ Chern-Simons gauge fluctuations can produce tree level attraction in channels with non-zero angular momentum, $\ell\neq0$, offering natural routes toward chiral topological superconductivity of the underlying \emph{holes}. Depending on the color/valley symmetry of the order parameter\footnote{We choose our conventions such that $\ell$ carries the sign of the \emph{Cooper pair} angular momentum, which opposes the sign of the thermal Hall effect and thus the chiral central charge, $c_-$~\cite{Alicea2012}. In other words, for $\ell$ odd, ${\sgn(\ell)=-\sgn(c_-)}$. We will refer to superconductors with $c_-=\pm1/2$ as $p\pm ip$.}, 
\begin{equation}
    \Delta^{\ell}_{\alpha\beta,IJ} = \langle \chi^\dagger_{\alpha,I} D_z^{\,-\ell} \chi^\dagger_{\beta,J} \rangle\, ,\qquad D_z=D_x^b-iD_y^b\,.
\end{equation}
These topological superconductors can correspond to the minimum angular momentum channel for pairing. As shown in Appendix~\ref{app:attr_vs_repuls_tree_level}, the tree level statistical interaction picks $\ell < 0$ for the color-antisymmetric channel and $\ell > 0$ for the color-symmetric channel. A summary of color superconducting phases is presented in Figure~\ref{fig:color_SC}.


\subsubsection{Topological superconductivity from color-valley-locking}\label{sec:CVL_sc}

The most natural color superconducting channel pairs distinct colors with distinct valleys, preserving translation symmetry while Higgsing the gauge group completely. Such a phase is the cousin of the famous ``color-flavor-locked'' phase that is the poster child for color superconductivity in QCD. We dub it the \emph{color-valley-locked} (CVL) superconductor (see Figure~\ref{fig:color_SC}),

\begin{equation}
    \Delta_{\alpha\beta,IJ}
    \propto \sum_{n=1}^3 \varepsilon_{\alpha\beta n}\, \varepsilon_{IJn}\, ,\qquad \ell=-1\,,
\end{equation}
This choice of pairing channel only has three independent components. The full order parameter may be expressed as
\begin{equation}
\label{eq: CVL VEV}
    \Delta^\alpha_I = \langle \varepsilon^{\alpha\beta\gamma} \varepsilon_{IJK} \chi^\dagger_{\beta,J} D_z \chi^\dagger_{\gamma,K} \rangle = \overline{\Delta} ~ (q_x - i q_y)\, \delta^\alpha_I\, ,
\end{equation}
where the constant $\overline\Delta$ is the order parameter strength. Antisymmetry of the color and valley indices has forced $\ell$ to be odd in cooperation with Pauli exclusion, and the sign of the tree level statistical interaction indicates that the $\ell<0$ is attractive while $\ell>0$ is repulsive (see Appendix~\ref{app:attr_vs_repuls_tree_level}). The minimal pairing channel thus has angular momentum $\ell = -1$, corresponding to $p+ip$ quark pairing. We note that in the QCD setting an $s$-wave channel is possible due to the existence of an additional helicity degree of freedom.

This choice of order parameter completely Higgses the gauge group by locking color and valley, annihilating any topological order. Under U$(3)$,  
\begin{equation}
    \Delta^\alpha_I \rightarrow (\det U)^{-1} U^\alpha_{\,\,\,\beta}\, \Delta^\beta_I\, .
\end{equation}
Evidently, the order parameter is preserved by $U=\pm \bs{1}_3$, corresponding to a $\mathbb{Z}_2$ subgroup of $\U(1)_{\mathrm{EM}}$ after implementation of the constraint due to $c_\mu$. Hence, we find charge-$2e$ superconductivity of the microscopic electrons. The quark condensate is also invariant under the $\mathbb{Z}_3$ valley permutation symmetry, provided each permutation is matched with a counteracting one from the permutation subgroup of the global part of the color SU$(3)$, indicating that the superconductivity is uniform. Furthermore, in situations where the valley symmetry is promoted to a global SU$(3)$, this symmetry will also be preserved through locking with a corresponding color rotation. Hence such a state may be energetically privileged if there is an approximate SU$(3)$ valley symmetry at low doping.

That condensation of pairs with fractional charge may be immediately interpreted as  charge-$2e$ superconductivity is a consequence of gauge invariance. Recall that in order to implement fractionalization of the global U$(1)_\mathrm{EM}$ charge, we promote the gauge group to U$(3)$, which is Higgsed down to SU$(3)$ by a constraint. It is convenient to factorize 
\begin{align}
\label{eq: U(3) as a quotient}
\U(3)=\frac{\SU(3)\times\U(1)}{\mathbb{Z}_3}\,,
\end{align}
where the quotient by a $\mathbb{Z}_3$ one-form symmetry identifies U$(1)$ elements related under the center element of $\SU(3)$. More explicitly, any U$(3)$ matrix may be decomposed as ${U=e^{i\theta/3} V}$, where ${e^{i\theta/3} = (\det U)^{1/3}\in \U(1)}$ and ${V\in\SU(3)}$. Importantly, $\det U=e^{i\theta}$ is a complex phase that must be $2\pi$ periodic, meaning that we must identify ${\theta/3 \sim \theta/3 +2\pi/3}$. Such a shift equivalently corresponds to acting with the center element of SU$(3)$. This identification reduces the residual symmetry preserving the order parameter to the $\mathbb{Z}_2$ group elements, $U=\pm\bs{1}_3$, which are identified with the $\mathbb{Z}_2$ subgroup of U$(1)_{\mathrm{EM}}$ under the Lagrange multiplier constraint, indicating charge-$2e$ superconductivity.

In Appendix~\ref{app:color_sc_eff_lags}, we translate the above discussion into the less abstract language of Ginzburg-Landau theory for the order parameter and derive the effective Lagrangian,
\begin{equation}
    \mathcal{L}_{\mathrm{eff}} = \frac{2}{2\pi} \alpha d A - \frac{2}{4\pi} AdA - 4\mathrm{CS}_g + \dots \,,
\end{equation}
where $\alpha$ is a $\U(1)$ gauge field, and the ellipses denote higher order terms in the fluctuation expansion for the order parameter. The mixed Chern-Simons term between $\alpha$ and $A$ clearly indicates charge-$2e$ superconductivity, and the background terms arrange neatly into a well-defined $\Spin_c$ Chern-Simons term.

A key conceptual advantage to working with color superconducting orders is that a gauge invariant Cooper pair may be constructed from local matter operators. In this case, the electronic Cooper pair is a sextet of quarks. In momentum space,
\begin{equation}
\label{eq: electron order param CVL}
    \Delta_{IJK}^\mathrm{el} = \varepsilon_{\alpha\beta\gamma} \Delta^\alpha_I \Delta^\beta_J \Delta^\gamma_K \sim \left(\overline{\Delta}\right)^3 (q_x - i q_y)^3\, \varepsilon_{IJK}\, .
\end{equation}
Note that although the left-hand-side is manifestly gauge invariant, the use of $\sim$ above is because have presumed the Higgsing of gauge fluctuations such that covariant derivatives may be replaced with ordinary ones.  

Equation~\eqref{eq: electron order param CVL} presents a clear window into the nature of the electronic superconductivity. Because each quark pair forms a $p+ip$ superconductor, we first note that the pairing symmetry of microscopic holes has $\ell=-3$ and corresponds to $f+if$. Furthermore, given the translation symmetry action in Eq.~\eqref{eq: translation action}, we can immediately observe that this superconductor is completely uniform, with no oscillating components: It is invariant under cyclic permutations, and the quark valley momenta are postulated to sum to a multiple of a reciprocal lattice vector. This will not be true for other color superconductors, which generically exhibit both uniform and oscillating components.

In Appendix~\ref{app:BdG_Majorana_vortices}, we present a detailed Bogoliubov-de Gennes (BdG) construction of the edge modes in the CVL superconductor. In the weak pairing limit, we find that there are nine Majorana zero modes (MZMs) capable of binding to a $h/2e$ vortex. These MZMs transform into one another under the residual $\mathbb{Z}_3\times\mathbb{Z}_3$ lattice translation symmetry. Each MZM contributes $c_-=+1/2$ to the total chiral central charge of the superconducting phase. However, the SU$(3)_{-1}$ quark theory, Eq.~\eqref{eq:SU(3) full Lagrangian}, exists on a background $c_-=-2$ thermal Hall effect, which can be read off from the $-4\mathrm{CS}_g$ term in the Lagrangian. Physically, this background contribution comes from combining the thermal Hall offset native to the ${\mathrm{U}(1)_{+3}\leftrightarrow\mathrm{SU}(3)_{-1}}$ level-rank duality with the Chern band addition associated with the particle-hole transformation needed to produce a theory valid at $\nu=2/3$. 

The final result for the chiral central charge of the CVL superconductor is therefore
\begin{equation}
    c_- = \frac{9}{2} - 2 = \frac{5}{2}\, .
\end{equation}
Of the proposals thus far for uniform topological superconductivity proximate to the ${\nu=2/3}$ state~\cite{Shi:2025arn,lotric2026,Fan:2026kay,Wang:2026ziw}, the CVL superconductor constitutes the most direct realization. It is also manifestly compatible with the approximate SU$(3)$ valley symmetry. In the Sections below, we demonstrate that the same mechanisms discussed in these works (with one intriguing exception) can also be embedded as instabilities of the quark Fermi surface model. 

In the strong pairing limit, the MZMs are absent, and the chiral central charge is determined entirely by the background contribution, resulting in $c_-=-2$. Although this phase is adiabatically connected to the $c_-=-2$ superconductor obtained by doping charge-$2e/3$ bound states in Ref.~\cite{Shi2025c}, the dynamics giving rise to the superconductor here rely on the simultaneous presence of both color and valley degrees of freedom. In contrast, the superconductor found in Ref.~\cite{Shi2025c} could also have been obtained starting from a field theory with no valley symmetry implemented at all -- this strongly suggests that the anyon dynamics in the two scenarios should be considered distinct, despite the equivalence of the final ground state superconducting phases. We develop a proper understanding of how the physics described in Ref.~\cite{Shi2025c} can be embedded into the quark Fermi surface model in Section~\ref{sec:bound_state}.

We also remark that a superconductor with $c_-=-13/2$ is also possible if the quarks choose to pair in the $p-ip$ channel. However, this requires subverting the tree level expectation that this channel is repulsive.


\subsubsection{SC$\star$ phases from intra-valley, color-symmetric pairing}

While the CVL superconductor corresponds to the simplest pairing instability of the quark Fermi surface, it is also possible for quarks to pair symmetrically. In fact, gauge fluctuations mediate attractive interactions in these channels as well, albeit with $p-ip$ ($\ell=+1$) pairing symmetry (see Appendix~\ref{app:attr_vs_repuls_tree_level}). Unlike the CVL, color-symmetric pairing Higgses the gauge group to a non-trivial subgroup. The resulting phase is a superconductor coexisting with topological order, often dubbed a SC$\star$ phase. If the pairing is egalitarian among the valleys, the superconductivity has a spatially uniform component; it is not a pair density wave in the strictest sense. However, in general these SC orders also carry spatially modulated components, which break translation symmetry.

We write the color-symmetric order parameter in momentum space as 
\begin{equation}
\label{eq: color sym order param}
    \Delta_{\alpha\beta,IJ} = \overline{\Delta}\, (q_x+iq_y)\, M_{\alpha\beta}\, \delta_{IJ}\, ,
\end{equation}
where we assume the pairing is within each valley. We focus on two possibilities for the symmetric matrix, $M_{\alpha\beta}$: (1) purely intra-color pairing and (2) inter-color symmetric pairing. 

We start with intra-color pairing,
\begin{equation}
    M^\mathrm{intra}_{\alpha\beta} = \delta_{\alpha\beta}\, .
\end{equation}
Because $M_{\alpha\beta}$ is in the symmetric representation of U$(3)$, it transforms as ${M\rightarrow UMU^T}$. Evidently, intra-color pairing Higgses the $\U(3)$ gauge group down to $\O(3) = \SO(3)\times\mathbb{Z}_2$, with the $\SO(3)$ factor generated by $(2iT^2, 2iT^5, 2iT^7 )$, while the $\mathbb{Z}_2$ factor corresponds to the transformations $U=\pm\bs{1}_3$. Following the discussion below Eq.~\eqref{eq: U(3) as a quotient}, we see that this $\mathbb{Z}_2$ subgroup of the $\U(3)$ gauge symmetry is identified with $\mathbb{Z}_2$ subgroups of $\U(1)_\mathrm{EM}$ via the Lagrange multiplier constraint, and we find charge-$2e$ superconductivity.

The three components of the $\SO(3)$ gauge field can be parameterized as 
\begin{equation}
    b' = \begin{pmatrix}
        0 & b'_1 & b'_2\\
        -b'_1 & 0 & b'_3\\
        -b'_2 & -b'_3 & 0
    \end{pmatrix}\, ,
\end{equation}
and the $\U(3)_{-1}$ Lagrangian gets Higgsed to $\SO(3)_{-2}$, supplemented with additional counterterms that depend on $A$, as shown in Appendix~\ref{app:color_sc_eff_lags}. A brief summary of Chern-Simons theories with orthogonal gauge groups can be found in Appendix~\ref{app:O(N)_andZ2_CS_theory}.

The quarks are charged under the residual $\SO(3)$ gauge symmetry, and each valley transforms under the vector representation. The chirality of quark pairing implies that an additional topological response is generated for the residual gauge field upon integrating out the Bogoliubov quasiparticles. For a single valley, integrating out quarks paired with angular momentum $\ell$ results in a $\SO(3)_{-\ell}$ response term, so that three valleys with $\ell = 1$ generates $\SO(3)_{-3}$. This shifts the level of the $\SO(3)$ Chern-Simons term to $-5$, 
\begin{equation}\label{eq:intra-color_eff_Lag}
    \mathcal{L}_\eff = - \frac{5}{8\pi} \Tr \left( b'db' + \frac{2}{3}b'^3 \right) + \frac{2}{2\pi} \alpha dA - \frac{2}{4\pi} AdA - 4\CS_g + \ldots\, .
\end{equation}
Note that the normalization of Chern-Simons terms for $\SO(N)$ theories differs from that of $\SU(N)$ theories by a factor of $2$ (see, e.g.,~\cite{Aharony2016b,Cordova2018}). This Lagrangian describes an SC$\star$ phase with a \textit{non-abelian} $\SO(3)_{-5}$ topological order, the TQFT part of which has chiral central charge $+5/2$.

In Appendix~\ref{app:BdG_Majorana_vortices}, we present a detailed BdG analysis of the superconductor, including a calculation of zero energy edge modes. We find nine MZMs that can bind to an $h/2e$ vortex, each of which contribute $-1/2$ to the chiral central charge due to the $p-ip$ pairing of quarks. These nine MZMs arrange themselves into three vectors that transform the residual $\SO(3)$ gauge symmetry -- one per valley. Combined with the TQFT part as well as the background contribution, the total chiral central charge of this phase is
\begin{equation}
    c_- = -4\, .
\end{equation}
This SC$\star$ phase is quite exotic. It is a testament to the breadth of superconducting phases possible, both with and without topological order, that can in principle arise as instabilities of an anyon gas.

The gauge invariant Cooper pair can be constructed from the color-symmetric order parameter by taking the determinant in color space. In momentum space,
\begin{equation}
    \Delta^\mathrm{el}_{II',JJ',KK'} = \frac{1}{6}\varepsilon^{\alpha\beta\gamma} \varepsilon^{\alpha'\beta'\gamma'} \Delta_{\alpha\alpha',II'} \Delta_{\beta\beta',JJ'} \Delta_{\gamma\gamma',KK'} \sim \left(\overline{\Delta}\right)^3 (q_x+iq_y)^3 \delta_{II'} \delta_{JJ'} \delta_{KK'}\, .
\end{equation}
It carries three independent valley indices, $I,J,$ and $K$. When the three indices are all distinct, the corresponding component of the gauge invariant order parameter carries no net momentum. However, when two out of three valley indices are identical, the wave vectors of the valley minima may not add to a reciprocal lattice vector, and the resulting phase is spatially modulated in addition to carrying a uniform SC component.

We can also consider the strong pairing regime. There the Higgsed Lagrangian remains unchanged, but the $\SO(3)_{-3}$ topological response generated by the paired quarks is no longer present, and neither are the MZMs. The result is then a different SC$\star$ phase with $\SO(3)_{-2}$ topological order and chiral central charge $0$.

We now move on to the case where quarks pair symmetrically between colors. This case is characterized by the matrix,
\begin{equation}
    M^\mathrm{inter} = \begin{pmatrix}
        0 & 1 & 1\\
        1 & 0 & 1\\
        1 & 1 & 0
    \end{pmatrix}\, = U_\Delta \begin{pmatrix}
        2 & 0 & 0\\
        0 & -1 & 0\\
        0 & 0 & -1
    \end{pmatrix} U_\Delta^T\, ,
\end{equation}
where we have diagonalized the matrix using a $\U(3)$ transformation $U$ in the second equality to make the residual symmetry more obvious. Evidently, this pairing matrix Higgses the $\U(3)$ gauge symmetry to $\O(2)\times\mathbb{Z}_2$, with the $\mathbb{Z}_2$ factor acting on the first diagonal component, given by the matrices $\mathrm{diag}(\pm 1,1,1)$ in the diagonalized basis, while the $\O(2)$ factor acts on the eigenspace corresponding to eigenvalue $-1$. From here onward we will work exclusively in the diagonalized basis, renaming cyan as the eigenvector corresponding to eigenvalue $2$, and magenta-yellow the eigenspace for eigenvalue $-1$.

Whereas $\O(3)$ can be decomposed into a direct product of $\SO(3)$ and $\mathbb{Z}_2$, $\O(2)$ is instead a semi-direct product, $\SO(2)\rtimes\mathbb{Z}_2$. The $\SO(2)$ factor is generated by $2iT^5$ in the diagonalized basis, while the $\mathbb{Z}_2$ factor can be chosen to reflect either along the yellow or magenta axes, i.e., $\mathrm{diag}(1,\pm 1,1)$ or $\mathrm{diag}(1,1,\pm 1)$, respectively. The two choices are related by a $\pi$-rotation coming from the $\SO(2)$ factor, and we will assume the former~\cite{Cordova2018}.

The residual gauge symmetry is then $\left(\SO(2)\rtimes\mathbb{Z}^m_2\right)\times\mathbb{Z}^c_2$, where we use the superscripts to indicate which component the $\mathbb{Z}_2$ symmetry acts on (e.g. $m$ stands for the color ``magenta''). This residual symmetry
includes the $\mathbb{Z}_2$ subgroup of transformations $U=\pm\bs{1}_3$, which corresponds to the $\mathbb{Z}_2$ subgroup of $\U(1)_\mathrm{EM}$, confirming that this state is indeed a charge-$2e$ superconductor. 

Because the group O$(2)$ 
contains both $\mathbb{Z}_2^m$ and SO$(2)$ factors, the U$(3)_{-1}$ Chern-Simons term in the Lagrangian decomposes to a sum of SO$(2)$ and $\mathbb{Z}_2^m$ Chern-Simons terms, which need not have the same level.  Obtaining the SO$(2)$ level may be done straightforwardly by noting the $\SO(2)$ gauge field is embedded into the $\U(3)$ gauge field as
\begin{equation}
    a_{\SO(2)} = \begin{pmatrix}
        0 & 0 & 0\\
        0 & 0 & a\\
        0 & -a & 0
    \end{pmatrix}\, .
\end{equation}
Plugging this into the Lagrangian and noting the difference in level quantization produces a SO$(2)_{-2}$ Chern-Simons term. Obtaining the $\mathbb{Z}_2^m$ level is much more subtle, as it cannot be obtained from directly expanding the Chern-Simons term in the Lagrangian and is unrelated to the chiral central charge. In Appendix~\ref{app:color_sc_eff_lags}, we present an argument based on how $\mathbb{Z}_2^m$ is embedded into the U$(3)$ gauge group that implies its level should be $-4$ (after implementing the constraint from $c_\mu$ that reduces U$(2)$ to SU$(2)$). We therefore find that the Chern-Simons term in the Lagrangian is Higgsed to $\O(2)_{-2,-4}$, where the two levels in the subscript are for the $\SO(2)$ the $\mathbb{Z}_2$ subgroups (see Appendix~\ref{app:O(N)_andZ2_CS_theory} for a brief review of $\O(N)$ TQFTs).

Integrating out weakly paired quarks (see Appendix~\ref{app:BdG_Majorana_vortices}) produces separate responses for the SO$(2)$ and $\mathbb{Z}_2^m$ gauge fields as well, and these must also be kept track of. Specifically, the cyan quarks are charged under $\mathbb{Z}_2^c$ but neutral under $\O(2)$, while the magenta-yellow doublet of quarks is charged under $\mathbb{Z}_2^m$ and transforms in the vector representation of $\SO(2)$. Integrating out quark pairs carrying angular momentum $\ell$ then generates an $\O(2)_{-\ell,-\ell}$ response per valley. For $\ell=1$, the three valleys generate an additional $\O(2)_{-3,-3}$ response. Adding this response to the TQFT already present in the Lagrangian gives a O$(2)_{-5,1}$ theory, where we used the fact that the $\mathbb{Z}_2^m$ level is only defined modulo $8$. We finally conclude that the SC$\star$ phase for intra-valley, color-symmetric pairing can be characterized by the Lagrangian,
\begin{equation}
    \Lag_\eff = -\frac{5}{8\pi} \Tr \left( a_{\SO(2)} d a_{\SO(2)} \right) + \Lag_{(\mathbb{Z}_2)_1}[\widetilde{b}^m] + \frac{2}{2\pi} \alpha dA - 4\mathrm{CS}_g + \ldots\, .
\end{equation}
The first term is an $\SO(2)_{-5} = \U(1)_5$ theory. The second term describes a $(\mathbb{Z}_2)_1$ gauge theory, or equivalently a gauged fermionic SPT. Together, these two terms describe an abelian $\O(2)_{-5,1}$ topological order accompanying the charge-$2e$ superconductor. The ellipses may include additional couplings of $\widetilde{b}^m$ to background gauge fields, which our arguments in Appendix~\ref{app:color_sc_eff_lags} do not uniquely fix.

To complete the analysis of this superconductor and obtain its chiral central charge, we need to include the gravitational response. In Appendix~\ref{app:BdG_Majorana_vortices} we perform a detailed BdG analysis where we find nine MZMs (corresponding to the three colors per valley) capable of binding to $h/2e$ vortices, which contribute a net $-9/2$ to the chiral central charge. Of these nine modes, three (one per valley) are charged under $\mathbb{Z}_2^c$, but neutral under $\O(2)$, while the remaining six organize themselves into three $\O(2)$ vectors that are neutral under $\mathbb{Z}_2^c$. Adding these to the contribution from the topological sector, the total chiral central charge is
\begin{equation}
    c_- = -\frac{11}{2} = \frac{5}{2} \mod 8\, .
\end{equation}
The strong pairing regime of this phase has the Higgsed $\O(2)_{-2,-4}$ Lagrangian, but the additional $\O(2)_{-3,-3}$ response is not generated, and neither are the Majorana defects. The result is a different SC$\star$ phase with $\O(2)_{-2,-4}$ topological order and chiral central charge $-1$.

The gauge invariant Cooper pair can be constructed from the quark order parameter in the same way as for intra-color pairing. In momentum space,
\begin{equation}
    \Delta^\mathrm{el}_{II',JJ',KK'} = \frac{1}{6}\varepsilon^{\alpha\beta\gamma} \varepsilon^{\alpha'\beta'\gamma'} \Delta_{\alpha\alpha',II'} \Delta_{\beta\beta',JJ'} \Delta_{\gamma\gamma',KK'} \sim \left(2\overline{\Delta}\right)^3 (q_x+iq_y)^3 \delta_{II'} \delta_{JJ'} \delta_{KK'}\, .
\end{equation}
The dependence of the Cooper pair on valley indices is also identical to the case of intra-color pairing, and we conclude that the superconductor generally exhibits spatial modulation on top of a uniform component.


\subsection{Color ferromagnetism}

\begin{figure}[t]
    \centering
    \includegraphics[width=0.6\linewidth]{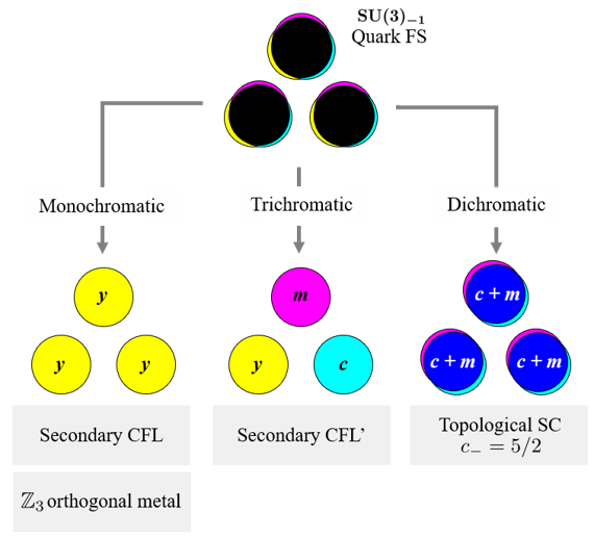}
    \caption{Three scenarios for color ferromagnetic instabilities of the black Fermi surface: (1) \textit{monochromatic valley polarization} leading to either a secondary composite Fermi liquid~\cite{Shi2025c} or a $\mathbb{Z}_3$ orthogonal metal \cite{Shi2026,senthil2026fractionalizedmetalsdopedanyons,han2026orthogonal} differentiated by fractionalization of the remaining quarks, (2) \textit{trichromatic valley polarization} leading to a different secondary composite Fermi liquid with no net color flux in mean field theory~\cite{Fan:2026kay}, and (3) \textit{dichromatic order in each valley} leading to a topological chiral superconductor~\cite{Shi2025c,lotric2026}.}
    \label{fig:colorFM}
\end{figure}

Having surveyed the possible pairing instabilities of the quark Fermi surface, we turn our attention to itinerant ferromagnetism, which causes the originally black (color unpolarized) Fermi surfaces to pick one or more colors. This effect may occur due to an interplay of gauge fluctuation effects with electrostatic interactions, and it competes with color superconductivity. We anticipate it to be preferred in lower density regimes or in the presence of strong Coulomb repulsion, where color superconductivity might be less energetically favored.

When the quark Fermi surface color polarizes, Eq.~\eqref{eq: non abelian Gauss general} tells us that color flux is generated, meaning that anyons in such a state experience one another as magnetic flux. As a result, they fill Landau levels as in Laughlin's original mechanism for anyon superconductivity. Surprisingly, through color ferromagnetism the quark Fermi surface can capture almost all of the earlier proposed anyon-driven phases associated with doping charge-$e/3$ anyons into the $\nu=2/3$ state (see Figure~\ref{fig:colorFM} for a summary of color ferromagnetic phases). 

The starting point for characterizing the various possible ferromagnetic instabilities is the color density in each valley, defined by
\begin{equation}
      C^a_I = \langle \chi^\dagger_I T^a \chi_I \rangle\, ,
\end{equation}
where $C_I^a$ are the components of the $3\times 3$ matrix $C_I$ in the standard basis for the adjoint representation. In spinor components,
\begin{align}
(C_I)^\alpha_{\,\,\,\beta} &= \sum_{a=1}^8 C_I^a T^a = \frac{1}{2} \langle \chi^\dagger_{I,\beta} \chi_I^\alpha \rangle - \frac{\sum_{\gamma} \langle \chi^\dagger_{I,\gamma} \chi_I^\gamma \rangle}{6} \delta^\alpha_{\,\,\,\beta}\,.
\end{align}
Since this matrix can always be diagonalized, the important components are $C^3$ and $C^8$, which can be expressed in terms of the quark densities, $\rho_I^\alpha = \langle \chi^\dagger_{I,\alpha} \chi_I^\alpha \rangle$, as
\begin{equation}
    C_I^3 = \frac{1}{2} \left( \rho_I^c - \rho_I^m \right)\, , \qquad C_I^8 = \frac{1}{2\sqrt{3}} \left( \rho_I^c + \rho_I^m - 2\rho_I^y \right)\, .
\end{equation}
The values of each color density in each valley determine the color-polarization pattern, as well as the fluxes for $b^3$ and $b^8$ which are pinned to the values of $\sum_I C_I^3$ and $\sum_I C_I^8$ respectively. The $\U(3)$ embedding of the $\SU(3)$ gauge group allows us to work in a color-resolved basis, $(\widetilde{T}^c, \widetilde{T}^m, \widetilde{T}^y)$, for the diagonal generators, $(T^3, T^8, \bs{1}_3)$, defined as follows:
\begin{align}
    \widetilde{b}^c\,\widetilde{T}^c=\begin{pmatrix} \widetilde{b}^c& 0 & 0 \\ 0&0&0\\0&0&0\end{pmatrix}\,,~~\widetilde{b}^m\,\widetilde{T}^m=\begin{pmatrix} 0& 0 & 0 \\ 0&\widetilde{b}^m&0\\0&0&0\end{pmatrix}\,,~~\widetilde{b}^y\,\widetilde{T}^y=\begin{pmatrix} 0& 0 & 0 \\ 0&0&0\\0&0&\widetilde{b}^y\end{pmatrix}\,.
\end{align}
Each quark color couples only to the corresponding color-resolved components. We can then define a set of 9 filling fractions via the charge-flux relation for each quark,
\begin{equation}
    \rho^c_I = \nu_I^c \frac{\varepsilon_{ij}\d_i\widetilde{b}^c_j}{2\pi}\, , \qquad 
    \rho^m_I = \nu_I^c \frac{\varepsilon_{ij}\d_i\widetilde{b}^m_j}{2\pi}\, , \qquad 
    \rho^y_I = \nu_I^c \frac{\varepsilon_{ij}\d_i\widetilde{b}^y_j}{2\pi}\, ,
\end{equation}
and different phases are characterized by different assignments of filling fractions.

Any generic set of filling fractions will Higgs the $\U(3)$ gauge symmetry down to $\U(1)^3$, where the three $\U(1)$ factors correspond to the color-resolved generators. 
In order to preserve lattice translation symmetry, we will assume that the total quark density per valley, ${\rho_I^c + \rho_I^m + \rho_I^y}$, is the same for all valleys. Hence, when two out of three filling fractions in a valley are identical, and assignment of filling fractions is valley-independent, the residual gauge symmetry is instead enlarged to ${\U(2)\times\U(1)}$.

Our focus will be on the simplest color-polarized phases, where the emergent quark Landau levels are either completely or half-filled. Which particular order is preferred at zero temperature depends on microscopic energetics. In particular, we consider the following three cases: \textit{monochromatic valleys}, where each valley polarizes into a single color, e.g.\ yellow, independent of the valley, \textit{trichromatic valleys}, where each valley polarizes into a single, distinct color, and \textit{dichromatic order in each valley}, where each valley polarizes into an equal mixture of two colors, e.g.\ cyan and magenta, independent of the valley.

For monochromatic valleys, we find that the quarks form Landau levels at half-integer filling, leading on further fractionalization of the quarks to a metallic state: either the secondary CFL of Ref.~\cite{Shi2025c} or the $\mathbb{Z}_3$ orthogonal metal introduced in Refs.~\cite{Shi2026,senthil2026fractionalizedmetalsdopedanyons,han2026orthogonal}. For trichromatic valleys, we find a ``color-valley-locked'' CFL that is identical to the theory of Ref.~\cite{Zhang2025a,Fan:2026kay}. These compressible phases can then experience pairing and ultimately produce superconductivity or other correlated phases at zero temperature. On the other hand, when each valley has dichromatic order, the system immediately forms a topological chiral superconductor with $c_- = 5/2$ without need of an intermediate compressible phase~\cite{Shi2025c,lotric2026}.

These three solutions already capture many of the phases that have been studied through distinct abelian models and parton constructions, and it is surprising that they can all be collected into different instabilities of a single nonabelian model with well-defined order parameters. The quark model hence allows a systematic study of the competition between these different proposed phases.


\subsubsection{Monochromatic valleys: Secondary CFL or orthogonal metal}

The simplest color ferromagnet has all three valleys take the same color, say yellow,
\begin{align}
\langle \chi^\dagger_{y,I}\chi_{y,I}\rangle = \delta \neq 0\,,\qquad \langle \chi^\dagger_{c,I}\chi_{c,I}\rangle=\langle \chi^\dagger_{m,I}\chi_{m,I}\rangle=0\,,
\end{align}
where no sum over the valley index, $I$ is assumed (we denote sums over valley indices explicitly throughout the discussion below). In terms of the order parameters $C_I^3$ and $C_I^8$, this requires $C_I^3 = 0$ and $C^8_I < 0$\footnote{Unlike $\SU(2)$ ferromagnets, the sign of $C^8$ is indeed invariant under $\SU(3)$ when $C^3=0$, being related to the sign of the determinant of the matrix $C^\alpha_\beta$. This makes $C^8<0$ and $C^8>0$ distinct configurations.}, with equal magnitude for all three valleys. This corresponds to the  order parameter matrix, ${C_I=C_I^aT^a}$,
\begin{equation}
\label{eq: monochrome C}
    C_I = \frac{\delta}{6}\begin{pmatrix}
        -1 & 0 & 0\\
        0 & -1 & 0\\
        0 & 0 & 2
    \end{pmatrix}\, ,
\end{equation}
which evidently preserves a $\U(2)\times\U(1)$ subgroup of the $\U(3)$ gauge symmetry, with the $\U(2)$ factor generated by the cyan-magenta subspace.

We assume that the depleted cyan and magenta quarks form a trivial band insulator, leaving only the yellow quarks, $\chi_I^y$. The Higgsed gauge field is parameterized as
\begin{equation}
    b = \begin{pmatrix}
        b' & 0\\
        0 & \widetilde{b}^y
    \end{pmatrix}\, , \qquad \Tr b = \Tr b' + \widetilde{b}^y\, ,
\end{equation}
where $b'$ is a $\U(2)$ gauge field and $\tilde{b}^y$ is abelian. The Higgsed Lagrangian then takes the following form:
\begin{equation}\label{eq:u(2)xu(1)_Higgsed_Lag}
    \Lag =
    \Lag_{\chi^y}[\widetilde{b}^y] - \frac{1}{4\pi} \Tr\left( b'db' + \frac{2}{3} b'^3 \right) - \frac{1}{4\pi} \widetilde{b}^y d \widetilde{b}^y + \frac{1}{2\pi}cd(\Tr b' + \widetilde{b}^y - A) + \frac{1}{4\pi}AdA-4\mathrm{CS}_g\, ,
\end{equation}
and Gauss' laws reduce to
\begin{equation}
    \begin{split}
        \sum_I \chi^\dagger_{y,I} \chi^y_{I} =~ &\frac{\varepsilon_{ij}\d_i\widetilde{b}^y_j}{2\pi} - \frac{\varepsilon_{ij}\d_i c_j}{2\pi}\, ,\\
        0 = \frac{\varepsilon_{ij}\d_i\Tr b'_j}{2\pi} - 2\frac{\varepsilon_{ij}\d_i c_j}{2\pi}\, &, \qquad 0 = \varepsilon_{ij}\d_i \left( \Tr b'_j + \widetilde{b}^y_j \right)\, .
    \end{split}
\end{equation}
Since each valley only has yellow quarks, the mean-field values of the gauge fields are related as ${2\langle\varepsilon_{ij}\partial_i c_j\rangle=\langle\varepsilon_{ij}\partial_i\Tr b_j'\rangle=-\langle\varepsilon_{ij}\partial_i \widetilde{b}^y_j\rangle}$, which turns the charge-flux relation for the yellow quarks into
\begin{equation}
    \sum_I \langle \chi^\dagger_{y,I} \chi^y_{I} \rangle = \frac{3}{2} \frac{\langle\varepsilon_{ij}\partial_i \widetilde{b}^y_j\rangle}{2\pi}\,.
\end{equation}

Proceeding further requires fractionalization of the yellow quark, $\chi_y$, which may be understood either through a parton decomposition or more traditional flux attachment. Choosing to distribute the emergent flux evenly among all three valleys, one natural possibility is for each valley to form a composite Fermi liquid (CFL) at half-filling in the ``background'' of $\widetilde{b}_y$. The $\U(2)_{-1}$ Chern-Simons term in the Lagrangian is a trivial theory that does not directly couple to any matter, meaning that it can be integrated out (see Appendix~\ref{app:CS_identities}). The effective Lagrangian is written in terms of composite fermions $\xi_I$, which couple to emergent $\Spin_c$ and $\U(1)$ gauge fields $\alpha_I$ and $\beta_I$,
\begin{equation}
    \begin{split}
        \Lag_\eff = ~&\sum_I \left( \mathcal{L}_{\xi_I}[\alpha_I] + \frac{1}{2\pi} (\widetilde{b}^y-\alpha_I) d\beta_I - \frac{2}{4\pi} \beta_I d\beta_I \right)\\
        &- \frac{1}{4\pi} \widetilde{b}^y d \widetilde{b}^y + \frac{2}{4\pi} cdc + \frac{1}{2\pi} \widetilde{b}^y d c - \frac{1}{2\pi} cdA + \frac{1}{4\pi} AdA - 4\CS_g\, .
    \end{split}
\end{equation}
Up to a sign flip, $c\rightarrow-c$, this Lagrangian is identical to the secondary CFL discovered in Ref.~\cite{Shi2025c}. Through pairing, the composite fermions, $\xi$, may realize various superconducting and non-superconducting descendant phases, which were considered in Refs.~\cite{Shi2025c,Shi2026}.

Alternatively, the yellow quarks may fractionalize e.g. as $\chi_I\sim \psi_I\varphi$, where $\varphi$ is a boson carrying EM charge and $\psi_I$ is a composite fermion carrying the valley index.  At $\nu_\varphi=3/2$, $\varphi$ forms the particle-hole conjugate of the $\nu=1/2$ bosonic Laughlin state. This leads also to a composite Fermi surface of the $\psi_I$ fermions, but the gauge field is Higgsed down to $\mathbb{Z}_3$, leading to an orthogonal metal phase discussed recently in Refs.~\cite{Shi2026,senthil2026fractionalizedmetalsdopedanyons,han2026orthogonal}.    


\subsubsection{Trichromatic valleys: A color-valley-locked metal}\label{sec:triochromatic_valleys}

Another possibility is to polarize each valley into a different color, say cyan for the first, magenta for the second, and yellow for the third,
\begin{align}
\langle\chi^\dagger_{c,1}\chi_{c,1}\rangle=\langle\chi^\dagger_{m,2}\chi_{m,2}\rangle=\langle\chi^\dagger_{y,3}\chi_{y,3}\rangle= \delta \neq0\,,
\end{align}
with all other densities vanishing. This configuration is captured by the following values of the order parameter matrices:
\begin{equation}
    C_1 = \frac{\delta}{6}\begin{pmatrix}
        2 & 0 & 0\\
        0 & -1 & 0\\
        0 & 0 & -1
    \end{pmatrix}\, , \qquad C_2 = \frac{\delta}{6}\begin{pmatrix}
        -1 & 0 & 0\\
        0 & 2 & 0\\
        0 & 0 & -1
    \end{pmatrix}\, , \qquad C_3 = \frac{\delta}{6}\begin{pmatrix}
        -1 & 0 & 0\\
        0 & -1 & 0\\
        0 & 0 & 2
    \end{pmatrix}\, .
\end{equation}
Unlike monochromatic coloring, this breaks the $\U(3)$ gauge group to ${\U(1)
\times\U(1)\times\U(1)}$, where each factor is generated by a color-resolved generator. The Higgsed gauge field is then parameterized by its diagonal components only,
\begin{equation}
    b = \begin{pmatrix}
        \widetilde{b}^c & 0 & 0\\
        0 & \widetilde{b}^m & 0\\
        0 & 0 & \widetilde{b}^y
    \end{pmatrix}\, ,
\end{equation}
and the Higgsed Lagrangian is given by
\begin{equation}
    \begin{split}
        \Lag &= \Lag_{\chi_1^c}[\widetilde{b}^c] + \Lag_{\chi_2^m}[\widetilde{b}^m] + \Lag_{\chi_3^y}[\widetilde{b}^y] - \frac{1}{4\pi} \sum_{n=c,m,y} \widetilde{b}^n d \widetilde{b}^n\\
        &\qquad + \frac{1}{2\pi} cd\left( \widetilde{b}^c + \widetilde{b}^m + \widetilde{b}^y - A \right) + \frac{1}{4\pi}AdA-4\mathrm{CS}_g\, .
    \end{split}
\end{equation}
Setting the external magnetic field to zero, the equations of motion may be expressed as
\begin{equation}
    \begin{split}
        \chi^\dagger_{c1} \chi^c_1 &= \frac{\varepsilon_{ij}\d_i\widetilde{b}^c_j}{2\pi} - \frac{\varepsilon_{ij}\d_i c_j}{2\pi}\, , \qquad
        \chi^\dagger_{m2} \chi^m_2 = \frac{\varepsilon_{ij}\d_i\widetilde{b}^m_j}{2\pi} - \frac{\varepsilon_{ij}\d_i c_j}{2\pi}\, ,\\
        \chi^\dagger_{y3} \chi^y_3 &= \frac{\varepsilon_{ij}\d_i\widetilde{b}^y_j}{2\pi} - \frac{\varepsilon_{ij}\d_i c_j}{2\pi}\, , \qquad
        0 = \varepsilon_{ij}\d_i\left( \widetilde{b}^c_j + \widetilde{b}^m_j + \widetilde{b}^y_j \right)\, .
    \end{split}
\end{equation}
Requiring the quark densities in all three valleys to be identical in order to preserve lattice translation invariance, and noting that the mean field value of $\langle\varepsilon_{ij}\partial_ic_j\rangle/2\pi=\delta$ is the doped charge density, we are forced to set $\langle \varepsilon_{ij} \d_i\widetilde{b}^c_j \rangle = \langle \varepsilon_{ij} \d_i\widetilde{b}^m_j \rangle = \langle \varepsilon_{ij} \d_i\widetilde{b}^y_j \rangle = 0$. The three charge-$e/3$ quark colors then mutually form Fermi surfaces in each valley and remain metallic.

The resulting theory is none other than the composite fermion theory introduced in Ref.~\cite{Zhang2025a,Fan:2026kay}. The Lagrange multiplier can be integrated out to obtain
\begin{equation}
\label{eq: Fan Lagrangian}
    \begin{split}
        \Lag &= \Lag_{\chi_1^c}[\widetilde{b}^c] + \Lag_{\chi_2^m}[\widetilde{b}^m] + \Lag_{\chi_3^y}[-\widetilde{b}^c-\widetilde{b}^m+A] - \frac{2}{4\pi} \widetilde{b}^c d \widetilde{b}^c - \frac{2}{4\pi} \widetilde{b}^m d \widetilde{b}^m - \frac{1}{2\pi} \widetilde{b}^c d \widetilde{b}^m\\
        &\qquad + \frac{1}{4\pi}AdA-4\mathrm{CS}_g\, .
    \end{split}
\end{equation}
Under weak $p+ip$ inter-pocket pairing, Ref.~\cite{Fan:2026kay} found that this state gives rise to a $f+if$ superconductor with chiral central charge $c_- = -1/2$. 

The physics of this state closely echoes the SU$(3)_{-1}$ quark metal: It is again a system of charge-$e/3$ fermions strongly coupled to gauge fields, but in this case there are fewer fluctuating gauge degrees of freedom. One may then envision a situation where a quark metal at high temperatures evolves into this phase in intermediate temperature regimes, leading eventually to superconductivity at the lowest temperature scales. In the future, it would be interesting to further flesh out how this crossover might look, since it requires no additional quark fractionalization. 


\subsubsection{Dichromatic order in each valley: $c_-=5/2$ topological superconductivity}\label{sec:dichromatic}

Lastly, we consider the phase obtained by polarizing each valley into an equal mixture of cyan and magenta, paralleling the projective parton construction carried out in Ref.~\cite{lotric2026}. This phase has the order parameter $C_I^8>0$, 
\begin{equation}
\label{eq: trichrome C}
    C_I = \frac{\delta}{6} \begin{pmatrix}
        1 & 0 & 0\\
        0 & 1 & 0\\
        0 & 0 & -2
    \end{pmatrix}\, .
\end{equation}
We have assumed that the quark densities are equal in each valley to preserve lattice translation invariance, so that the coefficient of the matrix is independent of the valley index $I$. Evidently, the order parameter Higgses the $\U(3)$ gauge symmetry down to $\U(2)\times\U(1)$, where the $\U(2)$ subgroup rotates the cyan-magenta subspace and the $\U(1)$ factor acts on yellow quarks. The Higgsed Lagrangian takes the same form as Eq.~\eqref{eq:u(2)xu(1)_Higgsed_Lag}:
\begin{equation}
    \Lag = \Lag_\psi[b'] + \Lag_{\chi^y}[\widetilde{b}^y] - \frac{1}{4\pi} \Tr \left( b'db' + \frac{2}{3}b'^3 \right) - \frac{1}{4\pi} \widetilde{b}^y d \widetilde{b}^y + \frac{1}{2\pi} cd(\Tr b' + \widetilde{b}^y - A) + \frac{1}{4\pi}AdA-4\mathrm{CS}_g\,,
\end{equation}
where $\psi$ is the cyan-magenta doublet and $b'$ is a $\U(2)$ gauge field. Gauss' laws reduce to
\begin{equation}
    \begin{split}
        \sum_I \psi^\dagger_I \psi_I = \frac{\varepsilon_{ij}\d_i\Tr b'_j}{2\pi} &- 2\frac{\varepsilon_{ij}\d_i c_j}{2\pi}\, ,\\
        0 = \frac{\varepsilon_{ij}\d_i\widetilde{b}^y_j}{2\pi} - \frac{\varepsilon_{ij}\d_i c_j}{2\pi}\, , \qquad 0 &= \varepsilon_{ij}\d_i \left( \Tr b'_j + \widetilde{b}^y_j \right)\, .
    \end{split}
\end{equation}
In contrast to monochromatic coloring, since $C^8_I > 0$, we look for mean-field solutions where the yellow quarks have vanishing density, which requires $\langle \varepsilon_{ij} \d_ic_j \rangle = \langle \varepsilon_{ij} \d_i\widetilde{b}^y_j \rangle = -\langle \varepsilon_{ij} \d_i\Tr b'_j \rangle$, turning the charge-flux relation for the cyan-magenta doublet into
\begin{equation}
    \sum_I \langle \psi^\dagger_I \psi_I \rangle = 3 \frac{\langle \varepsilon_{ij} \d_i\Tr b'_j \rangle}{2\pi}\,.
\end{equation}
Distributing the flux equally among the three valleys, each valley fills a single Landau level in $\Tr b'$, and integrating them out generates a $\U(2)_3$ Chern-Simons term in $b'$, modifying the level of the native Chern-Simons term in the Higgsed Lagrangian. After some algebra we find that the effective Lagrangian is a $\U(2)_{2,0}$ Chern-Simons term given by
\begin{equation}
    \Lag = \frac{2}{4\pi} \Tr \left( b'db' + \frac{2}{3} b'^3 \right) - \frac{1}{4\pi} (\Tr b')d(\Tr b') + \frac{1}{2\pi}(\Tr b')dA + \frac{4}{4\pi} AdA + 8\CS_g\, ,
\end{equation}
which describes a topological chiral superconductor with a non-abelian, invertible $\SU(2)_2$ neutral sector, and chiral central charge,
\begin{equation}
    c_- = \frac{5}{2}\, .
\end{equation}
To see that this is indeed a topological superconductor, one can note that the non-abelian sector of the $\U(2)_{2,0}$ theory captures the neutral Ising and Majorana defects, with the latter of which binding to $h/2e$ vortices. The charge-$2e$ Meissner effect can be seen from the mixed Chern-Simons term between $\Tr b'$ and $A$. 


\subsection{Technicolor anyons: Charge-$2e/3$ bound state formation}\label{sec:bound_state}

\begin{table}[t]
    \centering
    \begin{tabular}{c|c|c}
        Color polarization scenario & Doped state & $c_-$ \\ \hline\hline
        Monochromatic valleys & Chiral SC & $-2$\\ 
        Trichromatic valleys & Chiral SC & $-2$\\ 
        Dichromatic order per valley & Charge-$4e$ SC$\star$ & $0$
    \end{tabular}
    \caption{Primary superconducting descendants of the $2e/3$ bound states. We assume bound state formation occurs within each valley. \textit{Monochromatic valley polarization} reproduces the superconductor obtained in Ref.~\cite{Shi2025c} \textit{Trichromatic valley polarization} reproduces the same phase without activating any non-abelian fluxes. \textit{Dichromatic order per valley} results in the charge-$4e$ SC$\star$ found in Ref.~\cite{Shi:2025arn} and elaborated upon in Ref.~\cite{Han2026}.}
    \label{tab:2/3_bound_phases}
\end{table}

In the discussion above, we considered the possibilities for a gas of charge-$e/3$ anyons. However, it is also possible for anyons to form bound states, leading to a gas of charge-$2e/3$ anyons, which is naturally expected to lead to superconductivity~\cite{Shi2025c}. This may occur, for example, when Coulomb repulsion is sufficiently screened. Remarkably, the quark metal framework also allows us to consider this possibility and derive the corresponding superconducting phases.

Level-rank duality offers a straightforward path to build higher-charge anyon composites out of quark fields. The charge-$2e/3$ anyon, for example, is simply the color-antisymmetric combination of two $\chi$'s. We may therefore consider a scenario where the charge-$2e/3$ anyon becomes lighter than the charge-$e/3$ anyon, in the sense that quarks form spatially local charge-$2e/3$ pairs in each valley,
\begin{equation}\label{eq:pre-formed_pair_field}
    \Phi^\alpha_I(x) = \varepsilon^{\alpha\beta\gamma} \chi^\dagger_{\beta,I}(x) \chi^\dagger_{\gamma,I}(x)\, ,
\end{equation}
where no sum over $I$ is assumed on the right-hand-side. This operator may be understood as creating a charge-$2e/3$ anyon and, like $\chi$, resides in the fundamental representation of SU$(3)$. Note that the sign of the field $\Phi$'s EM charge has opposite sign to that of the field $\chi$ -- we make this choice to avoid working with fields in the \emph{anti}fundamental representation below. Nonetheless, we continue to take the physical vacuum as normal ordered with respect to $\chi$.

The scenario of bound state formation is distinct from the color superconductivity encountered in Section~\ref{sec:CVL_sc} in two key ways. First, we primarily focused on color superconductors in the BCS limit, where the coherence length is large and pairing occurs as a Cooper instability of the quark Fermi surface. The bound states above are strictly in the opposite limit, where the coherence length is taken to be vanishingly small (we can relax this for one example below). Second, the particular pairing channel we consider here is distinct: It is antisymmetric in color but diagonal in valley indices. As a result, each valley may be assigned a linear combination of colors, and SU$(3)$ flux can be generated depending on the polarization of $\Phi_I^\alpha$, an effect that we did not have to consider for the color superconductors. As a result, we find it most convenient to start by considering a case where bound states form -- in the sense that single quark excitations are gapped -- but they do not necessarily condense, i.e. $\langle\Phi_I^\alpha\rangle=0$.  

We introduce the bound state field via a Hubbard-Stratonovich transformation to the SU$(3)_{-1}$ quark model. It couples through the Lagrangian,
\begin{equation}\label{eq: HS Lagrangian}
    \mathcal{L}_{\chi\Phi} = \sum_I \Phi^\alpha_I\, \varepsilon_{\alpha\beta\gamma} \chi^\beta_I \chi^\gamma_I + \mathrm{h.c.} - \Phi^\dagger \Phi\, .
\end{equation}
The equation of motion for $\Phi$ enforces the relation in Eq.~\eqref{eq:pre-formed_pair_field}. If the quarks form tight bound states, we may integrate them out to obtain a local Ginzburg-Landau theory for $\Phi^\alpha_I$. In particular, $\Phi_I^\alpha$ should couple minimally to the U$(3)$ gauge field via the covariant derivative,
\begin{equation}
\label{eq: Phi covariant derivative}
    D_\mu^b\Phi_I = \d_\mu \Phi_I - i b_\mu\cdot\Phi_I + i(\Tr b_\mu)\, \Phi_I\, .
\end{equation}
The unusual structure of the covariant derivative follows from the fact that $\Phi^\alpha_I$ possesses an unusual transformation law under the determinant part of U$(3)$. The revised coupling to ${\Tr b = \widetilde{b}^c + \widetilde{b}^m + \widetilde{b}^y}$ can be deduced from the fact that $\Phi^c_I$, $\Phi^m_I$, and $\Phi^y_I$ respectively carry charge under the combination of color-resolved gauge fields, $\widetilde{b}^m+\widetilde{b}^y$, $\widetilde{b}^c+\widetilde{b}^y$, and $\widetilde{b}^c+\widetilde{b}^m$. We may verify that $\Phi^\alpha_I$ creates particles with EM charge-2/3 by integrating out the Lagrange multiplier, $c_\mu$, to pin $\Tr b = A$ to $A$ and obtain
\begin{equation}
    D_\mu^b\Phi_I \rightarrow \d_\mu \Phi_I - i b_{\SU(3),\mu} \cdot \Phi_I + i \frac{2}{3}\,A_\mu\, \Phi_I\, ,
\end{equation}
where we applied the notation ${b=b_{\mathrm{SU}(3)}+\Tr b\,\bs{1}_3/3=b_{\mathrm{SU}(3)}+A\,\bs{1}_3/3}$. 

We therefore arrive at the bound state Lagrangian,
\begin{align}
\label{eq:pre-formed_pairs_Lag}
    \Lag&= \Lag_\Phi\left[ b-(\Tr b)\bs{1}_3 \right] - \frac{1}{4\pi} \Tr \left( bdb + \frac{2}{3}b^3 \right) + \frac{1}{2\pi} cd(\Tr b - A) + \frac{1}{4\pi}AdA-4\mathrm{CS}_g\, ,\\
    \Lag_\Phi&=\Phi^{\dagger}\left[iD^b_t+\frac{1}{2m_\Phi}(D^b_i)^2\right]\Phi-V[\Phi]\,.
\end{align}
This Lagrangian appears nearly identical to the quark metal Lagrangian, Eq.~\eqref{eq:SU(3) full Lagrangian}, with the crucial difference that $\Phi$ is a \emph{bosonic} field, whose covariant derivative is given in Eq.~\eqref{eq: Phi covariant derivative}. Proper bookkeeping of counterterms and couplings to the background Spin$_c$ connection is provided in Appendix~\ref{app:2e/3_gas_counterterms}.

Under doping, there are two natural possibilities for $\Phi^\alpha_I$: condensation and color polarization. Because superfluid orders for $\Phi^\alpha_I$ also generically produce color polarization, we will view the latter as the simpler order and organize our discussion around cases where ${\langle\Phi_I^\alpha\rangle=0}$ but ${\langle\Phi^\dagger_I T^a\Phi_I\rangle\neq0}$. Furthermore, due to its valley index, condensation of $\Phi_I^\alpha$ generically leads to breaking of translation symmetry, and our focus is primarily on uniform phases. Table~\ref{tab:2/3_bound_phases} summarizes the three phases considered in this Section.


\subsubsection{Monochromatic valleys: Recovering the Laughlin mechanism}

We first imagine that each valley polarizes to the same color, causing the anyons to feel one another as color flux. As was the case when we considered color ferromagnetism of quarks, this situation can be captured through the order parameter,
\begin{equation}
    \mathcal{C}^8_I = \langle \Phi^\dagger_I T^8 \Phi_I \rangle<0\, ,\qquad\mathcal{C}^3_I= \langle \Phi^\dagger_I T^3 \Phi_I \rangle=0\,,
\end{equation}
where $\mathcal{C}_1=\mathcal{C}_2=\mathcal{C}_3$. We orient the order parameter such that there is complete yellow polarization of each valley. 

The $\U(3)$ gauge group is again Higgsed down to $\U(2)\times\U(1)$, which we parameterize as 
\begin{equation}
\label{eq:U(2)U(1)_Higgsed_gauge_field}
    b = \begin{pmatrix}
        b' & 0\\
        0 & \widetilde{b}^y
    \end{pmatrix}\, , \qquad \Tr b = \Tr b' + \widetilde{b}^y\,.
\end{equation}
Because we take the cyan and magenta components to be depleted, we integrate them out and focus only on the yellow bound states, $\Phi^y_I$, which carry charge exclusively under the determinant part of the U$(2)$ factor,
\begin{equation}
    D_\mu \Phi^y_I = \d_\mu \Phi^y_I + i \,(\Tr b'_\mu)\, \Phi^y_I\, .
\end{equation}
But the Lagrange multiplier constraint due to $c_\mu$ identifies $\Tr b' + \widetilde{b}^y = A$. We may therefore replace $\Tr b'$ with $-\widetilde{b}^y+A$ to obtain
\begin{equation}\label{eq:pre-formed_Higgsed_Lag}
    \begin{split}
        \Lag & = \Lag_{\Phi_y} \left[  \widetilde{b}^y-A \right]- \frac{1}{4\pi} \Tr \left( b'db'+\frac{2}{3}b'^3 \right)\\
        &\qquad  - \frac{1}{4\pi} \widetilde{b}^y d \widetilde{b}^y + \frac{1}{2\pi} cd \left( \Tr b' + \widetilde{b}^y - A \right) + \frac{1}{4\pi}AdA-4\mathrm{CS}_g\, .
    \end{split}
\end{equation}
The remaining calculation proceeds identically to the quark case. The Gauss' laws of the Higgsed theory are the equations of motion,
\begin{equation}\label{eq:pre-formed_Higgsed_Gauss}
    \begin{split}
        \sum_I \Phi^\dagger_{y,I}\, \Phi^y_{I} =~ &\frac{\varepsilon_{ij}\partial_i \widetilde{b}_j^y}{2\pi} - \frac{\varepsilon_{ij}\partial_i c_j}{2\pi}\, ,\\
       0 = \frac{\varepsilon_{ij}\partial_i\Tr b'_j}{2\pi} - 2\frac{\varepsilon_{ij}\partial_i c_j}{2\pi}&\, ,\qquad 0 = \varepsilon_{ij}\partial_i \left( \Tr b'_j + \widetilde{b}_j^y \right)\, ,
    \end{split}
\end{equation}
where we have again assumed zero external magnetic field, $\varepsilon_{ij}\partial_i A_j=0$. A mean field solution where only yellow $\Phi$'s are at finite density requires $\Tr b'=2c=-\widetilde{b}^y$, meaning that the yellow bound states find themselves at total filling $\nu_{\Phi^y}=3/2$,
\begin{equation}
    \sum_I \langle \Phi^\dagger_{y,I} \Phi^y_I \rangle = \frac{3}{2}\frac{\langle \varepsilon_{ij}\partial_i \widetilde{b}_j^y\rangle}{2\pi}\,.
\end{equation}

We assume the $\Phi^y$'s prefer to form a gapped, topologically ordered state. A natural candidate preserving translation invariance is a valley-symmetric Halperin state defined by the $K$-matrix and charge vector,
\begin{equation}
\label{eq: Halperin K matrix}
K = \begin{pmatrix}
        0 & 1 & 1\\
        1 & 0 & 1\\
        1 & 1 & 0
    \end{pmatrix}\, , \qquad q = \matleft{l} 1\\1\\1\matright \,.
\end{equation}
After application of level-rank duality to integrate out the decoupled U$(2)_{-1}$ theory, the remaining dynamical U$(1)$ gauge fields can be integrated out to obtain
\begin{equation}
\label{eq: c- = -2 Lag}
    \Lag_\eff = \frac{2}{2\pi} cdA - \frac{2}{4\pi}AdA-4\mathrm{CS}_g\, ,
\end{equation}
which describes a superconductor with chiral central charge
\begin{equation}
    c_- = -2\, .
\end{equation}

The same result was obtained in Ref.~\cite{Shi2025c} through a fermionic parton construction with abelian gauge fields. There, the basic physical picture for the onset of superconductivity resembled the original proposals from the 1980s~\cite{Laughlin1988,Fetter1989,Chen1989}: Doped charge-$2e/3$ anyons witness one another as flux, filling an integer number of Chern bands to produce a superconductor with chiral central charge $c_-=-2$. However, the connection of these composites to the more elementary charge-$e/3$ anyons is not explicit in the abelian approach.  Level-rank duality makes these charge-$2e/3$ anyons' status as bound states of two charge-$e/3$ anyons explicit, and the analysis here gives a direct, dual embedding of the physics found in Ref.~\cite{Shi2025c} as color ferromagnetism. In fact, the theory in Eq.~\eqref{eq:pre-formed_pairs_Lag} is level-rank dual to three fermion species coupled to a U$(1)_{+3}$ abelian Chern-Simons gauge field~\cite{Aharony2016a,Hsin2016}, which is the traditional way of obtaining charge-$2e/3$ anyons via flux attachment.

We remark that the analysis here did not rely in an essential way on the interplay with valley symmetry, as was the case for e.g. the CVL superconductor. This is despite the use of the valley-symmetric Halperin state in Eq.~\eqref{eq: Halperin K matrix}. Had we considered a single valley, the analysis would not have changed significantly: The $\Phi^y$ bosons would have simply formed a $\nu=3/2$ bosonic Laughlin state (the particle-hole conjugate of the $\nu=1/2$ bosonic Laughlin state). The final superconductor in this case is identical.


\subsubsection{The same superconductor from trichromatic valleys}

We have already encountered the $c_-=-2$ chiral superconductor through two distinct mechanisms. In the first, we obtained it as the BEC limit of the CVL superconductor. In the second, we reached it through formation of anyonic bound states, which subsequently color polarize without becoming phase coherent. Here we consider a third path, where each valley takes a different color. Like in Section~\ref{sec:triochromatic_valleys}, where we considered the analogous color ferromagnet for quarks, this state Higgses the gauge group down to U$(1)^3$ and corresponds to the order parameter matrices,
\begin{equation}
    \mathcal{C}_1 = \frac{\delta}{12}\begin{pmatrix}
        2 & 0 & 0\\
        0 & -1 & 0\\
        0 & 0 & -1
    \end{pmatrix}\, , \qquad \mathcal{C}_2 = \frac{\delta}{12}\begin{pmatrix}
        -1 & 0 & 0\\
        0 & 2 & 0\\
        0 & 0 & -1
    \end{pmatrix}\, , \qquad \mathcal{C}_3 = \frac{\delta}{12}\begin{pmatrix}
        -1 & 0 & 0\\
        0 & -1 & 0\\
        0 & 0 & 2
    \end{pmatrix}\, .
\end{equation}
The equations of motion for the various color-resolved gauge fields are also the same as in Section~\ref{sec:triochromatic_valleys}, meaning that a solution where each valley experiences vanishing flux is possible. As a result, the bound state fields necessarily condense, 
\begin{align}
\langle\Phi_I^\alpha\rangle=\overline\Phi\,\delta^\alpha_I\neq0\,.
\end{align}
Higgsing the combinations of color-resolved gauge fields, $\widetilde{b}^c+\widetilde{b}^m,\widetilde{b}^y+\widetilde{b}^m,\widetilde{b}^c+\widetilde{b}^y$. Plugging in the constraint to eliminate ${\Tr b = \widetilde{b}^c+\widetilde{b}^m+\widetilde{b}^y=A}$ from the Lagrangian directly reproduces the action in Eq.~\eqref{eq: c- = -2 Lag}. Similar to the color-valley-locked superconductor, the charge-$2e$ order parameter can be produced directly by taking 
\begin{align}
\varepsilon_{\alpha\beta\gamma}\Phi^\alpha_I\Phi^\beta_J\Phi^\gamma_K=\left(\overline{\Phi}\right)^3\varepsilon_{IJK}\, ,
\end{align}
from which we can verify that the resulting superconductor is completely uniform.


\subsubsection{A charge-$4e$ SC$\star$ from charge-$2e/3$ bound states}

For the third possibility, we consider each valley polarizing uniformly to equal mixture of cyan and magenta. As for the unbound quarks, the valley-invariant order parameters $\mathcal{C}^8_I>0$ and $\mathcal{C}^3_I=0$ characterize this phase.
We assume that the phase completely depletes yellow bosons in each valley.

Once again, the $\U(3)$ gauge symmetry is Higgsed down to $\U(2)\times \U(1)$, enabling us to work in the same basis as Eq.~\eqref{eq:U(2)U(1)_Higgsed_gauge_field}. In the absence of yellow quarks, $\Phi_I^y$ can be integrated out to leave a trivial effective theory.
The remaining two-component boson in the cyan-magenta subspace is denoted by $\varphi_I = (\Phi_I^c\ \Phi_I^m)^T$, and couples to the $\U(2)$ gauge field $b'$ as well as the $\U(1)$ fields through the covariant derivative,
\begin{equation}
    D_\mu \varphi_I = \partial_\mu \varphi_I - ib'_\mu \cdot \varphi_I
    +i \left( \Tr b'_\mu +\widetilde{b}^y_\mu \right)  \varphi_I\ .
\end{equation}
The constraint $\Tr b' + \widetilde{b}^y = A$ from the Lagrange multiplier simplifies this to a coupling to $b'-A\bs{1}_2$, and the Higgsed Lagrangian is
\begin{equation}
    \begin{split}
        \Lag&= \Lag_\varphi
        \left[ b'-A\bs{1}_2 \right] 
        - \frac{1}{4\pi} \Tr \left( b'db'+\frac{2}{3}b'^3 \right)\\
            &\qquad  - \frac{1}{4\pi} \widetilde{b}^y d \widetilde{b}^y + \frac{1}{2\pi} cd \left( \Tr b' + \widetilde{b}^y - A \right) + \frac{1}{4\pi}AdA-4\mathrm{CS}_g\ .
    \end{split}
\end{equation}
The flux attachment constraints turn into
\begin{equation}
    \begin{split}
        & \qquad \qquad \sum_I  \varphi^\dagger _I \varphi_I = \frac{\varepsilon_{ij}\partial_i \Tr b'_j}{2\pi} - 2 \frac{\varepsilon_{ij}\partial_i c_j}{2\pi}\,, \\
        & 0 = \frac{\varepsilon_{ij}\partial_i \widetilde{b}^y_j}{2\pi} -\frac{\varepsilon_{ij}\partial_i c_j}{2\pi}\, ,
        \qquad
        0 = \varepsilon_{ij}\partial_i(\Tr b'_j + \widetilde{b}_j^y)\ .
    \end{split}
\end{equation}
A mean field solution satisfying these constraints requires $\widetilde{b}^y = c = -\Tr b'$, which leads to the charge-flux relation
\begin{align}
\begin{split}
   \sum_I  \langle\varphi^\dagger _I \varphi_I \rangle
   = 3
   \frac{\langle\varepsilon_{ij}\partial_i \Tr b'_j\rangle}{2\pi}\ .
\end{split}
\end{align}
If translation symmetry is to remain unbroken, the fluxes of $\Tr b'$ must be assigned to the three valleys equally, and thus the bosons in each valley experience a unit flux of $\Tr b'$.
Since $\varphi_I$ is coupled to $b' = b'_{\mathrm{SU(2)}}+\Tr b' \bs{1}_2/2$ with a half charge under $\Tr b'$, it finds itself at filling $\nu_{\varphi_I}=2$.

The most natural candidate for the two-component bosons at filling $\nu_{\varphi_I}=2$ is the bosonic IQH state~\cite{Senthil2013}. Integrating out the $\varphi_I$'s leads to a $\U(2)_{-1,1}$ Chern-Simons response from each valley. Proceeding as in Section~\ref{sec:dichromatic}, we arrive at a theory of a uniform charge-$4e$ superconductor with intrinsic topological order, i.e. a charge-$4e$ SC$\star$,
\begin{equation}
\begin{split}
    \Lag_{\mathrm{eff}} &= ~ -\frac{4}{4\pi} \Tr \left( b'db'+\frac{2}{3}b'^3 \right)
     + \frac{2}{4\pi}(\Tr b')d (\Tr b') 
     - \frac{2}{2\pi} (\Tr b')dA \\
     &\qquad- \frac{2}{4\pi}AdA - 4\CS_g\, .
\end{split}
\end{equation}
The chiral central charge of this exotic superconductor is
\begin{equation}
    c_-=0\, ,
\end{equation}
and it hosts parafermion zero modes due to the neutral $\SU(2)_{-4}$ sector of the $\U(2)_{-4,0}$ Lagrangian above. This superconductor was also found in Ref.~\cite{Shi:2025arn} by doping charge-$2e/3$ anyons using a U$(2)$-invariant parton ansatz, and its properties were studied extensively in Ref.~\cite{Han2026}. The fact that the superconductor has charge-$4e$ instead of $2e$ follows from the doubled flux quantization of $\Tr b'$, which lacks a self Chern-Simons term. Integrating over $\Tr b'$ therefore breaks the EM charge conservation symmetry down from U$(1)_{\mathrm{EM}}$ to $\mathbb{Z}_4$, leading to $h/4e$-vortices.


\section{Discussion}
\label{sec: discussion}

We have presented a unifying parent phase for anyon-driven phases emerging on doping the $\nu=2/3$ FQAH state, which we have dubbed the $\SU(3)_{-1}$ \emph{quark metal}. Through competing color superconducting and ferromagnetic instabilities, the quark metal can produce the broad range of anyon-driven phases discussed in the literature thus far, as well as providing access to new kinds of SC and SC$\star$ phases that have not been proposed before. The quark metal theory offers a concrete analytic setting where the overwhelming landscape of anyon-driven phases may be studied and their physics clarified, as each may be associated with a distinct order parameter and physical mechanism. 

Our approach also provides an avenue toward developing a universal theory of the finite-temperature normal state of anyon-driven superconductors (or other phases), as it clearly encodes information about both fusion and (to a lesser extent) braiding. It would be interesting to explore thermodynamic properties of the theory, which at sufficiently high temperatures might optimistically provide predictions that are independent of the specific anyon-driven phase chosen at zero temperature, following up on our earlier work in this direction~\cite{Nakajima2025}. Signatures of a quark metal state may also be visible experimentally through classic probes of fractional charge like shot noise and tunnelling, as was also noted recently in Ref.~\cite{senthil2026fractionalizedmetalsdopedanyons,han2026orthogonal} in the orthogonal metal context. 

In the future, it could also be very fruitful to use the quark metal theory as an analytic setting to gain insight into the competition between anyonic phases. Essential to this competition is the interplay of electrostatic repulsion with SU$(3)$ gauge field-mediated attraction. In particular, we expect as Coulomb repulsion is screened that color superconductivity and charge-$2e/3$ bound state formation may become preferred, consistent with recent geometric quantization approaches to the few-anyon problem~\cite{Li2026a}. On the other hand, poorly screened Coulomb interactions may privilege the formation of color ferromagnets through a Stoner-like mechanism. Techniques for studying non-Fermi liquid superconductivity and ferromagnetism may be applied to our model, as the SU$(k)_N$ Chern-Simons-fermion theory has a very well known Eliashberg-like deformation in the form of the 't Hooft limit (a simultaneous large-$N,k$ limit), which was adapted to finite density in Ref.~\cite{Geracie2016}. To our knowledge, this approach has never been generalized to study (non-Fermi liquid) superconductivity in Chern-Simons-matter theory.

Finally, we lament that although a SU$(k)_{1}$ quark metal theory is possible for all values of $k$ and thus all $\nu=1/k$ Laughlin states (and their particle-hole conjugates), a unified metallic parent theory does not seem available for arbitrary Jain states. Application of standard level-rank dualities to generic Jain $K$-matrix theories seems to always lead to outcomes where matter variables always feel flux upon doping and fill Landau levels. This suggests that in these cases no single choice of variables can monopolize the phase diagram.


\section*{Acknowledgements}

We are especially grateful to Fiona Burnell, Vladimir Calvera, Andrey Chubukov, Hao-Ran Cui, Ho Tat Lam, and Zhengyan Darius Shi for many enlightening discussions over this work's development, collaboration on related projects, and comments on the manuscript. We also thank Aleksey Cherman,  Eduardo Fradkin, Sal Pace, Sri Raghu, Raman Sohal, T. Senthil, Dam Thanh Son, Alex Thomson, Xiao-Chuan Wu, and Yue Yu for helpful discussions. We are also grateful to Ashvin Vishwanath, Zijian Wang, and Zheng-Duo Fan for pointing out an error in the initially posted version of the manuscript. YN is supported by the Funai Overseas Scholarship. HG and UM are supported by startup funds at the University of Minnesota. 


\appendix


\section{Level-rank duality with counterterms}\label{app:duality_counterterms}

Whether a Chern-Simons theory depends on a spin structure depends intricately on the gauge group as well as the Chern-Simons level. In order to get all the counterterms right, we need to keep track of the $\Spin_c$-ness of gauge fields carefully. In this section, we will closely follow Ref.~\cite{Hsin2016} and make sure that all fluctuating abelian gauge fields are $\U(1)$, and there's only one background $\Spin_c$. The background $\Spin_c$ connection is required solely to keep track of the spin-charge relation, wherein operators with odd electric charge must have half-integer spin while those with even electric charge must have integer spin \cite{Metlitski2015a}.

For the abelian theory, we have
\begin{equation}
    \Lag_{\U(1)_k}[a;A] = \frac{k}{4\pi} ada + \frac{1}{2\pi} adA\, .
\end{equation}
For even $k$, $\Lag_{U(1)_k}$ is bosonic and requires $A$ to be $\U(1)$, whereas for odd $k$, $A$ must be $\Spin_c$. In order to keep track of both these cases simultaneously, we define two different background gauge fields, a $\U(1)$ connection $B$, and a $\Spin_c$ connection $\cA$. We replace $A$ by $B+k\cA$, which is $\U(1)$ for even $k$ and $\Spin_c$ for odd $k$. Electric charge is determined solely by the coupling to $B$, while $\cA$ only exists to keep track of spin modulo integers, i.e., fermion parity.

For even $k$, even though the theory is purely bosonic, its level-rank dual contains fermionic matter and hence requires a spin structure to be well-defined. To this end, we will also add a trivial fermion theory, often denoted $\{1,\psi\}$, described by the Lagrangian
\begin{equation}
    \Lag_0[x,y;\cA] = \frac{1}{4\pi} xdx + \frac{1}{2\pi}xd(y+\cA)\, ,
\end{equation}
to the abelian theory, and define
\begin{equation}
    \hat{\Lag}_{\U(1)_k} = \frac{k}{4\pi} ada + \frac{1}{2\pi} ad(B+k\cA) + \Lag_0[\cA]\, .
\end{equation}

On the dual side, we have a non-abelian Chern-Simons matter theory with fermionic matter $\chi$. Unlike the abelian theory, pure $\SU(k)_{-1}$ Chern-Simons theory is purely bosonic and cannot be coupled to a background $\Spin_c$ connection. It's Lagrangian is given by
\begin{equation}
    \Lag_{\SU(k)_{-1}}[b,c;B] = -\frac{1}{4\pi} \Tr \left( bdb + \frac{2}{3}b^3 \right) - \frac{1}{4\pi} (\Tr b) d (\Tr b) + \frac{1}{2\pi} cd(\Tr b + B)\, ,
\end{equation}
where $b$ is $\U(k)$ gauge field, and $c$ is a Lagrange multiplier that implements the constraint $\Tr b = -B$, thereby turning $b$ into an $\SU(k)$ gauge field. We emphasize that even though the Lagrange multiplier implements this constraint, one cannot replace the $(\Tr b)d(\Tr b)$ term above by a $BdB$ counterterm. To remedy the lack of a background $\Spin_c$ connection, we add $\Lag_0[\cA]$ to the Lagrangian and perform some field redefinitions to arrive at
\begin{equation}
    \hat{\Lag}_{\SU(k)_{-1}} = - \frac{1}{4\pi} \Tr \left( bdb + \frac{2}{3} b^3 \right) + \frac{1}{2\pi} (c-\cA) d(\Tr b + B) - \frac{1}{4\pi} BdB + \Lag_0[x,y;\cA]\, .
\end{equation}
With these definitions, we can now write down the duality
\begin{equation}\label{eq:gen_duality_cts}
    \begin{split}
        \Lag_\phi[a] + \hat{\Lag}_{\U(1)_k}[a;B+k\cA] \longleftrightarrow ~ \Lag_\chi[b+\cA\bs{1}] + \hat{\Lag}_{\SU(k)_{-1}}[b,c;-B,\cA]\\ + \frac{1}{4\pi} BdB - \frac{1}{2\pi} Bd\cA - k\CS[\cA,g]\, ,
    \end{split}
\end{equation}
where $\CS[\cA,g] = \frac{1}{4\pi}\cA d\cA + 2\CS_g$ is a level $1$ Chern-Simons term for a $\Spin_c$ connection. Any additional counterterms can be added on both sides.

Note that although the left-hand side of the duality depends on background gauge fields only through the linear combination $B+k\cA$, this is not manifest on the right-hand side. In fact, the duality guarantees that any physical quantity computed from the right-hand side will repackage itself into this linear combination, as we will see for the various phases we consider.

Recall that the coupling to $B$ is what determines the electric charge of a particle, and observe that the fermions on the right-hand side do not couple directly to $B$ at all. Instead they couple to $\Tr b$ with fractional charge $1/k$, which in turn becomes a coupling to $B$ with fractional charge $1/k$ upon implementing the constraint $\Tr b = B$.


\subsection{Counter terms for the $\SU(3)$ model}\label{app:nu=2/3_counterterms}

The abelian Lagrangian describing the $\nu=2/3$ state takes the form
\begin{equation}
    \Lag_{\nu=2/3} = \Lag_\phi[a] + \frac{3}{4\pi} ada + \frac{1}{2\pi} adA + \CS[A,g]\, .
\end{equation}
To bring this to the form of the left-hand side of Eq.~\eqref{eq:gen_duality_cts}, we replace $A\rightarrow B+3\cA$ and add a trivial fermion theory $\Lag_0[\cA]$ to define
\begin{equation}
    \hat{\Lag}_{\nu=2/3} = \Lag_\phi[a] + \hat{\Lag}_{\U(1)_3}[a,B+3\cA] + \CS[B+3\cA,g]\, .
\end{equation}
Upon dualizing this theory we are then left with
\begin{equation}
    \Lag = \Lag_\chi[b+\cA\bs{1}_3] + \hat{\Lag}_{\SU(3)_{-1}}[b,c;-B,\cA] + \frac{1}{4\pi} BdB - \frac{1}{2\pi} Bd\cA + \CS[B+3\cA,g] - 3\CS[\cA,g]\, .
\end{equation}
This expression can be slightly simplified by expanding the terms in the Chern-Simons Lagrangian to obtain
\begin{equation}
    \begin{split}
        \Lag &= \Lag_\chi[b+\cA\bs{1}_3] - \frac{1}{4\pi} \Tr\left( bdb + \frac{2}{3}b^3 \right) + \frac{1}{2\pi} cd(\Tr b - B) - \frac{1}{2\pi}(\Tr b)d\cA + \Lag_\mathrm{ct}\, ,\\
        \Lag_\mathrm{ct} &= \CS[B+3\cA,g] - 3\CS[\cA,g] = \frac{1}{4\pi} BdB + \frac{3}{2\pi} Bd\cA + \frac{6}{4\pi} \cA d\cA - 4\CS_g\, .
    \end{split}
\end{equation}
Enriching with lattice translations doesn't change the form of the counterterms, and only adds a valley multiplicity to the fermions
\begin{equation}\label{eq:nu=2/3_gas_counterterms}
    \begin{split}
        \Lag &= \sum_{I=1}^3 \Lag_{\chi_I}[b+\cA\bs{1}_3] - \frac{1}{4\pi} \Tr\left( bdb + \frac{2}{3}b^3 \right) + \frac{1}{2\pi} cd(\Tr b - B) - \frac{1}{2\pi}(\Tr b)d\cA + \Lag_\mathrm{ct}\, ,\\
        \Lag_\mathrm{ct} &= \CS[B+3\cA,g] - 3\CS[\cA,g] = \frac{1}{4\pi} BdB + \frac{3}{2\pi} Bd\cA + \frac{6}{4\pi} \cA d\cA - 4\CS_g\, .
    \end{split}
\end{equation}


\subsection{Counterterms for the bound state Lagrangian}\label{app:2e/3_gas_counterterms}

To derive the counterterms for the theory of the $2e/3$ bound states, we start with Eq.~\eqref{eq:nu=2/3_gas_counterterms} and perform a Hubbard-Stratonovich transformation
\begin{equation}
    \begin{split}
        \Lag = \sum_{I=1}^3 ~ &\Lag_{\chi_I}[b+\cA\bs{1}_3] - \frac{1}{4\pi} \Tr \left( bdb + \frac{2}{3}b^3 \right) + \frac{1}{2\pi} cd(\Tr b - B) - \frac{1}{2\pi} (\Tr b)d\cA + \Lag_\ct\\
        &+ \Phi^\alpha_I \varepsilon_{\alpha\beta\gamma} \chi^\beta_I \chi^\gamma_I + \mathrm{h.c.} - \Phi^\dagger \Phi\, .
    \end{split}
\end{equation}
The field $\Phi^I$ describes intra-valley $s$-wave bound states of two quarks. Integrating out the quarks leaves us with a Lagrangian for $\Phi$ coupled to $b$, and the only ingredient needed to write down this Lagrangian is the form of the covariant derivative of $\Phi$. This can be determined from the equation of motion for $\Phi$, which determines its transformation properties under the $\U(3)$ group into which the $\SU(3)$ gauge symmetry is embedded, as well as the $\Spin_c$ global symmetry whose background gauge field is $\cA$. The equation of motion is
\begin{equation}
    \Phi^\alpha_I = \varepsilon^{\alpha\beta\gamma} \chi^\dagger_{\beta,I} \chi^\dagger_{\gamma,I}\, .
\end{equation}
The transformation properties of $\Phi$ under the $\U(3)$ gauge symmetry imply that its covariant derivative can be written as follows:
\begin{equation}
    D_\mu \Phi_I = \d_\mu \Phi_I - i b_\mu \cdot \Phi_I + i(\Tr b) \Phi_I + 2i\cA_\mu \Phi_I\, .
\end{equation}
The Lagrangian for $\Phi$ can then be straight-forwardly written down as
\begin{equation}\label{eq:pre-formed_pairs_counterterms}
    \begin{split}
        \Lag &= \sum_I \Lag_{\Phi_I} \left[ b - (\Tr b + 2\cA) \bs{1}_3 \right] - \frac{1}{4\pi} \Tr\left( bdb + \frac{2}{3}b^3 \right)\\
        &\qquad + \frac{1}{2\pi} cd(\Tr b - B) - \frac{1}{2\pi}(\Tr b)d\cA + \Lag_\ct\, ,\\
        \Lag_\ct &= \CS[B+3\cA,g] - \CS[\cA,g] = \frac{1}{4\pi} BdB + \frac{3}{2\pi} Bd\cA + \frac{6}{4\pi} \cA d\cA - 4\CS_g\, .
    \end{split}
\end{equation}

Using the Lagrangians~\eqref{eq:nu=2/3_gas_counterterms} and \eqref{eq:pre-formed_pairs_counterterms}, the effective Lagrangians for the various phases described in Section~\ref{sec:nu=2/3_doping} can be derived. In all these cases, we find that the effective Lagrangians re-package themselves as functions of the electromagnetic gauge field, $A=B+3\cA$, despite having to split it up into $B$ and $\cA$ to properly define the non-abelian theory.


\subsection{Some useful identities}\label{app:CS_identities}

In the derivation of the effective Lagrangians for the color ferromagnetic phases, we find it useful to integrate out trivial theories using the identities listed in this Section. Firstly, a $\U(1)_{\pm 1}$ theory is trivial, and can be replaced by Lagrangian that only depends on a background $\Spin_c$ connection $\cA$:
\begin{equation}
    \Lag_{\U(1)_1} \equiv \frac{1}{4\pi} ada + \frac{1}{2\pi} cd\cA \longleftrightarrow - \CS[\cA,g]\, .
\end{equation}
A change in overall sign similarly allows us to replace a $\U(1)_{-1}$ theory with a background counterterm. This generalizes to $\U(N)$ theories as well, with
\begin{equation}
    \Lag_{\U(N)_1} \equiv \frac{1}{4\pi} \Tr \left( bdb+\frac{2}{3}b^3 \right) + \frac{1}{2\pi} (\Tr b)d\cA \longleftrightarrow - N\,\CS[\cA,g]\, ,
\end{equation}
where $b$ is a $\U(N)$ gauge field.


\section{Tree level analysis of color pairing channels}\label{app:attr_vs_repuls_tree_level}

The various color superconducting and color ferromagnetic instabilities discussed in the main text are mediated by the gauge fluctuations coupled to the quark Fermi surface. In this Appendix, we compute the sign of the gauge-field-mediated interaction at tree level in the various channels corresponding to these instabilities, closely following Ref.~\cite{Bishara2007}.

We start with the Lagrangian for the $\SU(3)$ quark metal. Since the gauge field doesn't couple to the valley index, $I$, the sign of the interaction is independent of the valleys, and we will suppress this index in what follows for simplicity. At tree level, integrating out the gauge field is equivalent to solving the linearized version of Gauss' law,
\begin{equation}
    \rho^a_{\chi} = \varepsilon_{ij} \partial_i b_j^a\, ,
\end{equation}
where $f^{abc}$ are the $\SU(3)$ structure constants, to derive $b_i^a$ as a function of the color density $\rho_\chi^a = \chi^\dagger T^a \chi$. At this order we can drop the $b^2$ term in Gauss' law. Solving for $b^a_i$ is then easy to do in momentum space and leads to the interaction part of the Hamiltonian,
\begin{equation}\label{eq:H_int}
    H_{\mathrm{int}} = - \frac{2\pi i}{m_\chi} \left[ \sum_a (T^a)^\alpha_{\ \delta} (T^a)^\beta_{\ \gamma} \right] \sum_{p,p',q} \frac{\vec{q}\times\vec{p}}{q^2} \chi^\dagger_\alpha(p+q) \chi^\dagger_\beta(p'-q) \chi^\gamma(p') \chi^\delta(p)\, .
\end{equation}
The sign of the interaction picks up multiiplicative contributions from two sources: the tensor, $\sum_a T^a T^a$, as well as the Fourier decomposition of the momentum-dependent form factor of the interaction. The tensor can be decomposed into irreducible representations of $\SU(3)$, and the sign of the interaction will depend on the representation.

Let us first focus on the particle-particle (BCS) channel, which accounts for pairing instabilities of the Fermi surface. For this configuration, the momenta of the quarks can be restricted to the Fermi surface:
\begin{equation}\label{eq:H_int_BCS}
    H_{\mathrm{int}}^\mathrm{BCS} = - \frac{2\pi i}{m_\chi} \left[ \sum_a (T^a)^\alpha_{\ \delta} (T^a)^\beta_{\ \gamma} \right] \sum_{p,p'} \frac{\vec{p}\times\vec{p}'}{(\vec{p}-\vec{p}')^2} \chi^\dagger_\alpha(p) \chi^\dagger_\beta(-p) \chi^\gamma(-p') \chi^\delta(p')\, .
\end{equation}
The quarks live in the fundamental $(F)$ representation of $\SU(3)$, and the particle-particle channel decomposes into symmetric $(S_2)$ and antisymmetric $(A_2)$, where the latter is the same as the anti-fundamental $(\bar{F})$ representation. The tensor factor, restricted either the symmetric or antisymmetric representation, simplifies to
\begin{equation}
    \sum_a \left(T^a\right)^\alpha_{\ \delta} \left(T^a\right)^\beta_{\ \gamma}
    =\frac{1}{2}\left(c_2(\bs{R}) - 2c_2(F)\right)
    \delta^\alpha_{\ \delta} \delta^\beta_{\ \gamma}\, ,
\end{equation}
where $\bs{R}$ is the symmetric or antisymmetric representation, and $c_2(\bs{R})=\sum_aT^a_{(\bs{R})} T^a_{(\bs{R})}$ is the quadratic Casimir for the representation $\bs{R}$. The values of the quadratic Casimir in the relevant representations are $c_2(F) = c_2(A_2) = 4/3$, and $c_2(S_2) = 10/3$, and we find
\begin{equation}
    \frac{1}{2} \left[ c_2(A_2) - 2c_2(F) \right] = -\frac{2}{3}\, , \qquad \frac{1}{2} \left[ c_2(S_2) - 2c_2(F) \right] = +\frac{1}{3}\, .
\end{equation}
The momentum-dependent form factor in Eq.~\eqref{eq:H_int_BCS} can be decomposed in a Fourier series expansion:
\begin{equation}    
    \begin{split}
        -\frac{2\pi i}{m_\chi} &\frac{\vec{p}\times\vec{p}'}{(\vec{p}-\vec{p}')^2} \chi^\dagger_\alpha(p) \chi^\dagger_\beta(-p) \chi^\gamma(-p') \chi^\delta(p')\\
        &\propto - \frac{1}{m_\chi} \sum_{\ell} e^{i\ell\theta_{pp'}} \mathrm{sign}(\ell) \left(\lambda-\sqrt{\lambda^2-1} \right)^{|\ell|} \chi^\dagger_\alpha(p) \chi^\dagger_\beta(-p) \chi^\gamma(-p') \chi^\delta(p')\, ,
\end{split}
\end{equation}
where $\lambda \equiv (p/p'+p'/p)/2\geq 1$ and $\theta_{pp'} = \phi_p -\phi_{p'}$ is the angle between the two momenta, making $\ell$ the angular momentum of the Cooper pair, and we have defined $\mathrm{sign}(\ell)$ to vanish when $\ell = 0$.

Combining the two contributions, we find that the sign of the interaction is proportional to the following
\begin{equation}
    \mathrm{sign}(H_\mathrm{int}^\mathrm{BCS}) = - \frac{1}{2} \left[ c_2(\bs{R}) - 2 c_2(F) \right] \mathrm{sign}(\ell)
\end{equation}
We find that when quark pairing is anti-symmetric in color, $\ell < 0$ is preferred at tree level, making the quarks pair in the $p+ip$ channel, while for color-symmetric pairing, $\ell>0$, so that quarks pair in the $p-ip$ channel. The fact that the interaction vanishes for $s$-wave ($\ell = 0$) is an artifact of having dropped the Yang-Mills term, which introduces dynamics to the gauge field and renders the $\ell = 0$ channel attractive. 

The same analysis can be also applied to the color-ferromagnetic instabilities as well, for which we need to work in the particle-hole channel, which decomposes into the trivial and adjoint representations. Following a similar calculation there, we find that the interaction channel that mediates the ferromagnetic instability is repulsive.


\section{Doping a chiral spin liquid}\label{app:semion_superconductors}

We can illustrate the utility of the dual non-abelian theory through the example of a chiral spin liquid (CSL)~\cite{Kalmeyer1987,Wen1989}. The anyons of the CSL are semions possessing $\pi/2$ statistics. One realization of this is obtained from doping a chiral spin liquid on a triangular lattice with staggered flux, \`a la \cite{Divic2025,Pichler2025}, described by the abelian Lagrangian
\begin{equation}
    \Lag = \Lag_\phi[a] + \frac{2}{4\pi} ada + \frac{2}{2\pi} adA + 2\CS[A,g]\, .
\end{equation}
Observe that the coefficients of the $ada$ and $adA$ term dictates that the semions have statistics $\pi/2$ and electric charge $e$, unlike the CS-GL theory in Eq.~\eqref{eq: abelian CSGL}. This is because the microscopic lattice model comprises local bosons that carry charge $2e$, as opposed to local fermions.

The abelian Lagrangian is of the form of the left-hand side of Eq.~\eqref{eq:gen_duality_cts}, but with $B=0$ and $\cA\rightarrow A$ and a missing trivial fermion Lagrangian $\Lag_0[A]$. Upon adding this trivial Lagrangian to the left-hand side, we can dualize the theory to find the $\SU(2)_{-1}$ Lagrangian
\begin{equation}\label{eq:semion_gas_counterterms}
    \Lag = \Lag_\chi[b+A\bs{1}_2] - \frac{1}{4\pi} \Tr \left( bdb + \frac{2}{3} b^3 \right) + \frac{1}{2\pi} (c-A)d \Tr b + \Lag_0[A]\, .
\end{equation}
The last term can then be dropped. The quarks $\chi^\alpha$ form an $\SU(2)$-doublet with two possible colors: \textit{cyan} (c) and \textit{magenta} (m). The conventionally normalized generators for $\SU(2)$ are half-Pauli matrices, $T^a = \sigma^a/2$ with $a = 1,2,3$.

Furthermore, the lattice model used to describe the chiral spin liquid has a $\pi$-flux threaded throught each unit cell, requiring the unit cell to be doubled in order to faithfully realize lattice translations. This means that in the low energy effective theory, lattice translations are realized projectively, obeying the algebra $\tau_x \tau_y \tau_x^{-1} \tau_y^{-1} = -1$. As discussed in the main text, this results in a two-fold valley degeneracy for the quarks, which must carry an additional flavour index: $\chi^\alpha_I$, with $I=1,2$. Lattice translations are realized as permutations and valley-dependent phase shifts:
\begin{equation}
    \tau_x ~:~ \chi_{1,2} \rightarrow \chi_{2,1}\, , \qquad \tau_y ~:~ \chi_I \rightarrow (-1)^I \chi_I\, .
\end{equation}
Gauss' law in the non-abelian description pins the color density to non-abelian flux,
\begin{equation}
    \chi^\dagger_I T^a \chi_I = \frac{1}{2\pi} f^a\, ,
\end{equation}
so that, unlike in the abelian description, there are no emergent fluxes that couple to the quarks unless there is a net color charge.


\subsection{Color superconductor from baryon condensation}

The color-singlet superconductor is a result of condensation of the intra-valley baryon operator,
\begin{equation}
    \Delta = \sum_{I=1}^2 \langle \varepsilon^{\alpha\beta} \chi^\dagger_{\alpha,I}\chi^\dagger_{\beta,I} \rangle\, ,
\end{equation}
which is gauge invariant. As a result, the $\SU(2)$ gauge field remains unHiggsed and simply decouples from the matter sector. Since the baryon carries charge-$2e$, its condensation spontaneously breaks the electromagnetic $\U(1)$ symmetry down to $\mathbb{Z}_2$. We will assume for simplicity that quark pairing is $s$-wave and hence generates no additional chiral central charge.

Phase fluctuations of the order parameter can be dualized to an emergent $\U(1)$ gauge field $\alpha$. The baryon couples to $\Tr b + 2A$, making the result a mixed Chern-Simons term between $\alpha$ and $\Tr b + 2A$. The effective Lagrangian is then the following:
\begin{equation}
    \Lag_\eff = -\frac{1}{2\pi} \alpha d (\Tr b + 2A) - \frac{1}{4\pi} \Tr \left( bdb+\frac{2}{3}b^3 \right) + \frac{1}{2\pi} cd(\Tr b-A)\, .
\end{equation}
A shift of the Lagrange multiplier, $c\rightarrow c+\alpha$, allows us to write this as a decoupled trivial superconductor accompanied by an $\SU(2)_{-1}$ Chern-Simons theory:
\begin{equation}
    \Lag_\eff = -\frac{2}{2\pi} \alpha dA - \frac{1}{4\pi} \Tr\left( bdb+\frac{2}{3}b^3 \right) + \frac{1}{2\pi}(c-A)d\Tr b\, .
\end{equation}

This theory describes a charge-$2e$ superconductor co-existing with $\SU(2)_{-1}$ semion topological order, i.e., an SC$\star$ phase. The chiral central charge of this phase is that of the dark semion, supplied entirely by the $\SU(2)_{-1}$ Chern-Simons term, and takes the value
\begin{equation}
    c_- = 1\, .
\end{equation}

This SC$\star$ phase is entirely distinct from the one obtained from Laughlin's mechanism, which is topologically trivial. It is easy to understand this from the point of view of anyon condensation. The underlying chiral spin liquid consists of two anyons: the semion carrying charge $e$ and the local boson carrying charge $2e$. Condensing the local boson spontaneously breaks the electromagnetic $\U(1)$ symmetry and hence invalidates the charge assignment, but the semion remains otherwise untouched, leaving us with a topological order consisting of an uncharged semion.


\subsection{Color ferromagnet: Recovering Laughlin's mechanism}

The key property of Laughlin's mechanism in the abelian description is that the anyons experience each other as flux tubes thanks to the flux attachment that is implemented by the abelian Gauss' law. To reproduce this phase in the non-abelian description, we need to activate a flux in the non-abelian gauge field $b$ at finite quark density. This necessarily means a non-zero color density because of Gauss' law, which suggests that a color-ferromagnetic instability of the blue Fermi sea is likely to result in the same phase.

Consider aligning the color-ferromagnetic order parameter with the ``3'' direction in the color-triplet representation:
\begin{equation}
    C^3_I = \langle \chi^\dagger_I T^3 \chi_I \rangle = \frac{\langle \chi^\dagger_{c,I} \chi^c_I \rangle - \langle \chi^\dagger_{m,I} \chi^m_I \rangle}{2}\, .
\end{equation}
When $C^3 \ne 0$, the $\U(2)$ gauge symmetry is Higgsed down to $\U(1)^{\bs{1}_2}\times\U(1)^{T^3}/\mathbb{Z}_2$, where the superscripts denote the generators of the respective $\U(1)$ factors, allowing us to set $b^1=b^2=0$. The remaining abelian gauge fields are $\Tr b$ and $b^3$, but these are not independent because of the $\mathbb{Z}_2$ quotient. This is remedied by switching to a color-resolved basis for the diagonal generators, $T_c=\mathrm{diag}(1,0), T_m=\mathrm{diag}(0,1)$,
in which the gauge field components,
\begin{equation}
    \widetilde{b}^c = \frac{\Tr b}{2} + b^3, \qquad \widetilde{b}^m = \frac{\Tr b}{2} - b^3\, ,
\end{equation}
have independent flux quantization conditions. In this color-resolved basis, $\chi^c$ couples to $\widetilde{b}^c + A$ and $\chi^m$ to $\widetilde{b}^m + A$. The Higgsed Lagrangian take the form
\begin{equation}
    \Lag = \Lag_{\chi^c}[\widetilde{b}^c+A] + \Lag_{\chi^m}[\widetilde{b}^m+A] - \frac{1}{4\pi} \widetilde{b}^cd\widetilde{b}^c - \frac{1}{4\pi} \widetilde{b}^md\widetilde{b}^m + \frac{1}{2\pi}(c-A)d(\widetilde{b}^c+\widetilde{b}^m)\, ,
\end{equation}
and Gauss' laws, in the absence of the background electromagnetic field, turn into the following:
\begin{equation}
    \begin{split}
        \sum_I \chi^\dagger_{c,I} \chi^c_I &= \frac{\varepsilon_{ij}\d_i\widetilde{b}^c_j}{2\pi} - \frac{\varepsilon_{ij}\d_i c_j}{2\pi}\, ,\\
        \sum_I \chi^\dagger_{m,I} \chi^m_I &= \frac{\varepsilon_{ij}\d_i\widetilde{b}^m_j}{2\pi} - \frac{\varepsilon_{ij}\d_i c_j}{2\pi}\, ,\\
        0 &= \varepsilon_{ij}\d_i(\widetilde{b}^c_j + \widetilde{b}^m_j)\, .
    \end{split}
\end{equation}

At this stage we need to pick a mean-field solution for these equations. The simplest possibility is one where the color polarizes maximally into, say, cyan, for which the mean-field values of the guage field are related by $b^m = c = -b^c$. The magenta quarks are fully depleted, whereas the charge-flux relation for cyans forces each valley into a fermionic integer quantum Hall state (IQH),
\begin{equation}
    \sum_I \langle \chi^\dagger_{c,I} \chi^c_I \rangle = 2 \frac{\langle\varepsilon_{ij}\d_i\widetilde{b}^c_j\rangle}{2\pi}\, .
\end{equation}
Integrating them out generates a level $2$ Chern-Simons term for $\widetilde{b}^c+A$ including an additional gravitational Chern-Simons term due to the fact that the IQH state is fermionic, and adding them to the Higgsed Chern-Simons Lagrangian results in the following effective theory after some simplification:
\begin{equation}
    \begin{split}
        \Lag_\eff &= \frac{2}{4\pi} (\widetilde{b}^c + A) d (\widetilde{b}^c + A) + 4\mathrm{CS}_g - \frac{1}{4\pi} \widetilde{b}^c d \widetilde{b}^c - \frac{1}{4\pi} \widetilde{b}^m d \widetilde{b}^m + \frac{1}{2\pi} cd(\widetilde{b}^c + \widetilde{b}^m)\, ,\\
        &= \frac{2}{2\pi} \widetilde{b}^c d A + 2\CS[A,g]\, .
    \end{split}
\end{equation}
This is the Lagrangian for the superconductor that arises from Laughlin's mechanism.

We can also consider other mean-field solutions to Gauss' laws, characterized by two filling fractions, $\nu_c$ and $\nu_m$, which determine the quark densities via
\begin{equation}
    \sum_I \langle \chi^\dagger_{c,I} \chi^c_I \rangle = \nu_c \frac{\langle \varepsilon_{ij}\d_i\widetilde{b}^c_j \rangle}{2\pi}\, , \qquad \sum_I \langle \chi^\dagger_{m,I} \chi^m_I \rangle = \nu_m \frac{\langle\varepsilon_{ij}\d_i\widetilde{b}^m_j\rangle}{2\pi}\, .
\end{equation}
Since the mean-field values are required to obey $b^c = -b^m$, the two fillings must also be related by
\begin{equation}
    \nu_c - \nu_m = 2\, .
\end{equation}
Various solutions to this are possible, and different choices of filling fractions lead to different quantum Hall states for the quarks, which can be integrated out to generate different types of superconductors. The cyclotron frequencies of other integer-valued solutions are proportional to $\delta/(\nu_c+\nu_m)$, where $\delta$ is the doping, implying that the fully polarized solution has the largest mean-field gap. We hence expect it to be the most likely color ferromagnetic phase.


\section{Effective Lagrangians for color superconductors}\label{app:color_sc_eff_lags}

We derive, in this Appendix, the effective Lagrangians for the three color superconducting phases considered in Section~\ref{sec:color_sc}.


\subsection{Color-valley-locked superconductor}

To translate the group-theoretic discussion in the main text into the less abstract language of Ginzburg-Landau theory for the order parameter, $\Delta^\alpha_I$ -- which is decidedly \emph{not} gauge invariant -- it is necessary to work with phase variables are all manifestly $2\pi$ periodic, along with gauge fields that satisfy the standard Dirac flux quantization condition. We introduce the color-resolved gauge fields,
\begin{align}
\widetilde{b}^c\,\widetilde{T}^c=\begin{pmatrix} \widetilde{b}^c& 0 & 0 \\ 0&0&0\\0&0&0\end{pmatrix}\,,\,\widetilde{b}^m\,\widetilde{T}^m=\begin{pmatrix} 0& 0 & 0 \\ 0&\widetilde{b}^m&0\\0&0&0\end{pmatrix}\,,\,\widetilde{b}^y\,\widetilde{T}^y=\begin{pmatrix} 0& 0 & 0 \\ 0&0&0\\0&0&\widetilde{b}^y\end{pmatrix}\,.
\end{align}
Unlike the determinant U$(1)$ factor in Eq.~\eqref{eq: U(3) as a quotient}, each of the color-resolved gauge fields has integer-quantized fluxes and can be treated as ordinary U$(1)$ gauge fields after Higgsing the gauge fields corresponding to the remaining six generators. The phase variables coupling to them -- $\theta_c,\theta_m,\theta_y$, respectively -- therefore also manifestly have $2\pi$ periodicity. Performing a fluctuation expansion of the CVL order parameter in these coordinates, dualizing the phase variables to $\U(1)$ gauge fields $\alpha_c,\alpha_m,\alpha_y$, and gauge fixing to Higgs the other gauge components gives
\begin{equation}
\begin{split}
    \Lag_\eff &= \frac{1}{2\pi} \alpha_c d\left( \widetilde{b}^y+\widetilde{b}^m \right) + \frac{1}{2\pi} \alpha_m d\left( \widetilde{b}^y+\widetilde{b}^c \right) + \frac{1}{2\pi} \alpha_y d\left( \widetilde{b}^c+\widetilde{b}^m \right)\\
    &\qquad+\frac{1}{2\pi}cd\left(\widetilde{b}^c+\widetilde{b}^m+ \widetilde{b}^y-A\right)-\frac{1}{4\pi}\sum_{n=c,m,y}\widetilde{b}^n d\widetilde{b}^n+\frac{1}{4\pi}AdA-4\mathrm{CS}_g+\dots\,,
\end{split}
\end{equation}
where the ellipses include higher-order terms in the fluctuation expansion, along with the terms associated with the remaining generators. The self-Chern-Simons terms appear due to the SU$(3)_{-1}$ theory native to the Lagrangian. We emphasize the above is only the part of the effective action depending on the order parameter phase fluctuations, which we use to set the stage for a discussion of the $p+ip$ quark response below.

Introducing a new U$(1)$ gauge field, $\beta=\widetilde{b}^c+\widetilde{b}^m+ \widetilde{b}^y$, and changing variables in the path integral to substitute for $\widetilde{b}^y$ allows us to solve the constraint as $\beta=A$. Then we obtain
\begin{equation}
\begin{split}
    \Lag_\eff &= \frac{1}{2\pi} \alpha_c d\left( \widetilde{b}^c-A \right) + \frac{1}{2\pi} \alpha_m d\left( \widetilde{b}^m-A \right) + \frac{1}{2\pi} \alpha_y d\left( \widetilde{b}^c+\widetilde{b}^m \right)\\
    &\qquad +\frac{1}{2\pi}cd\left(\widetilde{b}^c+\widetilde{b}^m+ \widetilde{b}^y-A\right)-\frac{1}{4\pi}(A-\widetilde{b}^c-\widetilde{b}^m)d(A-\widetilde{b}^c-\widetilde{b}^m)\\&\qquad-\frac{1}{4\pi}\sum_{n=c,m}\widetilde{b}^n d\widetilde{b}^n+\frac{1}{4\pi}AdA-4\mathrm{CS}_g+\dots\,,
\end{split}
\end{equation}
Using $\alpha_m$ and $\alpha_y$ to Higgs the combinations $\widetilde{b}^c+\widetilde{b}^m$ and $\widetilde{b}^m-A$, we arrive at 
\begin{align}
    \mathcal{L}_{\mathrm{eff}} &= \frac{2}{2\pi} \alpha_c dA -\frac{2}{4\pi}AdA-4\mathrm{CS}_g+\dots \,,
\end{align}
which clearly indicates charge-$2e$ superconductivity. The proper quantization of the background terms is ensured by our judicious choice of $2\pi$-periodic phase coordinates.


\subsection{Intra-valley, intra-color pairing}

For intra-color pairing, the residual symmetry $\O(3)=\SO(3)\times\mathbb{Z}_2$ has a discrete factor. A fluctuation expansion in Goldstone bosons fails to capture this. Since the values $(\theta_c,\theta_m,\theta_y)$ of the three color-resolved phases must be identified up to $\SO(3)$ transformations. In particular, $(\pi,0,0)$, $(0,\pi,0)$ and $(0,0,\pi)$ represent the same point in the $\U(3)/\O(3)$ coset. We can implement this, however, by Higgsing only one of the three color-resolved gauge field $(\widetilde{b}^c, \widetilde{b}^m, \widetilde{b}^y)$,  to $\mathbb{Z}_2$, and fully Higgsing the remaining two. Eliminating all components of the gauge field except the $\SO(3)$ components, $b'$, and $\tilde{b}^n$ leaves us with the Higgsed Lagrangian
\begin{equation}
    \begin{split}
        \Lag &= \frac{2}{2\pi} \alpha_c d\widetilde{b}^c + \frac{1}{2\pi} \alpha_m d\widetilde{b}^m + \frac{1}{2\pi} \alpha_y d\widetilde{b}^y + \frac{1}{2\pi} cd \left( \widetilde{b}^c+\widetilde{b}^m+\widetilde{b}^y - A \right)\\
        &\qquad - \frac{1}{4\pi} \sum_{n=c,m,y} \widetilde{b}^nd\widetilde{b}^n - \frac{2}{8\pi} \Tr \left( b'db' + \frac{2}{3}b'^3 \right) + \frac{1}{4\pi} AdA - 4\CS_g + \ldots\, ,
    \end{split}
\end{equation}
where we have chosen to Higgs $\widetilde{b}^c$ to a $\mathbb{Z}_2$ gauge field. Note that the normalization of $\SO(N)$ Chern-Simons terms differs from that of $\SU(N)$ Chern-Simons terms by a factor of 2 (see, e.g., Refs.~\cite{Aharony2016b,Cordova2018}).

We can set $\widetilde{b}^m$ and $\widetilde{b}^y$ to zero upon integrating out $\alpha_m$ and $\alpha_y$, and the Lagrange multiplier, $c$, this pins $\widetilde{b}^c$ to $A$, resulting in
\begin{equation}
    \mathcal{L}_{\O(3)} = - \frac{2}{8\pi} \Tr \left( b'db' + \frac{2}{3}b'^3 \right) + \frac{2}{2\pi} \alpha dA - \frac{2}{4\pi} AdA - 4\CS_g + \ldots\, .
\end{equation}
We see that the $\mathbb{Z}_2$ factor of the residual $\O(3)$ symmetry is simply the Higgsed electromagnetic symmetry, and we have a charge-$2e$ superconductor.

In addition to this Higgsed Lagrangian, the quarks in each valley organize into the vector representation of the residual $\SO(3)$ symmetry. As a result, weak, $p-ip$ pairing generates an $\SO(3)_{-1}$ response per valley, shifting the level of the Higgsed $\SO(3)$ Chern-Simons term by $-3$, so that the effective Lagrangian takes the form
\begin{equation}
    \mathcal{L}_\eff = - \frac{5}{8\pi} \Tr \left( b'db' + \frac{2}{3}b'^3 \right) + \frac{2}{2\pi} \alpha dA - \frac{2}{4\pi} AdA - 4\CS_g + \ldots\, ,
\end{equation}
describing an SC$\star$ phase with a \textit{non-abelian} $\SO(3)_{-5}$ topological order.


\subsection{Intra-valley, inter-color pairing}

For inter-color pairing, we will work in a basis where the order parameter has been diagonalized to $\mathrm{diag}(2,-1,-1)$. The residual symmetry that preserves this configuration is $\O(2)\times\mathbb{Z}_2^c$, where $\mathbb{Z}_2^c$ acts on the cyan component. The Higgsing of the native $\U(3)_{-1}$ term follows a sequence of steps very similar to the case of intra-color pairing with a few subtle differences. Firstly, the gauge field, $a$, for the $\SO(2)$ subgroup of the $\O(2)=\SO(2)\rtimes\mathbb{Z}_2$ symmetry is embedded into $\U(3)$ as
\begin{equation}
    a_{\SO(2)} = \begin{pmatrix}
        0 & 0 & 0\\
        0 & 0 & a\\
        0 & -a & 0
    \end{pmatrix}\, .
\end{equation}
Secondly, in the color-resolved basis for the diagonalized pairing matrix, the cyan component is Higgsed separately to $\mathbb{Z}_2$. The redundancy of $(\theta_c,\theta_m,\theta_y)$ only identifies $(0,\pi,0)$ with $(0,0,\pi)$, while $(\pi,0,0)$ is a distinct point in the coset. This redundancy is accounted for by only Higgsing $\widetilde{b}^m$ to a $\mathbb{Z}_2$ gauge field, but completely eliminating $\widetilde{b}^y$. This is described by the Lagrangian
\begin{equation}
    \begin{split}
        \Lag &= \frac{2}{2\pi} \alpha_c d\widetilde{b}^c + \frac{2}{2\pi} \alpha_m d\widetilde{b}^m + \frac{1}{2\pi} \alpha_y d\widetilde{b}^y\\
        &\qquad + \frac{1}{2\pi} cd \left( \widetilde{b}^c + \widetilde{b}^m + \widetilde{b}^y - A \right) - \frac{1}{4\pi} \sum_{n=c.m.y} \widetilde{b}^nd\widetilde{b}^n\\
        &\qquad\qquad - \frac{2}{4\pi} \Tr \left( a_{\SO(2)}da_{\SO(2)} \right) + \frac{1}{4\pi} AdA - 4\CS_g + \ldots\, ,
    \end{split}
\end{equation}
After integrating out Lagrange multipliers and shifting $\alpha_m \rightarrow \alpha_m + \alpha_c$, we are left with
\begin{equation}
    \begin{split}
        \Lag_{\O(2)} &= -\frac{2}{4\pi} \Tr \left( a_{\SO(2)} d a_{\SO(2)} \right) + \left( \frac{2}{2\pi} \alpha_m d\widetilde{b}^m - \frac{2}{4\pi} \widetilde{b}^md\widetilde{b}^m + \frac{1}{2\pi} \widetilde{b}^mdA \right)\\
        &\qquad + \frac{2}{2\pi} \alpha_c dA - 4\CS_g  + \ldots\, .
    \end{split}
\end{equation}
The first term is a $\SO(2)_{-2} = \U(1)_2$ Chern-Simons theory. The term in the parenthesis is the abelian presentation of a $(\mathbb{Z}_2)_{-4}$ gauge theory for the determinant factor of $\O(2)$, commonly denoted $-4f[\widetilde{b}^m]$ (see Appendix~\ref{app:O(N)_andZ2_CS_theory} for a precise definition). The last term is what Higgses the electromagnetic $\U(1)$ symmetry down to $\mathbb{Z}_2$, resulting in a charge-$2e$ supercondcutor.

Furthermore, the residual $\SO(2)$ symmetry of the weakly paired, $p-ip$ quarks generates an $\O(2)_{-3,-3}$ response per valley, shifting the level of the Higgsed $\SO(2)_{-2}$ term by $-3$. The effective Lagrangian then takes the form
\begin{equation}
    \Lag_\eff = -\frac{5}{4\pi} \Tr \left( a_{\SO(2)} d a_{\SO(2)} \right) - 7f[\widetilde{b}^m] + \frac{1}{2\pi} \widetilde{b}^m dA + \frac{2}{2\pi} \alpha_c dA - 4\CS_g + \ldots\, ,
\end{equation}
describing an SC$\star$ phase with abelian $\O(2)_{-5,-7}$ topological order, where the two subscripts denote the $\SO(2)$ and $\mathbb{Z}_2$ levels. Since the level of a $\mathbb{Z}_2$ gauge theory is only defined modulo $8$, this is equivalent to an abelian $\O(2)_{-5,1}$ topological order.


\section{Facts about Majorana path integrals and $\O(N)$ TQFTs}\label{app:O(N)_andZ2_CS_theory}


\subsection{Aspects of $\O(N)_{K,L}$ TQFTs}

We summarize the important properties of $\SO(N), \mathbb{Z}_2$, and $\O(N)$ Chern-Simons theories in this Appendix, as well as their relation to Majorana path integrals, closely following Ref.~\cite{Cordova2018}. We begin with the simplest case -- $\SO(N)$ Chern-Simons theory -- which is described by a single gauge field $b_\mu$ expanded in a basis of generators of $N\times N$ antisymmetric real matrices, i.e., the vector representation. The Lagrangian for this Chern-Simons theory is given by
\begin{equation}
    \Lag_{SO(N)_k} = \frac{k}{8\pi} \Tr\left( bdb + \frac{2}{3} b^3 \right)\, , \qquad k\in\mathbb{Z}\, .
\end{equation}
Note that the normalization of the Chern-Simons term differs from that of $\SU(N)$ theories by a factor of $2$. When $k$ is odd, the theory requires a spin structure to be well-defined. These theories have a global $\mathbb{Z}_2$ symmetry, which we will choose to be represented by the orthogonal matrix $\mathrm{diag}(-1,1,\ldots,1)$ for $N = $ even, and the orthogonal matrix $-\bs{1}_N$ for $N=$ odd. This symmetry is often referred to as charge conjugation. The Chern-Simons theory can hence be coupled to a background $\mathbb{Z}_2$ gauge field, $B$, that sources this symmetry.

When the Chern-Simons level is unit, the theory is invertible and the partition function, $Z_{\SO(L)_1}[B]$, can be used to define $\mathbb{Z}_2$ gauge theories, denoted by $(\mathbb{Z}_2)_L$ or $L f[B]$, as follows
\begin{equation}
    e^{-i L f[B]} = \frac{Z_{\SO(L)_1}[B]}{Z_{\SO(L)_1}[0]}\, .
\end{equation}
It is possible to show that the level of a $\mathbb{Z}_2$ gauge theory is only defined modulo $8$, meaning that $(\mathbb{Z}_2)_{L+8} = (\mathbb{Z}_2)_L$. In particular, when $L=0$ or $4$, the theory is bosonic. These are known as Dijkgraaf-Witten theories.

When $L$ is even, the Lagrangian can be obtained from Higgsing a $\Spin_c$ gauge field, which we also refer to as $B$, down to $\mathbb{Z}_2$ by a Lagrange multiplier $\alpha_\mu$:
\begin{equation}
    \exp \left( i L f[B] \right) = \int \mathcal{D}\alpha ~ \exp \left( i\, \frac{2}{2\pi} \int \alpha d B + i\, \frac{L/2}{4\pi} \int BdB \right)\, .
\end{equation}
There is no such relation for odd $L$, since $L/2=1$ is the minimal allowed level for a spin Chern-Simons theory.

Lastly, we consider theories with $\O(N)$ gauge groups. The parity of $N$ is important for this discussion, since for even $N$, $\O(N) = \SO(N)\rtimes\mathbb{Z}_2$, whereas for odd $N$, $\O(N) = \SO(N)\times\mathbb{Z}_2$. The gauge field for $\O(N)$ is a pair $(b_\mu,c_\mu)$, where $b_\mu$ is $\SO(N)$, while $c_\mu$ is $\mathbb{Z}_2$. The specific form of the theory depends intricately on how the $\mathbb{Z}_2$ factor is embedded in $\O(N)$. For odd $N$, the canonical choice is to pick $-\bs{1}_N$. For even $N$, there is no canonical choice, and we will choose to represent the $\mathbb{Z}_2$ element by the matrix $\mathrm{diag}(-1,1,\ldots,1)$.

The $\O(N)$ theory can then be obtained from the $\SO(N)$ theory by gauging the $\mathbb{Z}_2$ global symmetry. In addition to this, one can add additional $(\mathbb{Z}_2)_p$ gauge theory terms for $c_\mu$, as well as a $\mathbb{Z}_2$-valued coupling between the $\mathbb{Z}_2$ gauge field and the second Steifel-Whitney class, $w_2$, of $\SO(N)$. The latter determines the charge of the $\SO(N)$ monopole under the $\mathbb{Z}_2$ factor. For simplicity, we will ignore this coupling, and refer the reader to Ref.~\cite{Cordova2018}. The action for the $\O(N)$ Chern-Simons theory then takes the form,
\begin{equation}
    S_{\O(N)_{k,p}}[b,c] = \frac{k}{8\pi} \int \Tr\left( bdb+\frac{2}{3}b^3 \right) + p f[c]\, ,
\end{equation}
with $p\simeq p+8$.


\subsection{Majorana path integrals and $\O(N)$ responses}

Having defined Chern-Simons theories with orthogonal gauge groups, we can evaluate path integrals for free Majorana fermions coupled to background gauge fields. Let $\psi^\rho$, $\rho=1,\ldots,N$, be a relativistic Majorana fermion living in the vector representation of $\O(N)$. The free theory for this fermion has a global $\O(N)$ symmetry, and can hence be coupled to background gauge fields $(B_\mu,C_\mu)$, where $B_\mu$ is $\SO(N)$ and $C_\mu$ is $\mathbb{Z}_2$. We will use our conventions above for the embedding of $\mathbb{Z}_2$ into $\O(N)$. If $m$ is the mass of the Majorana fermion, the phase of the path integral depends on the sign of $m$. Furthermore, the counterterms generated upon integrating out the Majorana fermion depend upon the parity of $N$ due to our choice of embedding for $\mathbb{Z}_2$. For odd $N$, we will use the convention,
\begin{equation}
    \begin{split}
        Z_{m>0}[B,C] &= |Z|\, ,\\
        Z_{m<0}[B,C] &= |Z| \exp \left( - i S_{\O(N)_{1,N}}[B,C] - iN \int\CS_g + \ldots \right)\, ,
    \end{split}
\end{equation}
where $|Z|$ is the absolute value of the fermion determinant, $\CS_g$ is the gravitational Chern-Simons term, and the ellipses denote the additional discrete coupling between $c_\mu$ and $w_2(\SO(N))$ which we have ignored. For $N = $ even, on the other hand, we have
\begin{equation}
    \begin{split}
        Z_{m>0}[B,C] &= |Z|\, ,\\
        Z_{m<0}[B,C] &= |Z| \exp \left( - i S_{\O(N)_{1,1}}[B,C] - iN \int\CS_g + \ldots \right)\, .
    \end{split}
\end{equation}
Note the difference in the second level of the $\O(N)$ theory for odd and even $N$.


\section{BdG analysis of color superconductors}\label{app:BdG_Majorana_vortices}

In this Appendix we derive the topological responses as well as Majorana zero modes of the various color superconductors using a BdG analysis.


\subsection{Topological response from Majorana path integrals}

The topological response is best understood as a consequence of Majorana path integrals at finite density, so it helps to work with a relativistic UV regularization in which the quarks are Dirac fermions at finite mass and chemical potential, which we will also denote $\chi^\alpha$ (suppressing helicity). To obtain the same edge content as $p-ip$ pairing of non-relativistic quarks, the Dirac fermions must be paired in $s$-wave. We will focus on the $p-ip$ example, since the non-trivial topological responses we are interested in are generated only for intra-valley, color-symmetric, $p-ip$ pairing. The analysis for $p+ip$ pairing of non-relativstic quarks -- relevant for the CVL phase -- can be obtained straightforwardly by acting with particle-hole conjugation, $\mathcal{CT}$. 

We begin with the BdG Hamiltonian for the Dirac quarks,
\begin{equation}
    \begin{split}
        H &= \chi^\dagger_{\alpha,I} \left( -i\sigma^1\d_x - i\sigma^3\d_y \right) \chi^\alpha_I + m\, \chi^\dagger_{\alpha,I} \sigma^2 \chi^\alpha_I - \mu_\mathrm{rel}\, \chi^\dagger_{\alpha,I} \chi^\alpha_I\\
        &\qquad + \frac{1}{2} \Delta_{\alpha\beta,IJ}\, (\chi^T)^\alpha_I \sigma^2 \chi^\beta_J + \frac{1}{2} \Delta^{\alpha\beta}_{IJ}\, \chi^\dagger_{\alpha,I} \sigma^y \chi^*_{\beta,J}\, ,
    \end{split}
\end{equation}
where we have suppressed spinor indices, assumed that the components of the pairing matrix are always real, and adopted the Clifford algebra convention, {$\gamma^0 = \sigma^2, \gamma^1 = -i\sigma^3$, and $\gamma^2 = i\sigma^1$.} We have also introduced the relativistic chemical potential, $\mu_\mathrm{rel}$, which is defined to vanish at charge neutrality. We assume the Dirac fermions pair in $s$-wave, allowing us to drop momentum-dependent form factors.

The symmetry of the Hamiltonian is determined by the form of the order parameter. For the CVL phase, the residual symmetry is global, and is given by the diagonal $\SU(3)$ subgroup of $\SU(3)_\mathrm{color}\times\SU(3)_\mathrm{valley}$ when lattice translation symmetry is enhanced to a full $\SU(3)_\mathrm{valley}$ symmetry. Under this diagonal subgroup, the quarks transform by a simultaneous action of a unitary matrix $U$ in both color and valley space, $\chi^\alpha_I \rightarrow U^\alpha_\beta U_{IJ} \chi^\beta_J$.

For intra-valley, intra-color pairing, $\Delta_{\alpha\beta,IJ} = \Delta \delta_{\alpha\beta} \delta_{IJ}$, and the gauge symmetry is reduced to $\SO(3)$, while the global $\U(1)_\mathrm{EM}$ symmetry is Higgsed to $\mathbb{Z}_2$. Under the residual gauge symmetry, the quark transforms by an orthogonal, $3\times3$, unimodular matrix, $\chi^\alpha_I \rightarrow O^\alpha_\beta \chi^\beta_I$. For intra-valley, inter-color pairing, on the other hand, the order parameter can be diagonalized to $\Delta_{\alpha\beta,IJ} = \Delta M_{\alpha\beta} \delta_{IJ}$, where $M=\mathrm{diag}(2,-1,-1)$. This breaks the $\U(3)$ gauge symmetry to $\O(2)\times\mathbb{Z}_2^c$, which acts on the quarks as a block diagonal, $3\times3$ matrix $U=(\pm1,O_{2\times2})$, where $O_{2\times2}$ is a $2\times2$ orthogonal matrix.

Since the residual symmetry in both intra-valley pairing scenarios is an orthogonal group, it helps to decompose the Dirac quarks into Majorana quarks by writing $\chi^\alpha_I = \xi^\alpha_I + i\zeta^\alpha_I$. The difference between upper and lower color indices no longer matters, since the Majoranas transform in the vector representation of $\O(N)$, which is real. The BdG Hamiltonian then takes the form
\begin{equation}
    \begin{split}
        H &= \xi^T_{\alpha,I} \left( -i\sigma^1\d_x - i\sigma^3\d_y \right) \xi_{\alpha,I} + \left( m\, \delta_{\alpha\beta} \delta_{IJ} + \Delta_{\alpha\beta,IJ} \right) \xi^T_{\alpha,I} \sigma^2 \xi_{\beta,J}\\
        &\qquad + \zeta^T_{\alpha,I} \left( -i\sigma^1\d_x - i\sigma^3\d_y \right) \zeta_{\alpha,I} + \left( m\, \delta_{\alpha\beta} \delta_{IJ} - \Delta_{\alpha\beta,IJ} \right) \zeta^T_{\alpha,I} \sigma^2 \zeta_{\beta,J}\\
        &\qquad\qquad + i \mu_\mathrm{rel} \left( \xi^T_{\alpha,I} \zeta_{\alpha,I} - \zeta^T_{\alpha,I} \xi_{\alpha,I} \right)\, .
    \end{split}
\end{equation}

Let us first consider intra-color pairing, for which the residual gauge symmetry is $\SO(3)$. The Hamiltonian is
\begin{equation}
    \begin{split}
        H_{\SO(3)} &= \xi^T_{\alpha,I} \left( -i\sigma^1\d_x - i\sigma^3\d_y \right) \xi_{\alpha,I} + \left( m+\Delta \right) \xi^T_{\alpha,I} \sigma^2 \xi_{\alpha,I}\\
        &\qquad + \zeta^T_{\alpha,I} \left( -i\sigma^1\d_x - i\sigma^3\d_y \right) \zeta_{\alpha,I} + \left( m - \Delta \right) \zeta^T_{\alpha,I} \sigma^2 \zeta_{\alpha,I}\\
        &\qquad\qquad + i \mu_\mathrm{rel} \left( \xi^T_{\alpha,I} \zeta_{\alpha,I} - \zeta^T_{\alpha,I} \xi_{\alpha,I} \right)\, .\, .
    \end{split}
\end{equation}
The eigenvalues of this Hamiltonian are the dispersions
\begin{equation}
    \begin{split}
        \epsilon^{(+)}_\pm(q) &= \pm \left[ q^2 + m^2+\mu^2+\Delta^2 + 2 \left( m^2\Delta^2 + \mu^2(q^2+m^2) \right)^{1/2} \right]^{1/2}\, ,\\
        \epsilon^{(-)}_\pm(q) &= \pm \left[ q^2 + m^2+\mu^2+\Delta^2 + 2 \left( m^2\Delta^2 + \mu^2(q^2+m^2) \right)^{1/2} \right]^{1/2}\, ,
    \end{split}
\end{equation}
and we will refer to the corresponding eigenvectors as $\eta^{(\pm)}_{\alpha,I}$, which are Majorana fermions with effective masses
\begin{equation}
    m^{(+)} = m + \sqrt{\mu_\mathrm{rel}^2+\Delta^2}\, , \qquad m^{(-)} = m - \sqrt{\mu_\mathrm{rel}^2+\Delta^2}\, .
\end{equation}
The phase of the path integral, and hence the topological response of the superconductor, is determined by the signs of these masses. Assuming that $m,\mu_\mathrm{rel},$ and $\Delta$ are positive, the first mass, $m^{(+)}$, is always positive, whereas the second one, $m^{(-)}$, can change sign as $\mu_\mathrm{rel}$ increases. Specifically the transition occurs at $\mu_\mathrm{rel}^2 = m^2-\Delta^2$. In the nonrelativistic limit, where $m\rightarrow0$ with $\mu\equiv\mu_\mathrm{rel}-m$ and $\Delta$ fixed, this transition occurs at vanishing non-relativistic chemical potential, $\mu=0$, in agreement with Ref.~\cite{Read2000}. The phase where the two masses have opposite signs is then identified with weak pairing, whereas the phase where both masses are positive is strong pairing.

We couple to the ``background'' $\SO(3)$ gauge field $b'$, defined as
\begin{equation}
    b' = \begin{pmatrix}
        0 & b'_1 & b'_2\\
        -b'_1 & 0 & b'_3\\
        -b'_2 & -b'_3 & 0
    \end{pmatrix}\, ,
\end{equation}
by replacing partial derivatives in the Majorana Hamiltonian with covariant derivatives. Given our conventions, the Majorana path integrals take the following form:
\begin{equation}
    \begin{split}
        Z_{\mu>0} &= |Z| \exp \left( -3i \int \Lag_{\SO(3)_1}[b'] - 9i \int \CS_g \right)\, ,\\
        Z_{\mu<0} &= |Z|\, .\\
    \end{split}
\end{equation}
The first line corresponds to weak, $p-ip$ pairing in the non-relativistic limit, whereas the second line corresponds to strong pairing. The additional $-9\CS_g$ term in the first line encodes the nine Majorana edge modes of this phase. Later in this Appendix, we will demonstrate how to derive these by solving the BdG Hamiltonian on the edge.

Moving on to intra-valley, inter color pairing, the residual symmetry is $\O(2)$, and the BdG Hamiltonian is best written by separating the cyan quarks, $\xi_{c,I}, \zeta_{c,I}$ from the $\O(2)$ doublets of magenta and yellow quarks, $\xi_{\rho,I}, \zeta_{\rho,I}$,
\begin{equation}
    \begin{split}
        H_{\O(2)} &= \xi^T_{c,I} \left( -i\sigma^1\d_x - i\sigma^3\d_y \right) \xi_{c,I} + \left( m+2\Delta \right) \xi^T_{c,I} \sigma^2 \xi_{c,I}\\
        &\qquad+ \zeta^T_{c,I} \left( -i\sigma^1\d_x - i\sigma^3\d_y \right) \zeta_{c,I} + \left( m-2\Delta \right) \zeta^T_{c,I} \sigma^2 \zeta_{c,I}\\
        &\qquad\qquad + i \mu_\mathrm{rel} \left( \xi^T_{c,I} \zeta_{c,I} - \zeta^T_{c,I} \xi_{c,I} \right)\\
        &\qquad + \xi^T_{\rho,I} \left( -i\sigma^1\d_x - i\sigma^3\d_y \right) \xi_{\rho,I} + \left( m-\Delta \right) \xi^T_{\rho,I} \sigma^2 \xi_{\rho,I}\\
        &\qquad\qquad + \zeta^T_{\rho,I} \left( -i\sigma^1\d_x - i\sigma^3\d_y \right) \zeta_{\rho,I} + \left( m+\Delta \right) \zeta^T_{\rho,I} \sigma^2 \zeta_{\rho,I}\\
        &\qquad\qquad\qquad + i \mu_\mathrm{rel} \left( \xi^T_{\rho,I} \zeta_{\rho,I} - \zeta^T_{\rho,I} \xi_{\rho,I} \right)\, .
    \end{split}
\end{equation}
This Hamiltonian has more eiganvalues than the $\SO(3)$ Hamiltonian,
\begin{equation}
    \begin{split}
        \epsilon^{(c,+)}_\pm(q) &= \pm \left[ q^2 + m^2+\mu^2+(2\Delta)^2 + 2 \left( m^2(2\Delta)^2 + \mu^2(q^2+m^2) \right)^{1/2} \right]^{1/2}\, ,\\
        \epsilon^{(c,-)}_\pm(q) &= \pm \left[ q^2 + m^2+\mu^2+(2\Delta)^2 - 2 \left( m^2(2\Delta)^2 + \mu^2(q^2+m^2) \right)^{1/2} \right]^{1/2}\, ,\\
        \epsilon^{(my,+)}_\pm(q) &= \pm \left[ q^2 + m^2+\mu^2+\Delta^2 + 2 \left( m^2\Delta^2 + \mu^2(q^2+m^2) \right)^{1/2} \right]^{1/2}\, ,\\
        \epsilon^{(my,-)}_\pm(q) &= \pm \left[ q^2 + m^2+\mu^2+\Delta^2 - 2 \left( m^2\Delta^2 + \mu^2(q^2+m^2) \right)^{1/2} \right]^{1/2}\, ,\\
    \end{split}
\end{equation}
where the first superscript denotes the dispersions for the cyan, $(c)$, component or the magenta-yellow doublet, $(my)$. The eigenvectors with these dispersions are Majoranas $\eta^{(c,+)}_I, \eta^{(c,-)}_I, \eta^{(my,+)}_{\rho,I}$, and $\eta^{(my,-)}_{\rho,I}$. The masses of these Majoranas are given by
\begin{equation}
    m^{(c,\pm)} = m \pm \sqrt{\mu_\mathrm{rel}^2 + (2\Delta)^2}\, , \qquad m^{(my,\pm)} = m \pm \sqrt{\mu_\mathrm{rel}^2 + \Delta^2}\, .
\end{equation}
Once again, phase transitions occur when any one of the Majorana masses changes sign. Assuming $m,\mu_\mathrm{rel}$ and $\Delta$ to be positive, two of the four masses, $m^{(c,+)}$ and $m^{(my,+)}$, are always positive and the corresponding Majorana eigenvectors do not contribute to the path integral. The other two masses can change sign as $\mu_\mathrm{rel}$ decreases, with $m^{(my,-)}$ becoming positive first, followed by $m^{(c,-)}$. In the strict non-relativistic limit where $\mu=\mu_\mathrm{rel}-m$ and $\Delta$ are finite but $m=\infty$, both transitions merge into a single one at $\mu=0$.

We couple to a background $\O(2)$ gauge field $(a,\widetilde{b}^m)$, where $a$ is $\SO(2)$ and $\widetilde{b}^m$ is $\mathbb{Z}_2$, and $a$ couples to only the $\O(2)$ doublet through the matrix
\begin{equation}
    \begin{pmatrix}
        0 & 0 & 0\\
        0 & 0 & a\\
        0 & -a & 0
    \end{pmatrix}\, ,
\end{equation}
and $\widetilde{b}^m$ couples only to the magenta quarks. The cyan quarks do not couple to any gauge field, and integrating them out only generates gravitational Chern-Simons terms. The path integrals are finally given by
\begin{equation}
    \begin{split}
        Z_{\mu>0} &= |Z| \exp \left( -3i \int \Lag_{\O(2)_{1,1}}[a,\widetilde{b}^m] - 9i \int \CS_g \right)\, ,\\
        Z_{\mu<0} &= |Z|\, .\\
    \end{split}
\end{equation}
The top line corresponds to weak $p-ip$ pairing of all quarks, while the bottom one corresponds to strong pairing. The $-9\CS_g$ term in the top line counts the Majorana edge modes of the superconductor.


\subsection{Solving the BdG Hamiltonian on the edge}

For the edge modes, our strategy will be to solve the BdG Hamiltonian on an annular geometry with inner radius $R$ and an $h/2e$ vortex threaded through the center. We will return to the non-relativistic theory and focus only on zero energy solutions, since we are primarily interested in superconducting defects and the contribution of the Majorana edge modes to the chiral central charge, for which we only need to know the number of zero modes. For a full derivation of the higher energy spectrum of edge modes, we refer the reader to Ref.~\cite{Alicea2012}.

In all three color superconductors, the order parameter is a specific choice of the matrix $\Delta_{\alpha\beta,IJ}=\langle \chi^\dagger_{\alpha,I}(\d_x\mp i\d_y) \chi^\dagger_{\beta,J} \rangle$, where the upper sign corresponds to $p+ip$ pairing and the lower sign to $p-ip$ pairing. For simplicity we will focus on $p+ip$ pairing. The calculation is identical for $p-ip$ pairing, and the number of zero modes doesn't change. The Hamiltonian can be written as
\begin{equation}
    H = \int_x \chi^\dagger_{\alpha,I} \left( -\frac{\nabla^2}{2m} - \mu \right) \chi^\alpha_I + \frac{1}{2} \left[ \Delta_{\alpha\beta,IJ} \chi^\alpha_I (\d_x + i\d_y) \chi^\beta_J + \mathrm{h.c.} \right]\, .
\end{equation}
Working in polar coordinates, $(r,\theta)$, we will model the annulus by an isotropic, radially varying chemical potential $\mu(r)$ such that $\mu(r) < 0$ when $r<R$. Threading an $h/2e$ vortex through the center is achieved by replacing $\Delta_{\alpha\beta,IJ} \rightarrow \Delta_{\alpha\beta,IJ} e^{-i\theta}$. Focusing on the edge and assuming that the chemical potential varies slowly, we can drop the gradient term in the Hamiltonian and focus only on low energy modes, turning the edge Hamiltonian into
\begin{equation}
    H_\mathrm{edge} = \int_x - \mu(r) \chi^\dagger_{\alpha,I} \chi^\alpha_I + \frac{1}{2} \left[ \Delta_{\alpha\beta,IJ} \chi^\alpha_I \left( \d_r + \frac{i}{r}\d_\theta \right) \chi^\beta_J + \mathrm{h.c.} \right]\, .
\end{equation}
We can write this equivalently in terms of a two-component spinor $\Psi^\alpha_I = (\chi^\alpha_I, \chi^\dagger_{\alpha,I})$ to put the Hamiltonian in a BdG form
\begin{equation}
    H_\mathrm{edge} = \frac{1}{2} \int_x \Psi^\dagger_{\alpha,I} \mathcal{H}^\alpha_{\beta,IJ} \Psi^\beta_J\, , \qquad \mathcal{H}^\alpha_{\beta,IJ} = \begin{pmatrix}
        -\mu(r) \delta^\alpha_\beta \delta_{IJ} & (\Delta^*)^{\alpha\beta}_{IJ} \left( -\d_r + \frac{i}{r} \d_\theta \right)\\
        \Delta_{\alpha\beta,IJ} \left( \d_r + \frac{i}{r} \d_\theta \right) & \mu(r) \delta^\alpha_\beta \delta_{IJ}
    \end{pmatrix}\, ,
\end{equation}
where $\Delta^*$ is the complex conjugate of $\Delta$. We will also assume that the order parameter has real components. Edge modes correspond to eigenfunctions of $\mathcal{H}$ that are exponentially localized at the boundary, and we use the following ansatz for the solutions:
\begin{equation}
    \xi^\alpha_{I,n}(r,\theta) = e^{in\theta} \begin{pmatrix}
        f^\alpha_I(r) + i g^\alpha_I(r)\\
        f_{\alpha,I}(r) - i g_{\alpha,I}(r)
    \end{pmatrix}\, .
\end{equation}
In the above, $f$ and $g$ are real functions of the radial coordinate. Since the fundamental and anti-fundamental representations of $\SU(3)$ differ only by complex conjugation, the difference between upper and lower indices on $f$ and $g$ does not matter any longer. Furthermore, preiodicity of wavefunctions requires $n$ to be an integer. The eigenvalue equation for $\mathcal{H}$ reduces to the following pair of equations
\begin{equation}
    \begin{split}
        [E+\mu(r)] e^{in\theta}(f^\alpha_I+ig^\alpha_I) &= \Delta^{\alpha\beta}_{IJ} \left( -\d_r + \frac{n}{r} \right) (f^\beta_J - i g^\beta_J)\, ,\\
        [E-\mu(r)] e^{in\theta}(f_{\alpha,I}-ig_{\alpha,I}) &= \Delta_{\alpha\beta,IJ} \left( \d_r + \frac{n}{r} \right) (f_{\beta,J} - i g_{\beta,J})\, .
    \end{split}
\end{equation}
Zero modes correspond to $E=n=0$, and the equations simplify to
\begin{equation}\label{eq:reduced_edge_mode_eqns}
    \begin{split}
        \left[ \mu(r) \delta^\alpha_\beta \delta_{IJ} + \Delta^{\alpha\beta}_{IJ} \d_r \right] f^\beta_J &= 0\, ,\\
        \left[ \mu(r) \delta^\alpha_\beta \delta_{IJ} - \Delta^{\alpha\beta}_{IJ} \d_r \right] g^\beta_J &= 0\, .
    \end{split}
\end{equation}


\subsubsection{Color-valley-locking}

For color-valley-locking, the order parameter is a contraction of Levi-Civita symbols which can be further simplified to
\begin{equation}
    \Delta_{\alpha\beta,IJ} = \frac{\Delta}{6}\sum_{n=1}^3 \varepsilon_{\alpha\beta n} \varepsilon_{IJn} = \frac{\Delta}{6} \left( \delta_{\alpha I} \delta_{\beta J} - \delta_{\alpha J} \delta_{\beta I} \right)\, .
\end{equation}
Requiring the edge modes to be localized forces either $g=0$ for the inner edge or $f=0$ on the outer edge. Let us focus on the inner edge and separate the radial dependence of the wavefunction. Two classes of solutions exist: those with vanishing $g_{\alpha,I}$ and those with vanishing $f_{\alpha,I}$,
\begin{equation}
    \begin{split}
        f_{\alpha,I} &= C_{\alpha,I} \exp \left( -\frac{p}{\Delta} \int_R^r dr'\, \mu(r') \right)\, , \qquad g_{\alpha,I} = 0\, ,\\
        g_{\alpha,I} &= C'_{\alpha,I} \exp \left( -\frac{q}{\Delta} \int_R^r dr'\, \mu(r') \right)\, , \qquad f_{\alpha,I} = 0\, ,
    \end{split}
\end{equation}
where $C$ and $C'$ are real matrices, and $p$ and $q$ are positive real numbers. Substituting this form into Eq.~\eqref{eq:reduced_edge_mode_eqns}, we find the following solutions:
\begin{equation}
    \begin{split}
        C_{\alpha,I} \propto \delta_{\alpha,I}\, , \qquad p &= 3\, ,\\
        C_{\alpha,I} = - C_{I,\alpha}\, , \qquad p &= 6\, ,\\
        C'_{\alpha,I} = C'_{I,\alpha}\, , \quad \sum_\alpha C_{\alpha,\alpha = 0}\, , \qquad q &= 6\, .
    \end{split}
\end{equation}
The first class of solutions has one independent real component -- the trace of the matrix $C$. The second class of solutions is parameterized by the three independent real numbers. The third class of solutions has five independent real components. The total number of independent real numbers characterizing the solutions, which is the total number of Majorana zero modes, is hence nine. Since quark pairing occurs in the $p+ip$ channel, these zero modes contribute $+9/2$ to the chiral central charge.


\subsubsection{Intra-valley, intra-color pairing}

In this case, the order parameter is a product of Kronecker deltas, $\Delta_{\alpha\beta,ij} = \Delta \delta_{\alpha\beta}\delta_{ij}$, turning Eq.~\eqref{eq:reduced_edge_mode_eqns} into
\begin{equation}
    \begin{split}
        \left[ \mu(r) + \Delta \d_r \right] f_{\alpha,I} &= 0\, \\
        \left[ \mu(r) - \Delta \d_r \right] g_{\alpha,I} &= 0\, .
    \end{split}
\end{equation}
Exponential localization on the inner edge requires $g=0$, and the solution is given by
\begin{equation}
    f_{\alpha,I} = C_{\alpha,I} \exp \left( -\frac{1}{\Delta} \int_R^r dr'\, \mu(r') \right)\, ,
\end{equation}
with no additional contraints on the matrix $C$, which has 9 independent, real components. We hence have $9$ Majorana modes. The quark pairing channel for this phase is $p-ip$ instead of $p+ip$, so the Majorana modes contribute $-9/2$ to the chiral central charge.

Unlike the CVL phase, the Majorana operators, $\gamma_{\alpha,I}$ carry independent color and valley indices. This phase has residual $\SO(3)$ gauge symmetry, and the color index descends to an index in the vector representation of $\SO(3)$, so that each valley has a Majorana triplet that transforms under $\SO(3)$. The valley index makes the Majorana zero modes charged under lattice translations, which act as they would on a single valley index -- by permutations or valley dependent phases.


\subsubsection{Intra-valley, inter-color pairing}

Lastly, for this phase the order parameter is a symmetric, off-diagonal matrix in color indices, $\Delta_{\alpha\beta,IJ} = \Delta M_{\alpha\beta} \delta_{IJ}$, where
\begin{equation}
    M_{\alpha\beta} = \begin{pmatrix}
        0 & 1 & 1\\
        1 & 0 & 1\\
        1 & 1 & 0
    \end{pmatrix}\, .
\end{equation}
Eq.~\eqref{eq:reduced_edge_mode_eqns} then reduces to
\begin{equation}
    \begin{split}
        \left[ \mu(r)\delta_{\alpha\beta} + \Delta M_{\alpha\beta} \d_r \right] f^\beta_I &= 0\, ,\\
        \left[ \mu(r)\delta_{\alpha\beta} - \Delta M_{\alpha\beta} \d_r \right] g^\beta_I &= 0\, .
    \end{split}
\end{equation}
To solve this, we can diagonalize $M_{\alpha\beta}$ by an orthogonal transformation to $P=\mathrm{diag}(2,-1,-1)$. The equations for the yellow component look slightly different from those for the cyan and magenta components:
\begin{equation}
    \begin{split}
        \left[ \mu(r) - \Delta \d_r \right] f^\alpha_I = 0\, , \qquad \left[ \mu(r) + 2\Delta \d_r \right] f^c_I &= 0\, ,\\
        \left[ \mu(r) + \Delta \d_r \right] g^\alpha_I = 0\, , \qquad \left[ \mu(r) - 2\Delta \d_r \right] g^c_I &= 0\, ,
    \end{split}
\end{equation}
where $\alpha = m,y$ only. Requiring the solution to be localized on the inner edge implies that $f_{\alpha,I} = 0$ when $\alpha = c,m$, and $g_{c,I} = 0$. The wavefunctions are given by
\begin{equation}
    \begin{split}
        g_{\alpha,I} &= C_{\alpha,I} \exp \left( -\frac{1}{\Delta} \int_R^r dr'\, \mu(r') \right)\, , \qquad \alpha = m,y\, ,\\
        f_{c,I} &= C_I \exp \left( -\frac{1}{2\Delta} \int_R^r dr'\, \mu(r') \right)\, .
    \end{split}
\end{equation}
These are also parameterized by 9 independent, real components -- 6 from the $2\times3$ matrix $C_{\alpha,I}$ and 3 from $C_I$, resulting in 9 Majorana edge modes. Since the quark pairing channel is $p-ip$ for this phase as well, they contribute $-9/2$ to the chiral central charge.

Since the residual symmetry is $\O(2)\times\mathbb{Z}_2$, the Majoranas split into three $\O(2)$ singlets, $\gamma_I$, which are charged under the $\mathbb{Z}_2$ factor, and three $\O(2)$ vectors, $\gamma_{\alpha,I}$, $\alpha = m,y$, which are neutral under the $\mathbb{Z}_2$ factor. Each of these carry a single valley index, and are hence permuted or phase-shifted by lattice translations.


\nocite{apsrev41Control}
\bibliographystyle{apsrev4-1}
\bibliography{merged}
\end{document}